\documentclass[aip,reprint,floatfix]{revtex4-1}

\usepackage{graphicx}
\usepackage{color}
\usepackage{booktabs}
\usepackage{sidecap}
\usepackage{multirow}
\usepackage{array}
\usepackage{setspace}

\usepackage{subfig}
\usepackage{tabularx}

\usepackage{graphics}
\usepackage{hyperref}
\usepackage{amssymb}

\usepackage{lineno}

\sidecaptionvpos{figure}{c}

\newcommand{\beginsupplement}{%
	\setcounter{table}{0}
	\renewcommand{\thetable}{S\arabic{table}}%
	\setcounter{figure}{0}
	\renewcommand{\thefigure}{S\arabic{figure}}%
}
\draft

\begin{document}

\title{Sub-national levels and trends in contraceptive prevalence, unmet need, and demand for family planning in Nigeria with survey uncertainty} 

\author{Laina D. Mercer}%
\affiliation{Institute for Disease Modeling, Bellevue, WA}

\author{Fred Lu}%
\affiliation{Stanford University, Department of Statistics, Stanford, CA}

\author{Joshua L. Proctor}%
\email{JProctor@idmod.org}
\affiliation{Institute for Disease Modeling, Bellevue, WA}

\date{\today}

\begin{abstract}

Ambitious global goals have been established to provide universal access to affordable modern contraceptive methods. The UN's sustainable development goal 3.7.1 proposes satisfying the demand for family planning (FP) services by increasing the proportion of women of reproductive age using modern methods. To measure progress toward such goals in populous countries like Nigeria, it's essential to characterize the current levels and trends of FP indicators such as unmet need and modern contraceptive prevalence rates (mCPR). Moreover, the substantial heterogeneity across Nigeria and scale of programmatic implementation requires a sub-national resolution of these FP indicators. However, significant challenges face estimating  FP indicators sub-nationally in Nigeria. In this article, we develop a robust, data-driven model to utilize all available surveys to estimate the levels and trends of FP indicators in Nigerian states for all women and by age-parity demographic subgroups. We estimate that overall rates and trends of mCPR and unmet need have remained low in Nigeria:  the average annual rate of change for mCPR by state is 0.5\% (0.4\%,0.6\%) from 2012-2017. Unmet need by age-parity demographic groups varied significantly across Nigeria; parous women express much higher rates of unmet need than nulliparous women.  Our hierarchical Bayesian model incorporates data from a diverse set of survey instruments, accounts for survey uncertainty, leverages spatio-temporal smoothing, and produces probabilistic estimates with uncertainty intervals. Our flexible modeling framework directly informs programmatic decision-making by identifying age-parity-state subgroups with large rates of unmet need, highlights conflicting trends across survey instruments, and holistically interprets direct survey estimates.

\end{abstract}

\maketitle

\section{Introduction}
\label{S:1}

International support for improving access to family planning (FP) services has had a significant resurgence in the past decade.  
Ambitious, world-wide goals have been constructed by coalitions of governmental and non-governmental agencies.  
One such goal, colloquially referred to as {\it 120 by 20}, aims to increase access to modern contraceptives for $120$ million more women by $2020$;  architects of {\it 120 by 20} included principals from the Bill and Melinda Gates Foundation (BMGF), the United Kingdom Department for International Development, the United States Agency for International Development (USAID), and the United Nations Population Fund~\cite{brown2014developing}.  
Identified at the inception of this goal, a major barrier to achieving {\it 120 by 20} was the determination of baseline of modern contraceptive prevalence rates (mCPR) in 2012 and ability to track yearly progress~\cite{brown2014developing}.
Similarly, the more recently developed sustainable development goal (SDG) 3.7.1~\cite{sdgsWeb} of increasing demand satisfied, defined as the ratio of mCPR to total contraceptive prevalence and unmet need, requires reliable estimates of FP indicators.
To meet those challenges, substantial investments in new measurement instruments~\cite{PMA2020} and novel model-based estimate methodologies~\cite{alkema2013national,cahill2017modern} enabled the establishment of a baseline and a methodology for estimating yearly national progress.

Unfortunately, the progress toward {\it 120 by 20} is currently falling short with only $38.8$ million estimated additional users since 2012~\cite{Track20Progress}.
With few exceptions~\cite{cavallaro2016taking}, national mCPR rates have not increased nor accelerated as planned~\cite{brown2014developing}, especially in priority settings such as Nigeria which have not observed substantial national increases in mCPR despite explicit country goals to increase by more than $1.5\%$ per year~\cite{NigeriaProgress}. 
To enable within country decision-making around increasing mCPR to meet the current needs of women, a finer-scale approach to evaluating progress and the impact of FP programs is essential.
%
%
%
%
Achieving global goals, whether {\it 120 by 20} or SDG 3.7.1, will require continued international commitment with specific, targeted family planning programs.  
In this article, we develop a model-based estimation approach to characterize FP indicators at a {\it sub-national scale}.
Here, we focus on Nigeria to demonstrate how our Bayesian hierarchical model integrates all available cross-sectional survey data from a diverse set of measurement instruments and incorporates measurement uncertainties to produce state-level estimates and uncertainty intervals for FP indicators.

Previous analyses and model-based estimates by Alkema et al.~\cite{alkema2013national} have focused on national level estimates of mCPR and unmet need.  
Their innovative modeling methodology is the foundation of the family planning estimation tool (FPET)~\cite{cahill2017modern}, which is being used to track progress toward the {\it 120 by 20} goals~\cite{Track20Progress}.
FPET has also been utilized to estimate FP indicators for 29 states and union territories in India, where high-sample, cross-sectional surveys have been performed~\cite{new2017levels}. 
The model in \cite{alkema2013national} assumes a parametric form of a logistic growth curve for mCPR increases;  the assumption is founded in the theory of social diffusion of ideas for family planning services~\cite{lin1974diffusion} and has enabled long-term forecasts of FP indicators matching population growth estimates in the next century~\cite{kantorova2017setting}.  
For sub-national estimates, strong parametric assumptions, lower density of measurements, and higher in-sample uncertainty is a challenge for the model in \cite{alkema2013national} to produce near-term sub-national forecasts.

%
Our methodology combines survey and spatial statistics for FP indicators, and is mathematically related to small area estimation (SAE) techniques used in estimating sub-national childhood mortality from multiple household surveys and health and demographic surveillance system sites~\cite{mercer2015}. 
Here, the underlying Bayesian hierarchical model inherits several important aspects of modern estimation methodologies: uncertainty is quantified for each estimate~\cite{mercer:etal:14}, data is integrated from multiple surveys and designs~\cite{mercer2015}, and spatio-temporal random effects are included in the model~\cite{schrodleheld:2011}.
SAE techniques have been previously implemented for FP indicators restricted to data collected from Performance Monitoring and Accountability 2020 (PMA2020) surveys for ten low-income countries~\cite{SAEJH}.
%
However, the PMA2020 surveys were only performed in two to seven states in Nigeria depending on the survey round;  the modeling approach in this paper is substantially broader incorporating multiple surveys (the Demographic and Health Surveys (DHS), Multiple Indicator Cluster Surveys (MICS), National Nutrition and Health Surveys (NNHS), and PMA2020), including survey specific effects in the model, and producing estimates and uncertainty intervals for all 36 states and the Federal Capital Territory.
Moreover, the underlying model can be easily generalized to other FP priority countries.
Using this methodology, we produce state-level estimates of FP indicators mCPR, traditional contraceptive rates, unmet need, and demand satisfied for Nigeria revealing substantial heterogeneity by state, age, and parity.  

Significant international resources and investments are being directed to Nigeria to meet the current demand for family planning services.
For example, external donors, such as BMGF and USAID, fund non-governmental groups like the Nigerian Urban Reproductive Health Initiative (NURHI) to implement and support information, education, and communication programs and supply chain improvements.
Moreover, donors in collaboration with the Nigerian government are helping rollout new contraceptive technologies such as the new self-injection Depo-Medroxyprogesterone Acetate - subcutaneous (DMPA SC), popularly referred to as Sayana Press.
These FP programs and contraceptive technologies are largely implemented at fine spatial resolutions.
%
%
Our modeled estimates and uncertainty intervals provide specific feedback to country programs:  new survey data (i.e., the upcoming 2018 DHS) can be contextualized across all surveys avoiding mis-interpretation of uncertain estimates that may na\"{\i}vely suggest large increases or decreases;  estimates and uncertainty intervals offer a basis to interpret the expected effect of investments and programs over time;  and analysis of age-parity subgroups helps characterize the demand for services.  

The article is organized as follows.
The first section describes the input data and model.  
The subsequent section outlines the underlying rates of mCPR, traditional contraceptives rates, and unmet need. 
We illustrate the need for model-based estimates in this context with an analysis of how different survey instruments by themselves can produce substantially different narratives.
The model is also adapted to analyze rates by age-parity subgroups.  
We conclude with a discussion about the implications of these results.

\begin{figure*}
	\centering
	\includegraphics[width=0.95\textwidth]{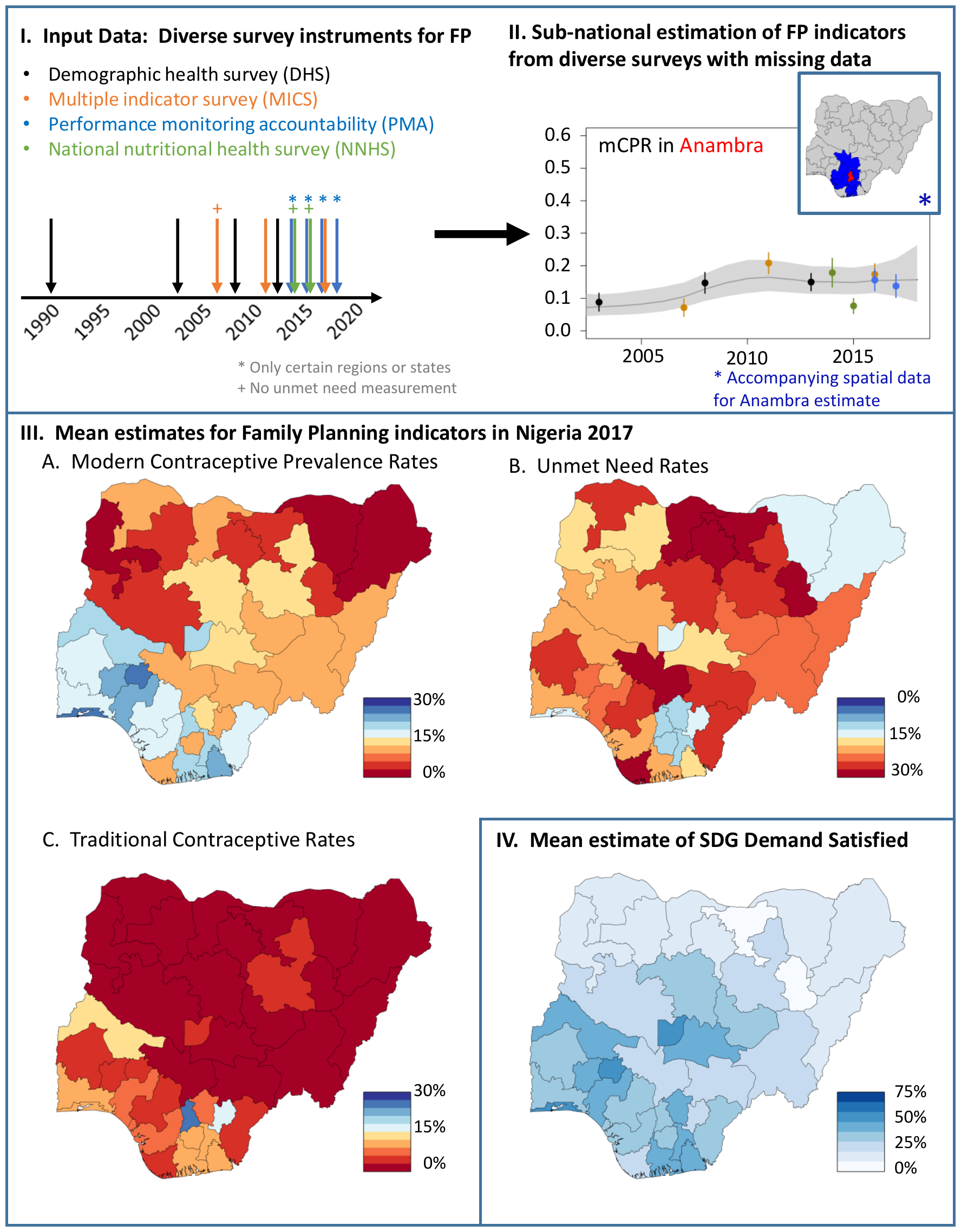}
	\caption{Description of data included (I), example of direct and model-based estimates of mCPR in Anambra state (II), maps of posterior median mCPR, unmet need, and traditional contraceptive prevalence rate for 2017 (III), and posterior median of demand satisfied for 2017 (IV).}
	\label{fig:Model}
\end{figure*}

\section{Data and methods}
\label{S:DataMethods}

We outline the available data and methods for analysis in this section.  
%
%
All code is publicly available in a github repository~\cite{MercerProctor2018}.

\subsection{Cross-sectional survey data}
%

Data comes from four DHS, three MICS, and two NNHS, each of which are nationally representative household surveys based on multi-stage cluster survey design.  
The four PMA2020 rounds which include two to seven states were also integrated into the analysis.  
We utilize all available DHS, MICS, NNHS, and PMA2020 surveys from Nigeria; see Appendix 1 for specific details about the survey, year, and complex design.
Fig.~\ref{fig:Model}(I) illustrates the year each survey was performed.  
For each survey, we extract individual level data describing modern or traditional contraceptive usage status, unmet need for FP, demographic information such as age and parity, survey design variables, and location information such as state.  
%
%
Appendix 1 outlines inconsistencies and unavailable data across surveys, i.e., unmet need cannot be computed from specific MICS and NNHS years.  
%

\subsection{Analyzing rates of contraceptive usage and unmet need}

Traditional and modern contraceptive prevalence rates (mCPR) are measured by computing the percentage of women who report themselves or partners as using at least one contraceptive method.  
Unmet need for family planning is measured by computing the percentage of women who do not want to become pregnant, either for ending childbearing or delaying the next pregnancy, and are not using a contraceptive method.
Here, we use the revised definition of unmet need~\cite{bradley2012revising}.
FP indicators are computed at the state level with respect to all women as well as four different demographic subgroups:  nulliparous women $15-24$, parous women $15-24$, nulliparous women $25$ years or older, and parous women $25$ years or older.  
%
%
%
All surveys included in our analyses provide state level estimates of the mCPR, but point estimates from different survey instruments show conflicting trends and a smoothing approach is necessary for jointly interpreting all available data.
For example, Fig.~\ref{fig:Model}(II) illustrates the mean estimate and confidence interval of mCPR for each survey in Anambra state for all women.   
%
%

\subsection{Model for sub-national estimation of family planning indicators}
%
Our statistical model assumes direct survey estimates are measurements with associated uncertainty and aims to estimate the underlying rates of FP indicators by state in Nigeria. 
This SAE modeling methodology integrates complex survey designs~\cite{lumley:10} and spatial statistics~\cite{schrodleheld:2011}.
Moreover, we incorporate a diverse set of survey instruments.
The result is a flexible, data-driven model structure that estimates the underlying rates of FP indicators in Nigeria in between surveys and in the presence of conflicting observations.
%


Given the detailed and diverse spatio-temporal information contained in the different cross-sectional survey instruments, we have developed a unifying framework to characterize the underlying sub-national trends of FP indicators.
Complex statistical models are required to simultaneously treat temporal trends, spatial effects, time-space interactions, survey-specific effects, and survey designs. 
Our framework is fundamentally a Bayesian hierarchical model allowing for a spatio-temporal smoothing which explicitly integrates multiple surveys, survey designs, and survey uncertainty.  
Some surveys do not include the variables required to estimate unmet need nor classify women into demographic subgroups, thus we have chosen to model FP indicators and age-parity subgroups independently.

The model estimates the levels and trends of FP indicators and enables short-term forecasts.
Data-driven model structures are proposed, fit to the data, and evaluated in a model selection procedure.  
%
In the first stage, the model assumes the pseudo-likelihood defined by the asymptotic distribution of the logit transformed design-based estimates and appropriately transformed design-based variance.
At the second stage, the logit transformation of the true FP indicator rates are modeled linearly as a function of independent random effects for short-term fluctuations, a random walk of order $2$ to capture national temporal trends, and temporally structured space-time interaction to account for sub-national temporal trends~\cite{schrodleheld:2011}. 
We also utilize independent and spatially structured random effects to account for effects by sub-national area and provide geographical smoothing~\cite{besag:etal:91,riebler2016intuitive}. 
The latter enables data and trends from physically adjacent states to support local estimates. 
The inset map of Fig.~\ref{fig:Model}(II) illustrates the location of Anambra in Nigeria (red) and the states where data is utilized in the model to support estimates for Anambra (blue).   
Survey type, survey-time, and survey-space random effects are also considered to account for systematic trends or biases across measurement instruments, time and space.
We report medians and 95\% credible intervals from the posterior distribution.  
See Appendix 2 for additional details on the random effects and hyperparameter priors.

\begin{figure*}[t]
	\centering
	\includegraphics[width=1.0\textwidth]{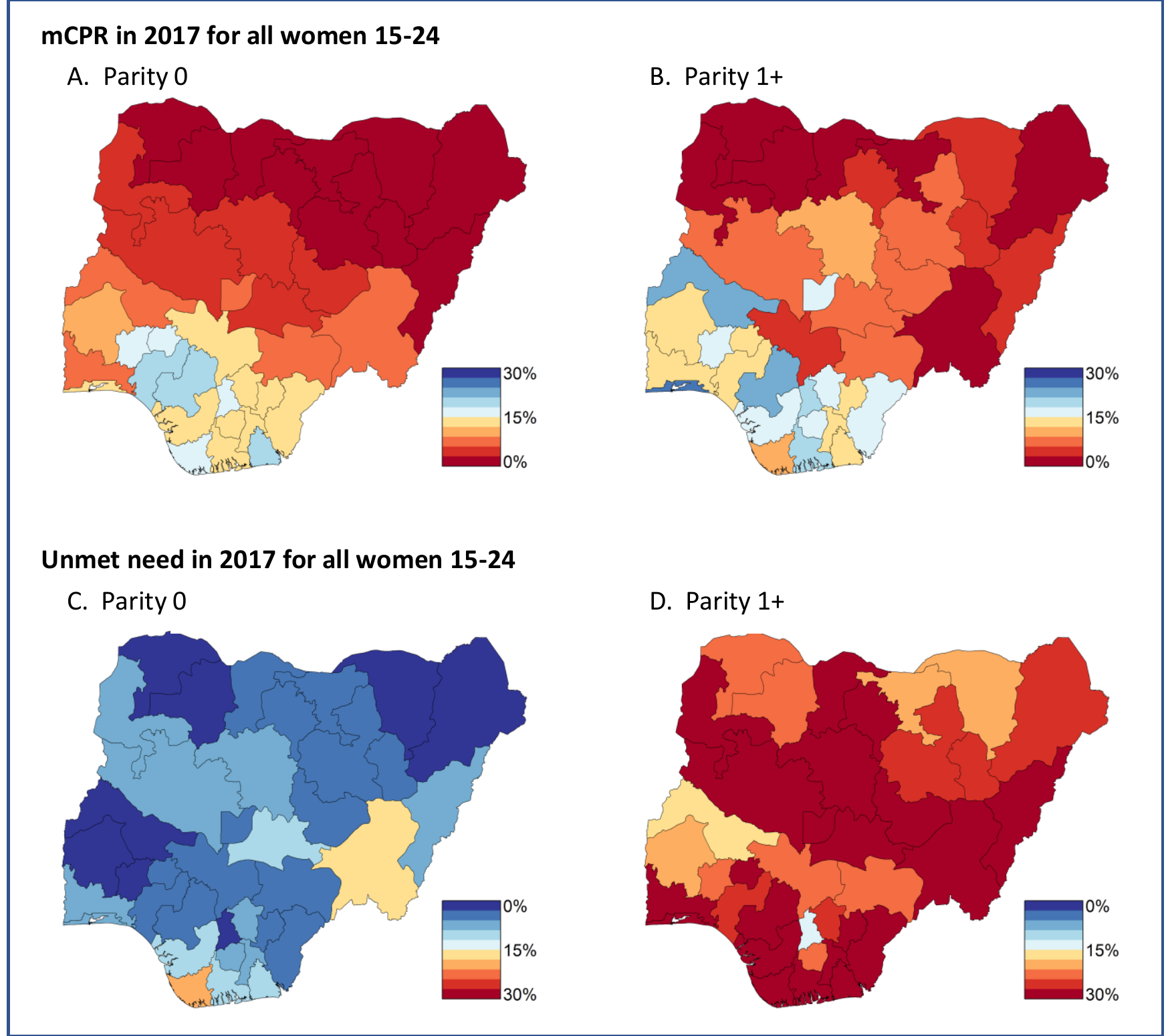}
	\caption{Posterior median estimates of mCPR (top) and unmet need (bottom) by demographic subgroups.}
	\label{fig:ParityAgeMaps}
\end{figure*}

\subsection{Fitting the model to data}

Models were fit using the R scientific computing language~\cite{citeR}.
Direct estimates and design-based variances were calculated using the survey package~\cite{package:survey:current}.  
Our hierarchical Bayesian space-time model is fit using the Integrated Nested Laplace Approximation (INLA)~\cite{rue2009approximate} as implemented in the INLA package in R~\cite{INLA:package}.  
INLA provides a fast, efficient, and accurate alternative to Markov Chain Monte Carlo (MCMC) and has been shown that the approximation is accurate for spatio-temporal smoothing models~\cite{fong2010bayesian,held2010posterior}.  
Code for all analyses is available at the github repository~\cite{MercerProctor2018}.


\subsection{Model selection procedure to evaluate}
\label{ss:modelselect}
We provide a detailed analysis of model structures including evaluating various random effects through principled model selection techniques.  
Three model selection procedures are considered:  the sum of the log conditional predictive ordinate (LCPO)~\cite{held2010posterior}, the deviance information criteria (DIC)~\cite{spiegelhalter2014deviance}, and the Watanabe-Akaike information criterion (WAIC)~\cite{watanabe2010asymptotic}.  
Each model selection criteria aims to evaluate goodness of fit and model complexity; we compute each measure since there has not been consensus on a single criteria~\cite{mercer2015} and DIC has been shown to under-penalize large models similar to ours~\cite{plummer2008penalized}.
Model selection criteria generally agree, but when in conflict final model selection relied on WAIC. 
The model selection procedure helps reveal the most parsimonious model structure while maintaining low model fitting error.
Appendix 3 illustrates each of the proposed models and their relative LCPO, DIC, and WAIC scores.  
%


%
\section{Results}
\label{s:Results}

\subsection{A Bayesian hierarchical model enables state-level estimates of primary FP indicators}

%
%
Posterior median estimates of the FP indicators for each Nigerian state are shown in Fig.~\ref{fig:Model}(III). 
Northern states in Nigeria have a much lower mCPR, less than 10\%, compared with many of the southern states which exceed 15\%.  
Rates of traditional contraceptive usage are low across most Nigerian states with notable exceptions such as Ebonyi.
State level estimates of unmet need are more highly variable across Nigeria.  
For example, states, such as Katsina (northern), Cross River (southern), and Niger (western), have rates of unmet need exceeding 20\%.  
Fig.~\ref{fig:Model}(IV) illustrates the SDG 3.7.1 indicator demand satisfied evaluated for each Nigerian state, combining mCPR, unmet need, and traditional contraceptive rates.
With the exception of Kaduna and Plateau, most northern states have rates of demand satisfied below 20\%.
The southern regions have higher rates of demand satisfied, but still do not surpass 75\%.

The Bayesian hierarchical model produces accompanying uncertainty intervals for each mean estimate.  
Figure~\ref{fig:Model}(II) illustrates the modeled uncertainty for the estimates in the state Anambra.  
The posterior median and $95\%$ credible interval (shaded region) show the difference between the modeled estimates with the different survey instruments.  
Moreover, the model enables projection of median estimates with corresponding uncertainty intervals past the most recent survey instrument.  
A detailed description of the selected models for each indicator is summarized in Appendix 3.
See Appendix 5 for visualizations of each PMA2020 state's temporal history of survey instruments, modeled estimates, and projections to 2018. 
 \begin{figure*}[t]
	\centering
	\includegraphics[width=1.0\textwidth]{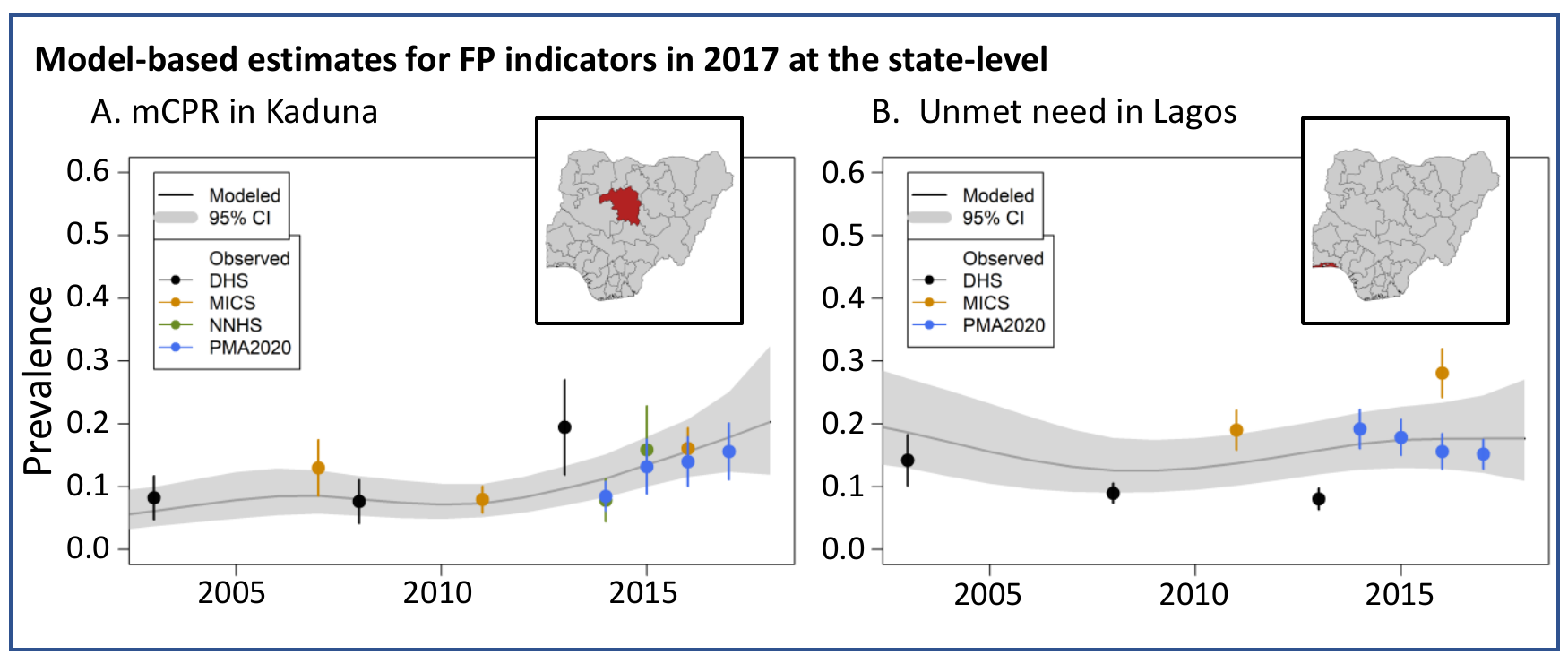}
	\caption{Direct and model-based estimates of mCPR in Kaduna (A) and unmet need in Lagos (B).}
	\label{fig:StateTrends}
\end{figure*}

\subsection{MCPR and unmet need varies by age, parity, and state}

State level estimates of unmet need vary significantly between nulliparous and parous women.
Fig.~\ref{fig:ParityAgeMaps} illustrates uniformly low rates of unmet need for nulliparous women in the $15-24$ age cohort.
For parous women, however, rates of unmet need vary significantly across the country;  for example, Ebonyi, Benue, Cross River, and Bayelsa have rates of unmet need near $30\%$.  
The northern states of Nigeria broadly have low rates of mCPR regardless of parity, however, rates of unmet need for parous women are relatively high in the north.
Qualitatively, patterns of unmet need and mCPR by state, age, and parity are similar for women older than $25$; see Appendix 6 for details.
Quantitatively, though, rates of unmet need for parous, older women are uniformly higher than the younger cohort across Nigerian states.
%
Maps of posterior medians of FP indicators by state, age, and parity can be found in Appendix 6.

\subsection{Differences across survey instruments in Nigeria}
The percent of the overall variation that is described by each set of random effects, defined as the variance divided by the sum of the variances, provides a description of the relative importance of each random effect.  
Between 70\% and 99\% of variation is described by the spatial and space-survey terms for nearly all outcomes and demographic subgroups, suggesting substantial spatial heterogeneity in overall rates and survey measurements; see Appendix 4 for more details.
The model for unmet need is an exception:  nearly 15\% of the variation is described by the survey type random effect, suggesting, on average, MICS observations are highest, DHS are lowest, and PMA2020 surveys are in between. 
This is clearly illustrated by examining direct survey and model-based estimates of unmet need in Lagos; see  Fig.~\ref{fig:StateTrends}(B). 
A similar trend can be found in Anambra, Kano, Nasarawa, Rivers, and Taraba states; Appendix 5 provides more detail for these states. 

\subsection{Kaduna and Lagos: a case study in conflicting trends}
Kaduna and Lagos states are current targets for large investments through the NURHI.
Figure~\ref{fig:StateTrends} displays the direct and model-based estimates for mCPR in Kaduna (A) and for unmet need in Lagos (B).
The 2013 DHS direct estimate of mCPR in Kaduna suggests a large increase relative to the 2011 MICS direct estimate, but given the large uncertainty in the 2013 direct estimate and subsequent mCPR observations, the model estimates a more gradual increase since 2011.
In Lagos, direct estimates from different survey instruments suggest conflicting trends in unmet need, with flat trends in DHS, increases in MICS, and a decline observed in PMA2020.
Due to this large variability by survey type, the model estimates a flat trend and large uncertainty in the underlying mean.
The direct and model-based estimates for all FP indicators in PMA2020 states can be seen in Appendix 5.

\subsection{Estimating mCPR changes from 2012 to 2017}

\begin{figure}[t]
	\centering
	\includegraphics[width=0.5\textwidth]{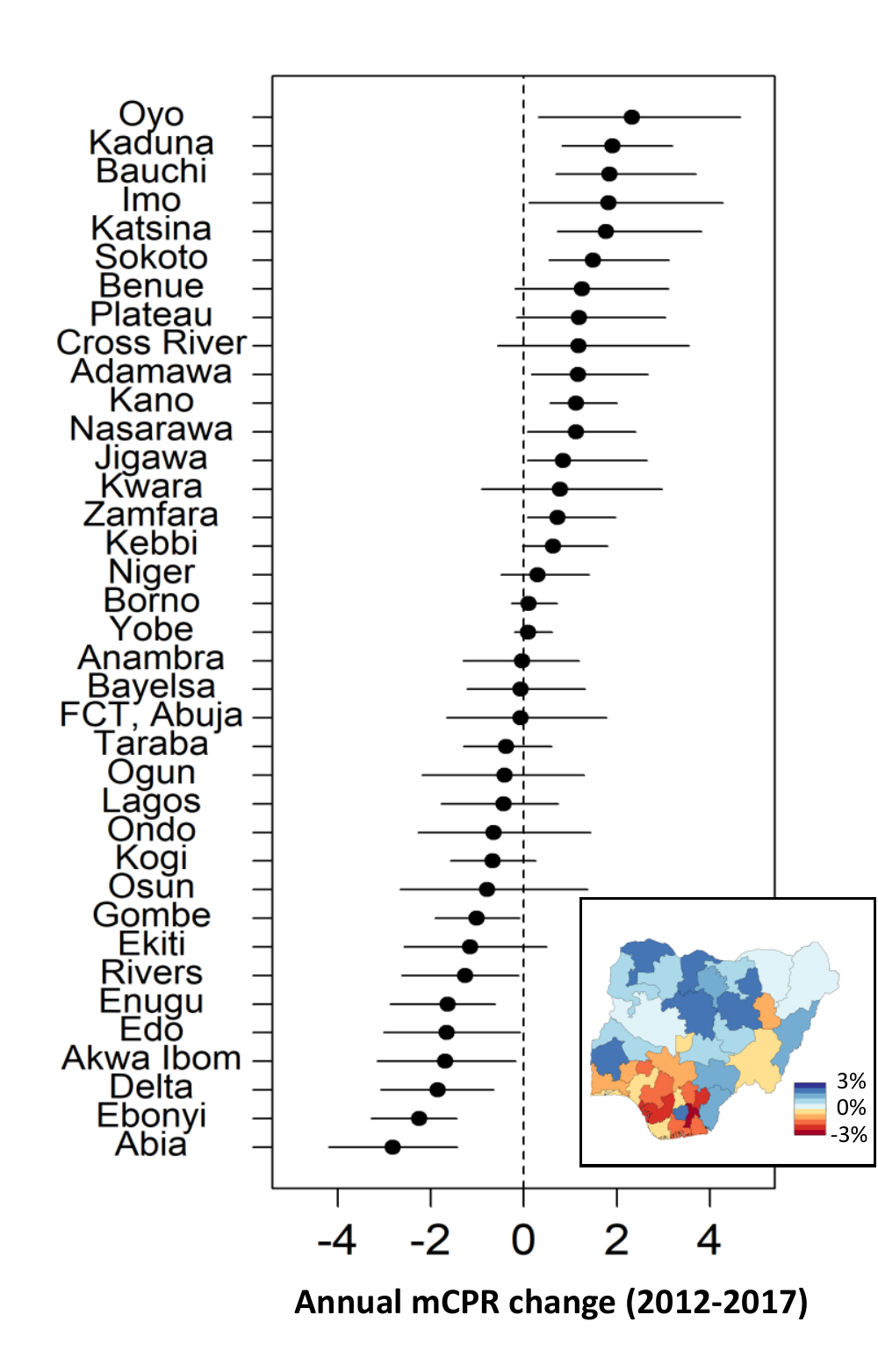}
	\caption{Posterior median and 95\% credible interval of annual change in mCPR from 2012 to 2017 and posterior median mCPR in 2017 (inset).}
	\label{fig:AnnualChange}
\end{figure}

Only eleven states have estimated changes significantly greater than zero.
Furthermore, only nineteen of thirty-seven states have positive estimates of annual change from 2012 to 2017.
Figure~\ref{fig:AnnualChange} displays the posterior median and 95\% credible intervals of the annual change in mCPR from 2012 to 2017.
Interpretation of annual changes should also consider population increases by state. 
 For example, according to the geo-referenced state level population estimates available for Nigeria (\textrm{https://nga.geopode.world/}), the estimated number of modern contraceptive users in Ondo state is approximately 150,000 women in both 2012 and 2017.  
Though not statistically significant, the stable number of users corresponds to an estimated 0.6\% annual decrease over the five-year period for Ondo state due to population growth.  
Furthermore, if current state level estimates are applied to the expected population of women between the ages of 15 to 49 in 2030 the total users of modern contraception would grow from approximately 5.0 to 6.7 million women, but overall national estimates would decrease from 11.1\% to 10.6\% due to faster population growth in states with lower mCPR.

\section{Discussion}
\label{s:discuss}

%
We have constructed a model to generate estimates for the underlying rates of contraceptive prevalence, unmet need, and demand satisfied at a sub-national resolution for Nigeria.
In a principled statistical framework, we have integrated spatial statistics, survey statistics, and multiple measurement instruments to produce sub-national annual estimates and short-term future forecasts.
Our statistical methodology is fundamentally data-driven; we incorporate all available survey data from Nigeria, include survey-specific designs and sampling uncertainty, and construct parsimonious, spatio-temporal statistical models for sub-national areas.
By utilizing a Bayesian hierarchical model, we can systematically incorporate sampling and non-sampling error for all available surveys. 
Our approach better characterizes the underlying landscape of FP indicators in Nigeria and enables public health officials to holistically interpret new survey results.

The generation of model-based estimates for FP indicators across Nigerian states revealed significant challenges to interpreting sub-national survey results.
For example, the 2013 DHS direct estimate for Kaduna state presented an approximately $10\%$ increase of mCPR over two years.
However, in the context of other subsequent surveys, including PMA2020, NNHS, and MICS, the DHS direct estimate appears unlikely; See Fig.~\ref{fig:StateTrends}(A) for reference.  
One advantage of our Bayesian hierarchical framework is identifying survey specific effects.  
Fig.~\ref{fig:StateTrends}(B) shows three different survey instruments, when taken individually, describe varying temporal trends, i.e., increasing, decreasing and staying uniform.  
The model-based estimate can integrate each of these estimates and identify systematic differences in survey types by FP indicator, such as relatively high estimates from MICS and relatively low estimates from DHS of unmet need.  
%
%
Without model-based estimates, it would be easy to overinterpret sub-national differences in direct estimates of FP indicators across a variety of survey instruments.


Our modeling framework can be an enabling technology for public health officials and policy-makers in Nigeria. 
Accounting for available survey data, integrating data into model-based estimates, and quantitatively identifying sources of variation in the data and estimates, we can provide estimates of the true {\it underlying} mCPR, traditional contraceptive rates, unmet need, and demand at a spatial resolution more aligned with FP programs and interventions. 
These state level estimates provide public health officials and policy makers a basis to design or interpret the impact of family planning programs or interventions.  
For example, mCPR is expectedly low in the northern states of Nigeria, generally aligning with other primary health indicators for routine services, such as routine immunization~\cite{gunnala2016routine}.
Rates of unmet need, however, vary significantly across the country providing a more nuanced perspective of family planning indicators within Nigeria; see Fig.~\ref{fig:Model}(III.B) for an illustration. 
Moreover, splitting women with unmet need according to age and parity, we see substantial heterogeneity sub-nationally and by parity.
Our approach can help inform the allocation of resources and program implementation, such as post-partum family planning programs in states with high rates of unmet need.  
In a country as large, diverse, and populous as Nigeria, interventions and programs will be implemented at much smaller spatial scales; using model-based estimates such as the one described in this article could enable better programmatic decisions.


On a broader note, our modeling framework and results are largely consistent with other modeling methodologies and help frame global family planning goals in a more local context.
Establishing national baselines in 2012 and tracking yearly progress was essential to characterizing progress toward {\it 120 by 20}~\cite{brown2014developing} in terms of required changes in national mCPR~\cite{NigeriaProgress}.
To date, however, estimated progress toward {\it 120 by 20} is lacking~\cite{Track20Progress} as well as country specific goals~\cite{NigeriaProgress}.
Our sub-national modeling efforts aim to characterize the underlying rates of FP indicators to help identify acceleration opportunities at the spatial scale of programs and interventions.  
Figure~\ref{fig:AnnualChange} illustrates Nigerian state level annual mCPR changes from 2012-2017; combined with population estimates, the absolute number of new users can be computed for each state.  
Our results confirm that at the national level mCPR has not changed significantly since 2012.  
However, our results have also expanded the scope of previous modeling and analyses by identifying which regions have statistically seen increases, decreases, or stayed neutral for mCPR.


Data availability poses serious constraints on our framework being used as a baseline and forecasting tool for small subregions in a country like Nigeria.
Despite the variety of survey instruments and known survey cluster locations, estimating FP indicators at even the LGA level produces substantial uncertainty intervals, shown in Figure 12 in the Appendix.  
Moreover, our forecasting methodology, using a data-driven, non-parametric, spatio-temporal model, generates rapidly growing uncertainty intervals over a relatively short three-year outlook.
Both types of uncertainty provide an important result for policy makers and large-scale donors to realistically frame the resolution available to evaluate the impact of investments at a fine spatial resolution.  
%
%
Our model selection procedure is consistent with research in the small area estimation community for under five mortality estimation~\cite{mercer2015}.
However, there are several open questions about constructing the spatio-temporal terms in the Bayesian hierarchical model:  what are the relevant geographic regions to include in the smoothing procedure?, which model selection criteria is most appropriate for these models?, and can routine data from the demographic health information systems be integrated and trusted in settings like Nigeria?
We expect to address many of these challenges in future work.


Notwithstanding many of these challenges and limitations, our study can enable public health officials and policy makers through better characterization of FP indicators at a sub-national and demographic subgroup scale.
Given the current lag in progress toward {\it 120 by 20}, a change of perspective is broadly required.
The FP community should shift from measuring national progress with an assumed broad increase of mCPR to a sub-national perspective focused on informing specific programs at the scale that programs are implemented.
Looking forward, the design and implementation of FP programs, funded by governmental and non-governmental organizations, need to have the most holistic and informed estimates to understand the current estimates and trends of FP indicators sub-nationally.
The global health community is poised to make significant gains to provide equitable access to family planning services; having a paired measurement and program strategy will be key to achieving these goals.  


\begin{acknowledgments}

The authors would like to thank Bill and Melinda Gates for their sponsorship through the Global Good Fund.  We would also like to thank the National Bureau of Statistics of Nigeria for providing access to the 2014 and 2015 National Nutritional Health Surveys.
\end{acknowledgments}
 \bibliographystyle{model1-num-names}
\bibliography{refs}

\newpage

\beginsupplement
\section*{Supplemental Information}

\section{Input Data}
\label{s:data}

In this section we describe the thirteen household surveys that were analyzed for the article `Sub-national levels and trends in contraceptive prevalence, unmet need, and demand for family planning in Nigeria with survey uncertainty.'
\subsection{Demographic and Health Surveys}

We processed the four available Demographic and Health Surveys (DHS) for Nigeria. The 1990 DHS selected 34 households from each of the 299 enumeration areas (EAs), which were stratified by urban/rural status and selected via probability proportional to size (pps), and was designed to produce national and regional (4 regions at that time) estimates~\cite{Nigeria:dhs:1990}.
A stratified two-stage sampling design was used for the 2003 DHS to provide estimates of health indicators at the national level, the six health regions, and by urban/rural status.  Households were select from a complete listing of households within the 365 selected EAs~\cite{Nigeria:dhs:2003}.
Similarly, the 2008 and 2013 DHS employed stratified multi-stage sampling designs consisting of 888 EAs and 904 EAs, respectively~\cite{Nigeria:dhs:2008,Nigeria:dhs:2013} and were intended to be representative at the national, regional, and state level.
 An additional survey was conducted in 1999, however the DHS Program was not centrally involved with the study and does not distribute the collected data. 

To calculate unmet need we relied on the recoded definitions used by the United Nations Population Division (UNPD), which applied the revised DHS unmet need definition~\cite{bradley2012revising} to previous surveys. To compute the revised unmet need in our analysis, we relied on the UNPD script available at \url{https://github.com/PhilUeff/PDU_FP-Indicators}. 

To ensure state-level values reflect current geographical divisions, we mapped the GPS coordinates associated with the sampled clusters from each survey onto the 2013 Nigeria DHS shapefile and used their corresponding state names. Each DHS survey contained a few clusters without recorded GPS coordinates (Table \ref{table-cluster}). For these, we assigned the clusters to the states that they were recorded as belonging to at the time. The assignment is ambiguous for one DHS 1990 cluster because the corresponding recorded state has split between the time of the survey and 2013; this is noted in Table~\ref{table-cluster}. 

Table~\ref{table1} compares national all-women estimates for contraceptive prevalence rate (CPR), modern contraceptive prevalence rate (mCPR), total unmet need (UnN), and demand satisfied by modern methods (DSM), as computed by the DHS survey official reports, the UN recode, and our analysis. While not presented, traditional contraceptive prevalence rate can be computed as CPR - mCPR.

Contraceptive prevalences match among all sources. The differences in unmet need between the DHS official reports and the UN computed values prior to 2012 reflect the revised unmet need calculation. Dashes indicate estimates that were not reported by the DHS.

\renewcommand{\arraystretch}{1.5}

\begin{table}[!ht]
\centering
\caption{Surveyed DHS clusters without GPS coordinates}
\begin{tabular}{lccc}
\toprule
DHS year & Cluster ID & Recorded state & Possible states \\
\midrule
\multirow{2}{*}{1990} & 1402 & Kwara & Kwara, Kogi \\
& 1554 & Lagos & Lagos \\
\midrule
\multirow{2}{*}{2003} & 170 & Katsina & Katsina \\
& 258 & Lagos & Lagos \\
\midrule
\multirow{1}{*}{2008} & 773 & Rivers & Rivers\\
\midrule
\multirow{7}{*}{2013} & 302 & Yobe & Yobe\\
& 373 & Kano & Kano\\
& 422 & Kebbi & Kebbi\\
& 514 & Anambra & Anambra\\
& 557 & Enugu & Enugu\\
& 569 & Enugu & Enugu\\
& 639 & Bayelsa & Bayelsa\\
\bottomrule
\end{tabular}
\label{table-cluster}
\end{table}

\begin{table*}[!ht]
\centering
\caption{DHS surveys: comparing DHS reported figures, UN computed figures, and our computed figures at the national level for all women. All values are reported in (\%).}
\begin{tabular}{lcccccccccccc}
\toprule
 & \multicolumn{4}{c}{DHS reported} & \multicolumn{4}{c}{UN computed} & \multicolumn{4}{c}{Our computed} \\
\cmidrule(lr){2-5}\cmidrule(lr){6-9}\cmidrule(lr){10-13}
 DHS year & CPR & mCPR & UnN & DSM & CPR & mCPR & UnN & DSM & CPR & mCPR & UnN & DSM \\
\midrule
2013 & 16.0 & 11.1 & 12.7 & 38.8 & 16.0 & 11.1 & 12.7 & 38.5 & 16.0 & 11.1 & 12.7 & 38.8 \\
2008 & 15.4 & 10.5 & 15.7 & -- & 15.4 & 10.5 & 16.1 & 33.2 & 15.4 & 10.5 & 16.1 & 33.2 \\
2003 & 13.3 & 8.9 & 13.6 & -- & 13.3 & 8.9 & 14.6 & 32.0 & 13.3 & 8.9 & 14.6 & 32.0 \\
1990 & 7.5 & 3.8 & -- & -- & 7.5 & 3.8 & 17.1 & 15.2 & 7.5 & 3.8 & 17.1 & 15.2 \\
\bottomrule
\end{tabular}
\label{table1}
\end{table*}

\subsection{Multiple Indicator Cluster Surveys}

We used three Multiple Indicator Cluster Surveys (MICS) surveys in our analysis. 
Thirty EAs were selected within each state for the 2007 MICS, which was designed to provide state and national level estimates of health indicators~\cite{Nigeria:mics:2007}.
A two-stage sampling design was used for the 2011 MICS, which was designed to provide estimates and the state and national level. Forty EAs were selected with equal probability within each state and all 1,480 EAs were included in field work~\cite{Nigeria:mics:2011}.
The 2016-17 MICS was designed to provide estimates at the national and state level and relied on a two-stage sampling design.  Sixty EAs were selected from each state, except for Lagos and Kano which were over sampled at 120 EAs.  Of the 2,340 sampled EAs only 2,239 were included in field work because 101 EAs from Borno, Yobe, and Adamawa states were excluded due to security concerns~\cite{Nigeria:mics:2016}.

The UNPD has recoded the 2007 and 2011 MICS surveys, but at the time of this writing, had not yet recoded the 2016 survey. We used the UNPD recode scripts from \url{https://github.com/PhilUeff/PDU_FP-Indicators} to process the 2007 and 2011 MICS survey data, then computed the survey estimates as with the DHS surveys. 
When we used the UN code on the data, we found a small difference in the 2011 estimated unmet need from the value reported by the UNPD.  
For the 2016 MICS survey, we checked that the variables were encoded in the same form as the 2011 MICS survey.
The unavailable estimates from the MICS official reports and 2016 UNPD recode are indicated with dashes. 
Finally, certain questions needed to calculate unmet need were not asked in the 2007 MICS survey. As a result, unmet need and demand satisfied for this survey are marked with 'NA'. 
The MICS official reports only present estimates over currently-married/in-union women. 
Table \ref{table2} displays the estimates reported by MICS, calculated by the UN, and calculated via our direct analysis of the microdata.

\begin{table*}[!ht]
\centering
\caption{MICS surveys: comparing MICS reported figures, UN computed figures, and our computed figures at the national level for all women. All values are reported in (\%).}
\begin{tabular}{lcccccccccccc}
\toprule
 & \multicolumn{4}{c}{MICS reported} & \multicolumn{4}{c}{UN computed} & \multicolumn{4}{c}{Our computed} \\
\cmidrule(lr){2-5}\cmidrule(lr){6-9}\cmidrule(lr){10-13}
 MICS year & CPR & MCPR & UnN & DSM & CPR & MCPR & UnN & DSM & CPR & MCPR & UnN & DSM \\
\midrule
2016 & -- & -- & -- & -- & -- & -- & -- & -- & 11.8 & 9.5 & 26.8 & 24.6 \\
2011 & -- & -- & -- & -- & 18.3 & 12.5 & 20.2 & 32.6 & 18.3 & 12.5 & 21.0 & 31.9 \\
2007 & -- & -- & NA & NA & 14.6 & 11.1 & NA & NA & 14.6 & 11.1 & NA & NA \\
\bottomrule
\end{tabular}
\label{table2}
\end{table*}


\subsection{Performance Monitoring and Accountability 2020}

We used four rounds of the Performance Monitoring and Accountability 2020 (PMA2020) surveys conducted in Nigeria.  We obtained the data for the 2014-2016 PMA2020 rounds~\cite{Nigeria:pma2020:2014,Nigeria:pma2020:2015,Nigeria:pma2020:2016} from IPUMS~\cite{Nigeria:pma:ipums}, which had recoded the PMA2020 surveys. We computed survey estimates using this data. The 2017 PMA2020 survey~\cite{Nigeria:pma2020:2017} had not been recoded at the time of this writing by IPUMS, so this survey was recoded manually.  The 2014 and 2015 rounds were designed to provide state level estimates for Kaduna and Lagos states and relied on a two-stage sampling design.  The 2016 and 2017 rounds were designed to provide national estimates relying on three-stage sampling design.  At the first stage seven states were selected and then within each stage two-stage sampling was implemented.  Post stratification was used to generate national estimates~\cite{Nigeria:pma2020:sampling}. Table \ref{table3} provides the point estimates reported by PMA2020 and those calcuated in our analysis of the microdata.

\begin{table*}[!ht]
\centering
\caption{PMA surveys: comparing PMA2020 reported figures and our computed figures at the national level for all women. All values are reported in (\%).}
\begin{tabular}{llcccccccc}
\toprule
 & & \multicolumn{4}{c}{PMA2020 reported} & \multicolumn{4}{c}{Our computed} \\
\cmidrule(lr){3-6}\cmidrule(lr){7-10}
 PMA year & State & CPR & MCPR & UnN & DS & CPR & MCPR & UnN & DS \\
\midrule
\multirow{7}{*}{2017} & Anambra & 24.2 & 13.9 & 12.0 & 38.3 & 24.1 & 13.8 & 12.0 & 38.2 \\ 
 & Kaduna & 17.4 & 15.6 & 22.4 & 39.2 & 17.4 & 15.6 & 22.4 & 39.2 \\ 
 & Kano & 6.2 & 4.4 & 24.6 & 14.4 & 6.2 & 4.4 & 24.6 & 14.4 \\ 
 & Lagos & 29.7 & 20.6 & 15.2 & 45.8 & 29.7 & 20.6 & 15.2 & 45.8 \\ 
 & Nasarawa & 18.9 & 16.9 & 21.0 & 42.3 & 18.9 & 16.9 & 21.0 & 42.3 \\ 
 & Rivers & 29.0 & 17.7 & 17.2 & 38.2 & 29.0 & 17.7 & 17.2 & 38.2 \\ 
 & Taraba & 14.1 & 10.5 & 24.4 & 27.3 & 14.1 & 10.5 & 24.4 & 27.3 \\ 
\midrule
\multirow{7}{*}{2016} & Anambra & 25.1 & 15.6 & 14.4 & 39.6 & 25.9 & 16.1 & 14.4 & 40.4 \\
 & Kaduna & 15.1 & 13.9 & 26.2 & 33.7 & 15.4 & 14.1 & 26.2 & 34.0 \\
 & Kano & 5.6 & 4.8 & 30.3 & 13.5 & 5.6 & 4.9 & 30.2 & 13.5 \\
 & Lagos & 26.4 & 19.7 & 15.6 & 46.8 & 26.0 & 19.7 & 15.6 & 46.9 \\
 & Nasarawa & 18.9 & 16.6 & 18.0 & 44.9 & 18.9 & 16.6 & 18.0 & 44.9 \\
 & Rivers & 27.5 & 19.4 & 16.4 & 44.2 & 28.8 & 19.6 & 16.4 & 43.7 \\
 & Taraba & 12.9 & 9.9 & 27.1 & 24.9 & 12.6 & 10.0 & 27.1 & 25.0 \\
\midrule
\multirow{2}{*}{2015} & Kaduna & 14.5 & 13.2 & 24.9 & 33.5 & 14.7 & 13.5 & 24.9 & 34.2 \\
 & Lagos & 27.8 & 21.0 & 17.8 & 46.1 & 27.8 & 21.4 & 17.8 & 46.4 \\
\midrule
\multirow{2}{*}{2014} & Kaduna & 8.7 & 8.4 & 28.2 & 22.8 & 8.7 & 8.5 & 28.2 & 22.9 \\
 & Lagos & 17.8 & 16.5 & 19.3 & 44.5 & 17.3 & 16.7 & 19.2 & 44.8 \\

\bottomrule
\end{tabular}
\label{table3}
\end{table*}

\subsection{National Nutrion and Health Surveys}

We analyzed the microdata from two National Nutrition and Health Surveys (NNHS) from 2014 and 2015. Both surveys were conducted using the Standardized Monitoring and Assessment of Relief and Transition (SMART) methods and relied on two-stage cluster and designed to provide estimates at the national and state level~\cite{Nigeria:nnhs:2014,Nigeria:nnhs:2015}. Unfortunately, the NNHS does not include the variables required to calculate unmet need nor parity, so the NNHS data was only included in the estimates of mCPR for all women.  Additionally, the survey reports highlight that in 9 Local Governmental Areas (LGAs) were excluded for security reasons and thus results for Borno are not representative of the whole state.

\section{Modeling family planning indicators}
\label{s:model}

Family planning indicators in Nigeria are estimated from multiple complex surveys, sometimes in the same year and can be quite noisy between surveys at the state level.  In an effort to understand the underlying population rates from which these survey observations were drawn, we are adapting a previously developed two-step space-time smoothing approach that acknowledge complex sampling designs to model the family planning indicators.

The first step  of our approach requires estimating the state ($i$), year ($t$), and survey ($s$) specific direct estimates of proportions and corresponding variance via the Hajek estimator~\cite{hajek:71} $\widehat{p}_{its}=\sum_{j}x_{j,its}w_{j,its}/\sum_{j}w_{j,its}$ where $x_{j,its}$ is the binary indicator for modern contraception, traditional contraception, unment need, or demand satisfied for woman $j$, in sampled in area $i$, time $t$, and survey $s$, and $w_{j,its}$ is her corresponding sampling weight.

In the second step we use a three-stage Bayesian hierarchichal model.  In the first stage we rely on the working likelihood based on the asymptotic distribution,
\[
Y_{its}\sim N\left(\mathbf{\eta_{its}},\widehat{V}_{its,DES} \right)
\]
where $Y_{its}=\textrm{logit}[\widehat{p}_{its}]$ and $\widehat{V}_{its,DES}$ is the design-based variance of $Y_{its}$. This pseudo-likelihood for a binary variable was initially developed to estimate zipcode-level smoking prevalence in Washington state~\cite{mercer:etal:14} relying on a single survey instrument.  This model was then adapted for the complex indicator of under-five child mortality~\cite{mercer2015} and extended the classic inseperable spatio-time models of~\cite{knorrheld:00,schrodleheld:2011} to include survey-specific random effects.

At the second stage we assume
\[
\eta_{its}=\mu+\gamma_{t}+\alpha_{t}+\theta_{i}+\nu_{s}+\delta_{it}+\phi_{is}+\psi_{ts},
\]
where $\mu$ is a shared interecept, the temporally structured random effects $\alpha$ are assigned second order random walk priors~\cite{rue:knorrheld:05,schrodleheld:2011}, the spatial random effects $\theta$ are assigned either the Besag, York, and Mollie (BYM)~\cite{besag:etal:91} or the scaled BYM~\cite{riebler2016intuitive}, the space-time interactions $\delta$ are assinged the `type II'  temporally structured interaction~\cite{schrodleheld:2011}, and the independent random effects $\gamma$, $\nu$, $\phi$, and $\psi$ are assigned independent mean-zero Normal distributions. Note, if $\theta\sim$BYM then $\theta_i=U_i+V_i$ where $U$ is assigned an intrinsic conditional autoregression prior (ICAR)~\cite{besag:kooperberg:95} and $V$ are assigned an independent mean-zero Normal prior.  At the third stage of the model we assume a default diffuse prior for $\mu$ and Gamma$(1,5e-5)$ priors were assigned to, $\sigma_{\alpha}^{-2}$, $\sigma_{\theta}^{-2}$ (or to both inverse variances for standard BYM), $\sigma_{\gamma}^{-2}$, $\sigma_{\delta}^{-2}$, $\sigma_{\nu}^{-2}$, $\sigma_{\phi}^{-2}$, and $\sigma_{\psi}^{-2}$.

Finally, to generate our estimates of the underlying rates we adopt the approach of Mercer (2015) and draw $k$=1,...1000 posterior samples for all parameters and calculate 
\[
\eta^{(k)}_{it}=\mu^{(k)}+\gamma^{(k)}_{t}+\alpha^{(k)}_{t}+\theta^{(k)}_{i}+\delta^{(k)}_{it}.
\]
The posterior median is used as the estimate and the uncertainty intervals are defined by the 2.5\% and 97.5\% quantiles.

\section{Model Selection}
\label{s:modelselection}
The twelve possible models considered for mCPR, traditional contraceptive prevalence, unmet need, and demand satisfied are described in Table \ref{possibleModels}.  Each model was fit for mCPR, traditional contraceptive prevalence, unment need and demand satisfied for all women and the four age and parity combinations.  For each model fit we calculated the sum of the log conditional predictive ordinate (LCPO)~\cite{held2010posterior}, the deviance information criteria (DIC)~\cite{spiegelhalter2014deviance}, and the Watanabe-Akaike information criterion (WAIC)~\cite{watanabe2010asymptotic}.  The results for mCPR and unment need for all women are shown in Table \ref{selectedModelsmCPR} and Table \ref{selectedModelsUnmet}, respectively with with the lowest DIC and WAIC and highest LCPO in bold.   Table \ref{selectedModels} shows the selected model for all outcomes and age-parity groups. 
\begin{table}[!ht]
\centering
\caption{Random effects models considered for time $t$, state $i$, and survey $s$. All models contain an intercept $\mu$, temporally independent $\gamma_t\sim N(0,\sigma_{\gamma}^2)$, and temporally structured effect $\alpha_t\sim RW2(\sigma_{\alpha}^2)$. The survey and survey interaction random effects were assigned mean-zero Normal priors.  In models designated with `a' $\theta_i\sim BYM(\sigma_{\theta}^2)$ and in models designated `b' $\theta_i\sim BYM2(\sigma_{\theta}^2)$.}
\begin{tabular}{l|l}
\toprule
 Model & outcome \\ 
\midrule
1& $\mu +\alpha_t + \gamma_t + \theta_i$ \\ 
 2& $\mu +\alpha_t + \gamma_t + \theta_i + \delta_{it}$ \\ 
 3& $\mu +\alpha_t + \gamma_t + \theta_i + \delta_{it} + \nu_s$ \\ 
 4&  $\mu +\alpha_t + \gamma_t + \theta_i + \delta_{it} + \nu_s+ \phi_{is}$ \\ 
 5& $\mu +\alpha_t + \gamma_t + \theta_i + \delta_{it} + \nu_s+ \psi_{ts}$\\ 
 6& $\mu +\alpha_t + \gamma_t + \theta_i + \delta_{it} + \nu_s+ \phi_{is}+ \psi_{ts}$ \\ 
\bottomrule
\end{tabular}
\label{possibleModels}
\end{table}

\begin{table}[!ht]
\centering
\caption{Model fit results for estimating mCPR for all women.}
\begin{tabular}{l|rrr}
\toprule
 Model & DIC & $\sum \log (CPO)$ & WAIC \\ 
\midrule
1a &1418.1&	-846.0&	1663.6\\
2a &514.7&	-535.9	&697.6\\
3a &483.7	&-525.9	&658.4\\
4a &419.2	&-515.0	&568.8\\
5a &477.6	&-507.6	&643.7\\
6a &402.7	&-490.5	&539.6\\
1b &1418.1&	-845.8	&1663.3\\
2b &514.3&	-535.4	&695.7\\
3b &483.2&	-525.1	&655.9\\
4b &418.6&	-513.5&	567.7\\
5b &476.6&	-506.9	&641.9\\
6b &\textbf{402.1}&	\textbf{-488.5}&	\textbf{538.4}\\
\bottomrule
\end{tabular}
\label{selectedModelsmCPR}
\end{table}

\begin{table}[!ht]
\centering
\caption{Model fit results for estimating unmet need for all women.}
\begin{tabular}{l|rrr}
\toprule
 Model & DIC & $\sum \log (CPO)$ & WAIC \\ 
\midrule
1a &1145.3&	-748.4&	1487.7\\
2a &27.6&	-298.9	&90.7\\
3a &5.0&	-283.3	&58.8\\
4a &-56.7&	-208.5&	-56.1\\
5a &4.4&	-281.4&	56.7\\
6a &\textbf{-56.8}&	\textbf{-208.0}&	\textbf{-56.4}\\
1b &1145.4	&-748.2	&1487.3\\
2b &26.3	&-299.3&	88.8\\
3b &3.8	&-283.2&	56.9\\
4b &-56.4	&-209.5	&-55.3\\
5b &4.2&	-281.8	&56.0\\
6b &-56.5&	-208.9	&-55.7\\
\bottomrule
\end{tabular}
\label{selectedModelsUnmet}
\end{table}

\begin{table}[!ht]
\centering
\caption{Selected models.}
\begin{tabular}{llc}
\toprule
 Age-parity group & outcome & selected model \\
\midrule
\multirow{4}{*}{All women} & mCPR & 6b\\ 
 & Unmet Need &6a\\ 
 & tCPR &  4b\\ 
 & Demand Satisfied & 6b\\ 
\midrule
\multirow{4}{*}{15-24yo, 0}& mCPR & 6a\\ 
 & Unmet Need &4a\\ 
 & tCPR &	 6a\\ 
 & Demand Satisfied &6a\\ 
\midrule
\multirow{4}{*}{15-24yo, 1+} & mCPR & 6b \\ 
 & Unmet Need & 4b\\ 
 & tCPR & 	 6b\\ 
 & Demand Satisfied &	6a \\ 
\midrule
\multirow{4}{*}{25-49yo, 0} &  mCPR & 6b	\\ 
 & Unmet Need &  6b\\ 
 & tCPR &6b\\ 
 & Demand Satisfied & 6a\\ 
\midrule
\multirow{4}{*}{25-49yo, 1+}& mCPR & 6b\\ 
 & Unmet Need &6b\\ 
 & tCPR & 4b\\ 
 & Demand Satisfied & 4b\\ 
\bottomrule
\end{tabular}
\label{selectedModels}
\end{table}

\section{Decomposition of Variance}
\label{s:variance}
To assess the relative importance of the individual random effects in explaining the variance in the observed data we calculated the percent of the total variance described by each random effect. Table \ref{variance} shows the percent of the variance described by each random effect where the total variance for model 6 is
\[
 \sigma_{\alpha}^2 +\sigma_{\gamma}^2+\sigma_{\theta}^2+\sigma_{\delta}^2+\sigma_{\nu}^2+\sigma_{\psi}^2+\sigma_{\phi}^2      
\]
and for outcomes and age-parity groups using model 4, the $\sigma_{\phi}^2$ term is removed.

\begin{table*}[!ht]
\centering
\caption{Percent of total variance described by each random effect, where the interpretation is $\sigma_{\theta}^2$ for spatial, $\sigma_{\gamma}^2$ for independent temporal, $\sigma_{\alpha}^2$ for structured (RW2) temporal, $\sigma_{\delta}^2$ for space-time interaction, $\sigma_{\nu}^2$ for surveys, $\sigma_{\phi}^2$ for survey-space interactions, and  $\sigma_{\psi}^2$ for survey-time interaction.}
\begin{tabular}{llcrrrrrrr}
\toprule
 Age-parity group & outcome & $\sigma_{\theta}^2$ & $\sigma_{\gamma}^2$ & $\sigma_{\alpha}^2$ & $\sigma_{\delta}^2$  & $\sigma_{\nu}^2$ & $\sigma_{\phi}^2$ & $\sigma_{\psi}^2$ \\
\midrule
\multirow{4}{*}{All women} & mCPR &74.5 & $<$0.05&	$<$0.05	&0.5	&$<$0.05	&23.8	&1.2
 \\ 
 & Unmet Need & $<$0.05	&$<$0.05	&$<$0.05	&0.6&	14.8&	79.3&	5.3\\ 
 & tCPR & 62.2	&$<$0.05		&12.5&	0.6&$<$0.05		&24.7&	- \\ 
 & Demand Satisfied & 43.4	&$<$0.05		&$<$0.05		&0.2	&$<$0.05		&55.7&	0.7 \\ 
\midrule
\multirow{4}{*}{15-24yo, 0}& mCPR &97.5&	$<$0.05	&	$<$0.05		&$<$0.05		&$<$0.05		&2.0	&0.5 \\ 
 & Unmet Need & $<$0.05		&$<$0.05		&7.1&	$<$0.05	&	$<$0.05	&	92.9&	-\\ 
 & tCPR & $<$0.05		&$<$0.05		&$<$0.05		&0.1&	$<$0.05	&	99.9	&$<$0.05	 \\ 
 & Demand Satisfied &$<$0.05		&$<$0.05		&$<$0.05		&$<$0.05		&$<$0.05		&93.0	&7.0\\ 
\midrule
\multirow{4}{*}{15-24yo, 1+} & mCPR & 79.1	&$<$0.05	&$<$0.05		&$<$0.05		&$<$0.05		&20.0	&0.8 \\ 
 & Unmet Need &56.5	&$<$0.05	&24.3&	$<$0.05		&$<$0.05		&19.2&- \\ 
 & tCPR & 80.3	&$<$0.05		&$<$0.05		&$<$0.05		&$<$0.05		&19.7&	$<$0.05	 \\ 
 & Demand Satisfied &$<$0.05		&$<$0.05		&$<$0.05		&$<$0.05		&$<$0.05		&99.9&	$<$0.05	 \\ 
\midrule
\multirow{4}{*}{25-49yo, 0} &  mCPR & 69.1	&$<$0.05		&2.8	&$<$0.05		&$<$0.05		&28.2	&$<$0.05	\\ 
 & Unmet Need & $<$0.05		&$<$0.05		&$<$0.05		&$<$0.05		&$<$0.05		&32.7	&67.2 \\ 
 & tCPR &99.9	&$<$0.05		&$<$0.05		&$<$0.05		&$<$0.05	&	$<$0.05	\\ 
 & Demand Satisfied &$<$0.05		&$<$0.05		&$<$0.05	&$<$0.05	&$<$0.05	&99.9	&$<$0.05	 \\ 
\midrule
\multirow{4}{*}{25-49yo, 1+}& mCPR &97.7	&$<$0.05		&$<$0.05		&0.1&	$<$0.05		&2.1	&0.1 \\ 
 & Unmet Need & 1.4	&$<$0.05		&$<$0.05		&0.3&	19.8&	70.1&	8.4\\ 
 & tCPR & 62.1&	$<$0.05	&13.6&	0.3&	0&	24.0&	-\\ 
 & Demand Satisfied & 46.2&$<$0.05&	0.1&	0&	0&	53.6	&-\\ 
\bottomrule
\end{tabular}
\label{variance}
\end{table*}

\section{Trends in family planning indicators in PMA2020 states}
\label{s:pma2020}

\begin{figure*}[t]
	\centering
\def\tabularxcolumn#1{m{#1}}

\begin{tabular}{cc}
\subfloat[mCPR]{\includegraphics[width=9cm]{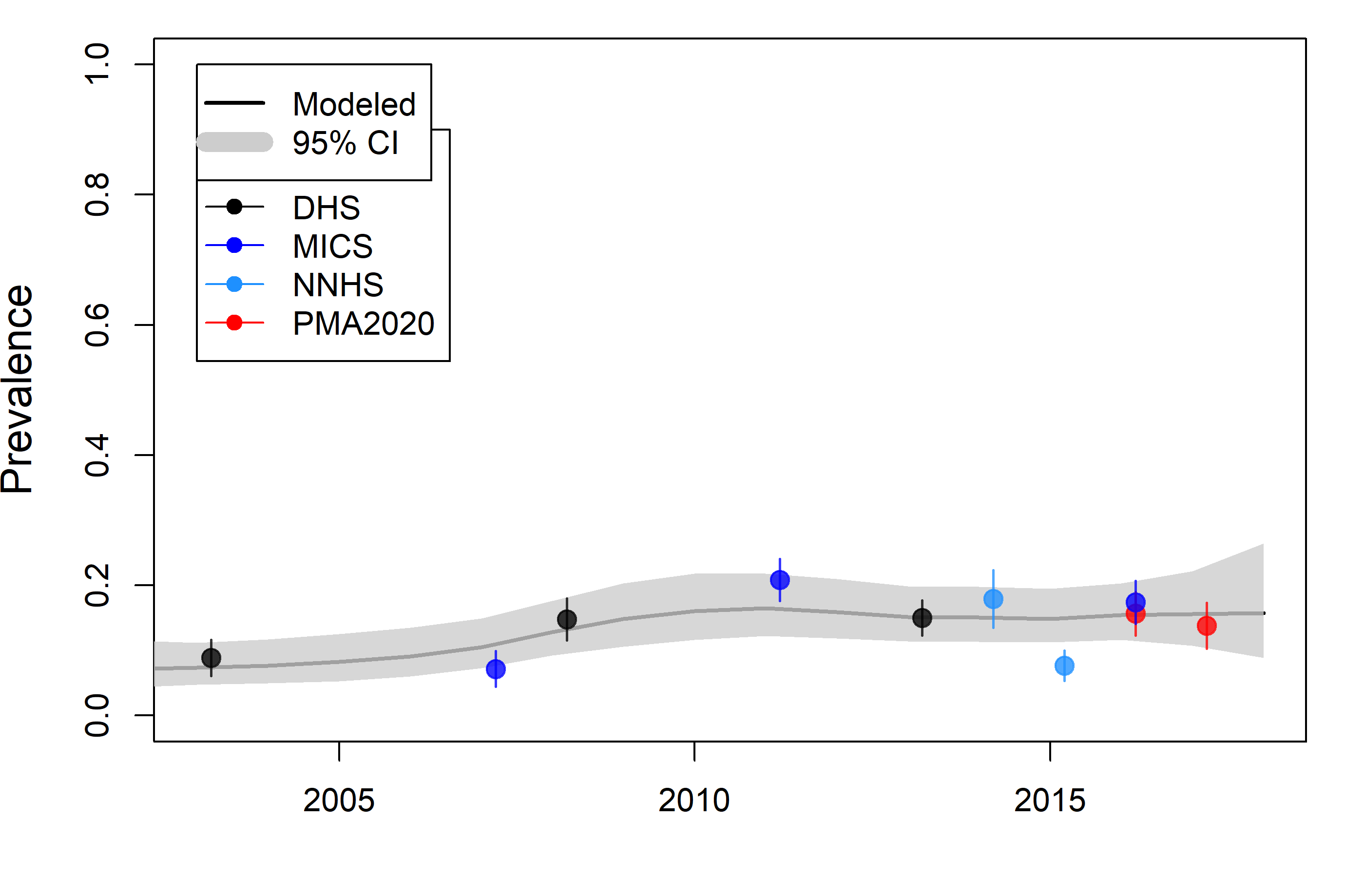}} 
   & \subfloat[Unmet Need]{\includegraphics[width=9cm]{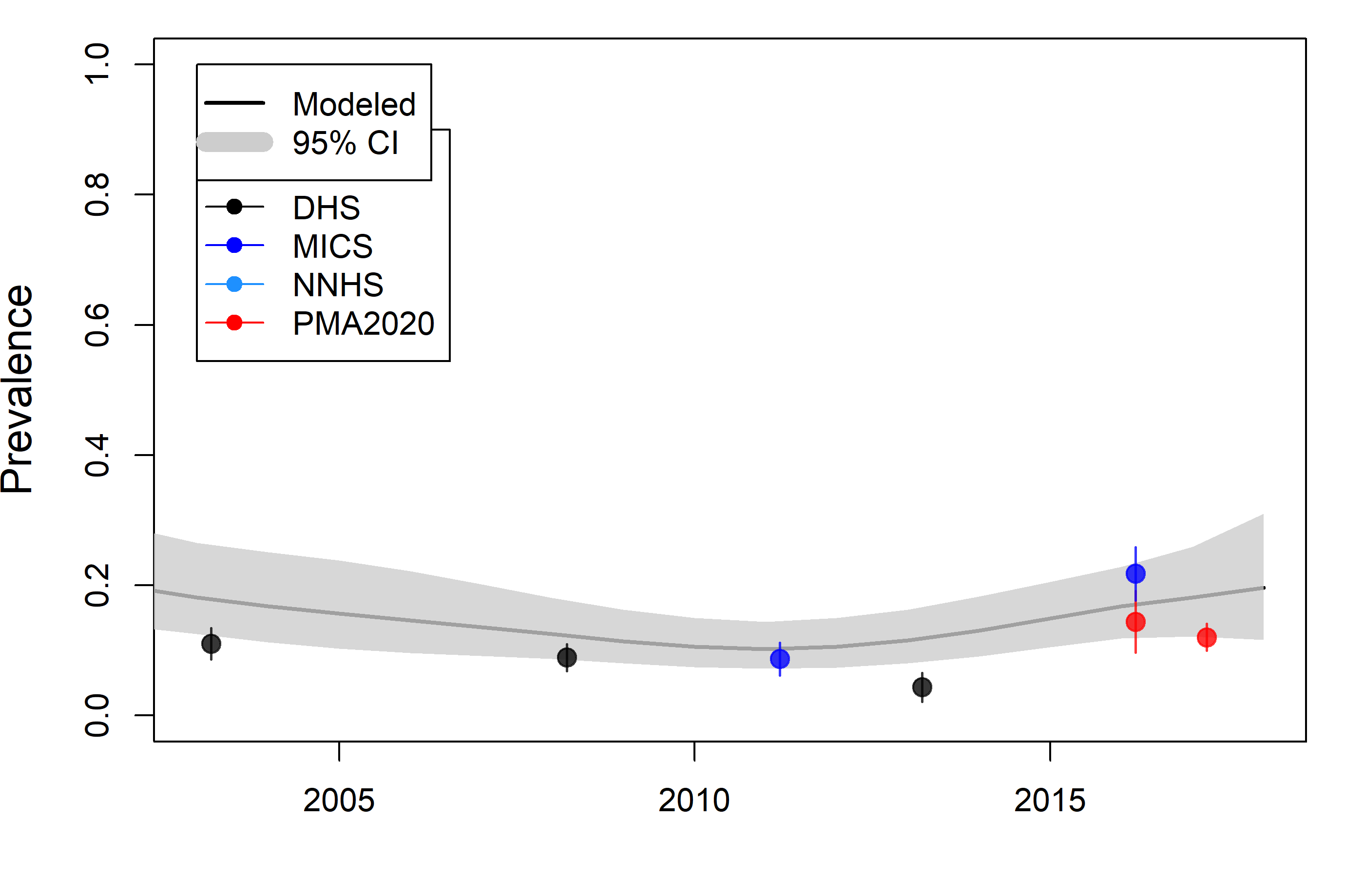}}\\
\subfloat[Traditional CPR]{\includegraphics[width=9cm]{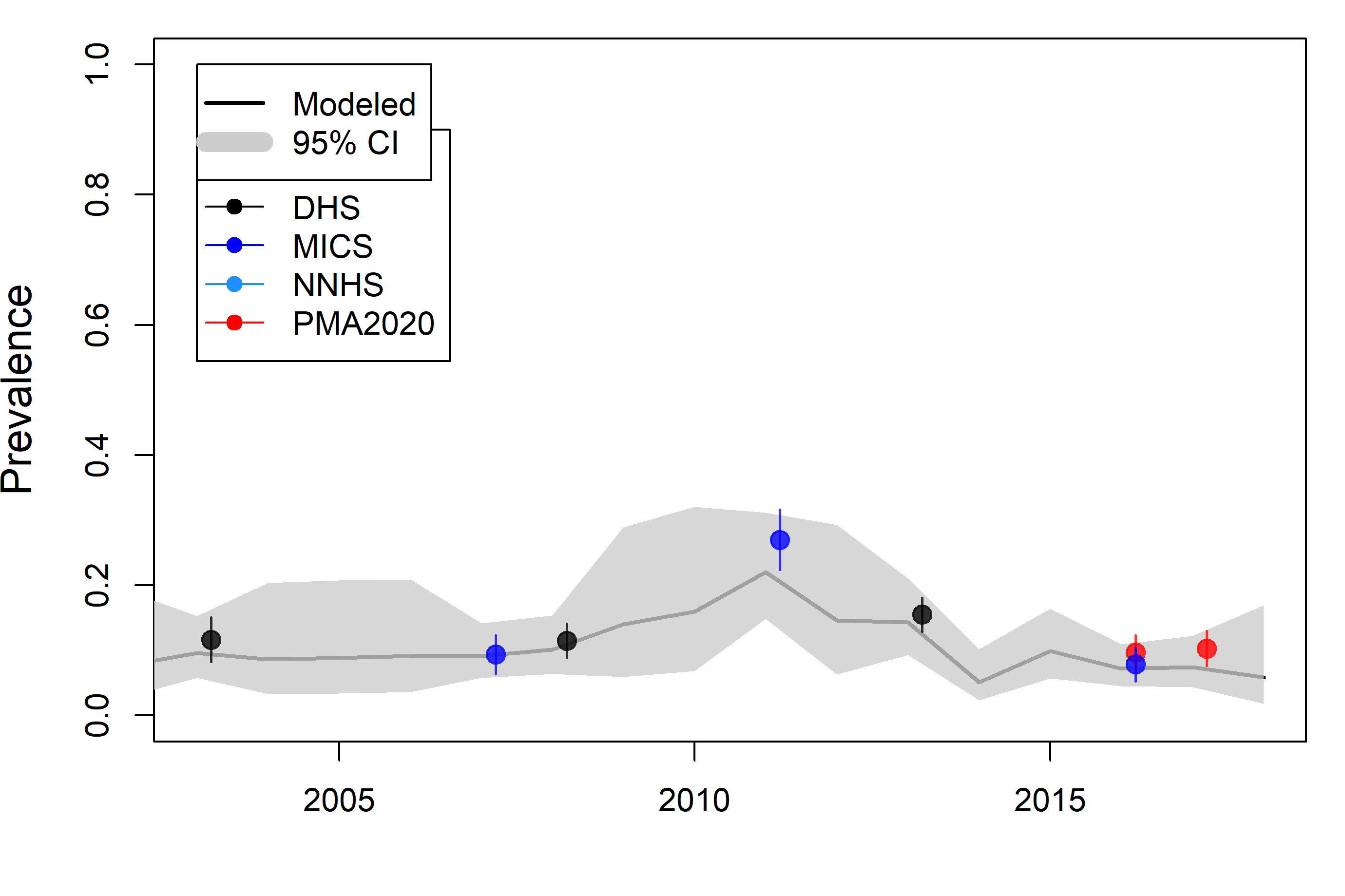}} 
   & \subfloat[Demand Satisfied]{\includegraphics[width=9cm]{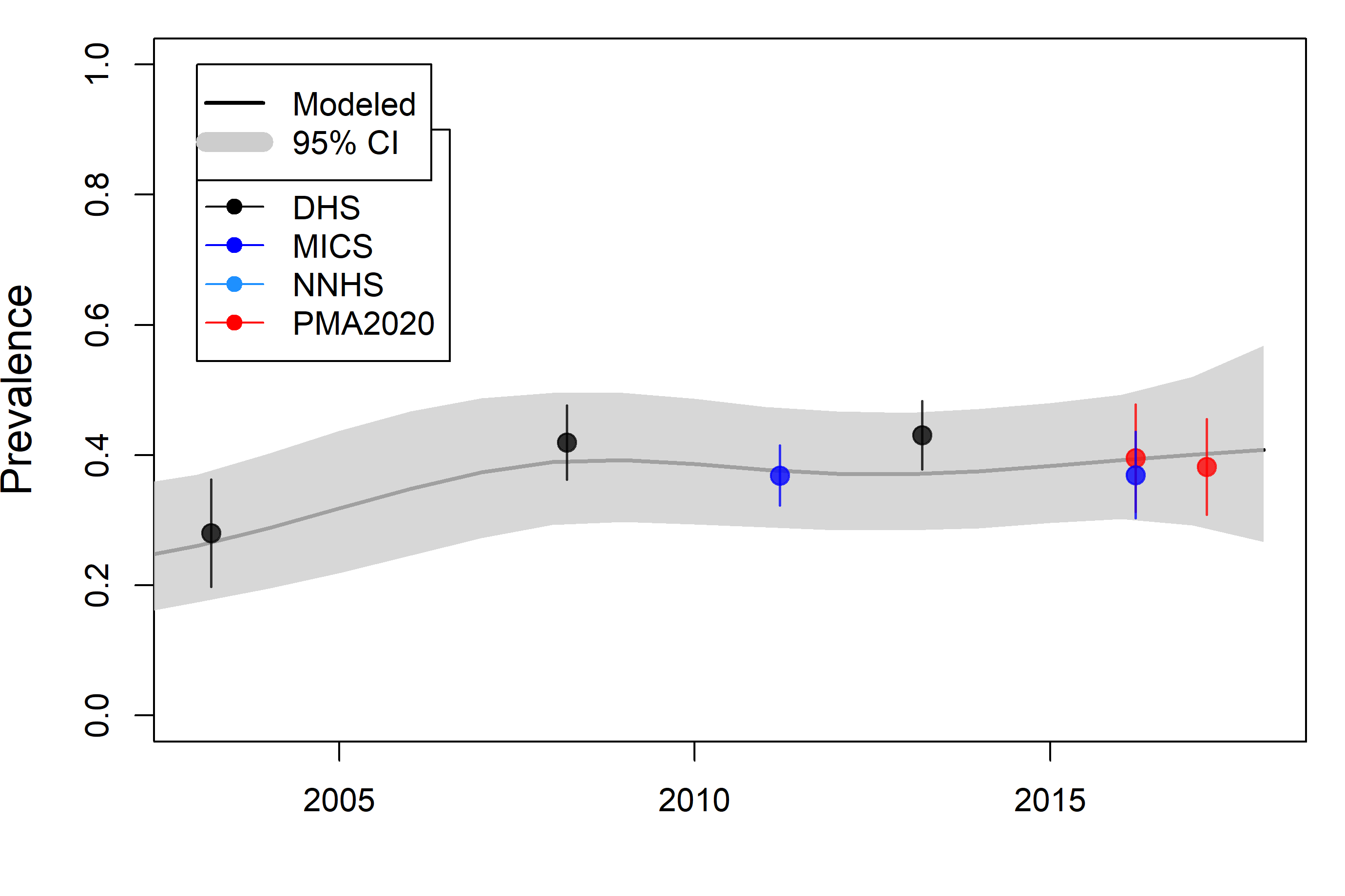}}\\
\end{tabular}

\caption{Data and smoothed estimates for family planning indicators in Anambra state.}\label{anambra}
\end{figure*}

\begin{figure*}[t]
	\centering
\def\tabularxcolumn#1{m{#1}}

\begin{tabular}{cc}
\subfloat[mCPR]{\includegraphics[width=9cm]{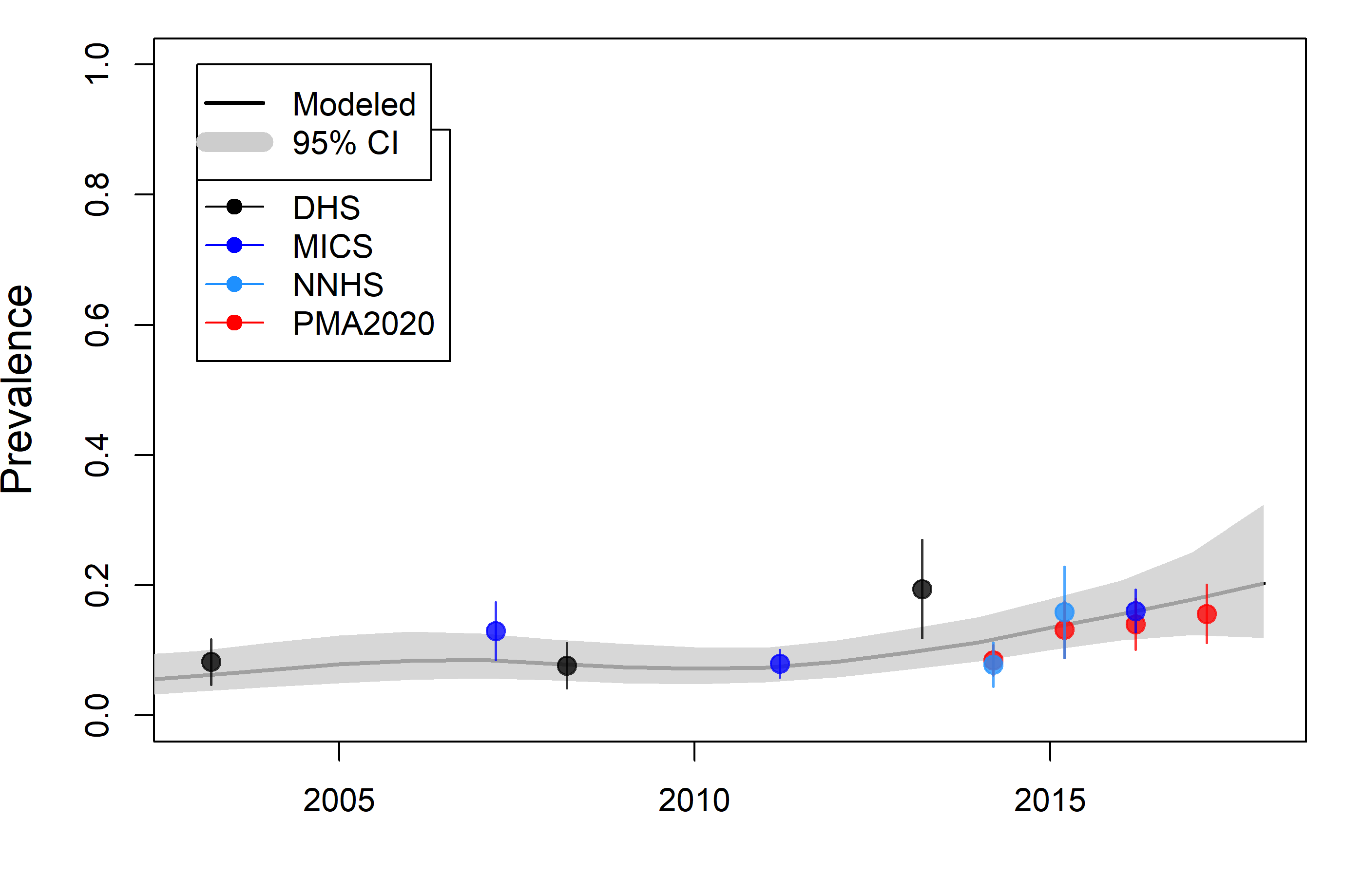}} 
   & \subfloat[Unmet Need]{\includegraphics[width=9cm]{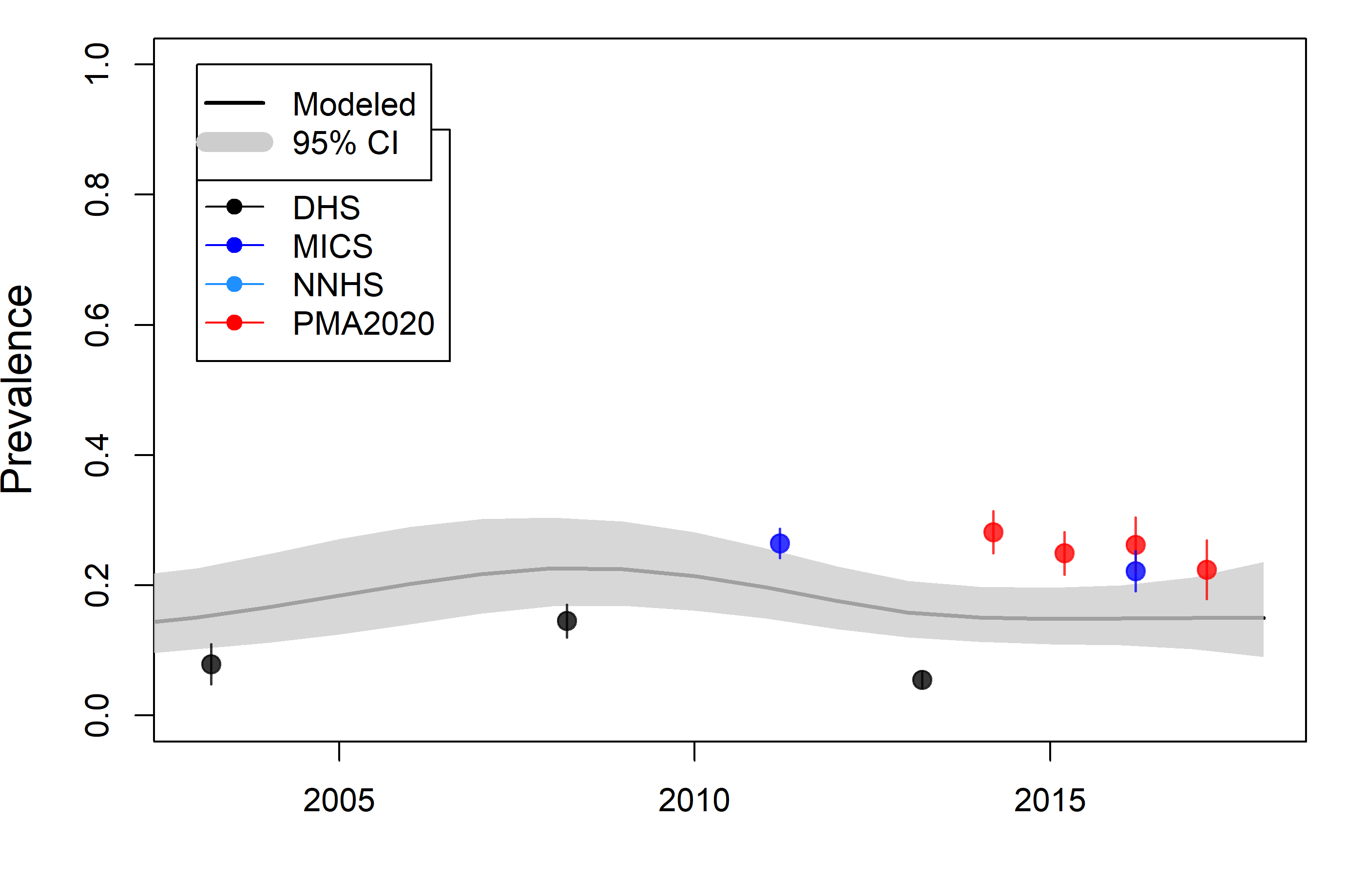}}\\
\subfloat[Traditional CPR]{\includegraphics[width=9cm]{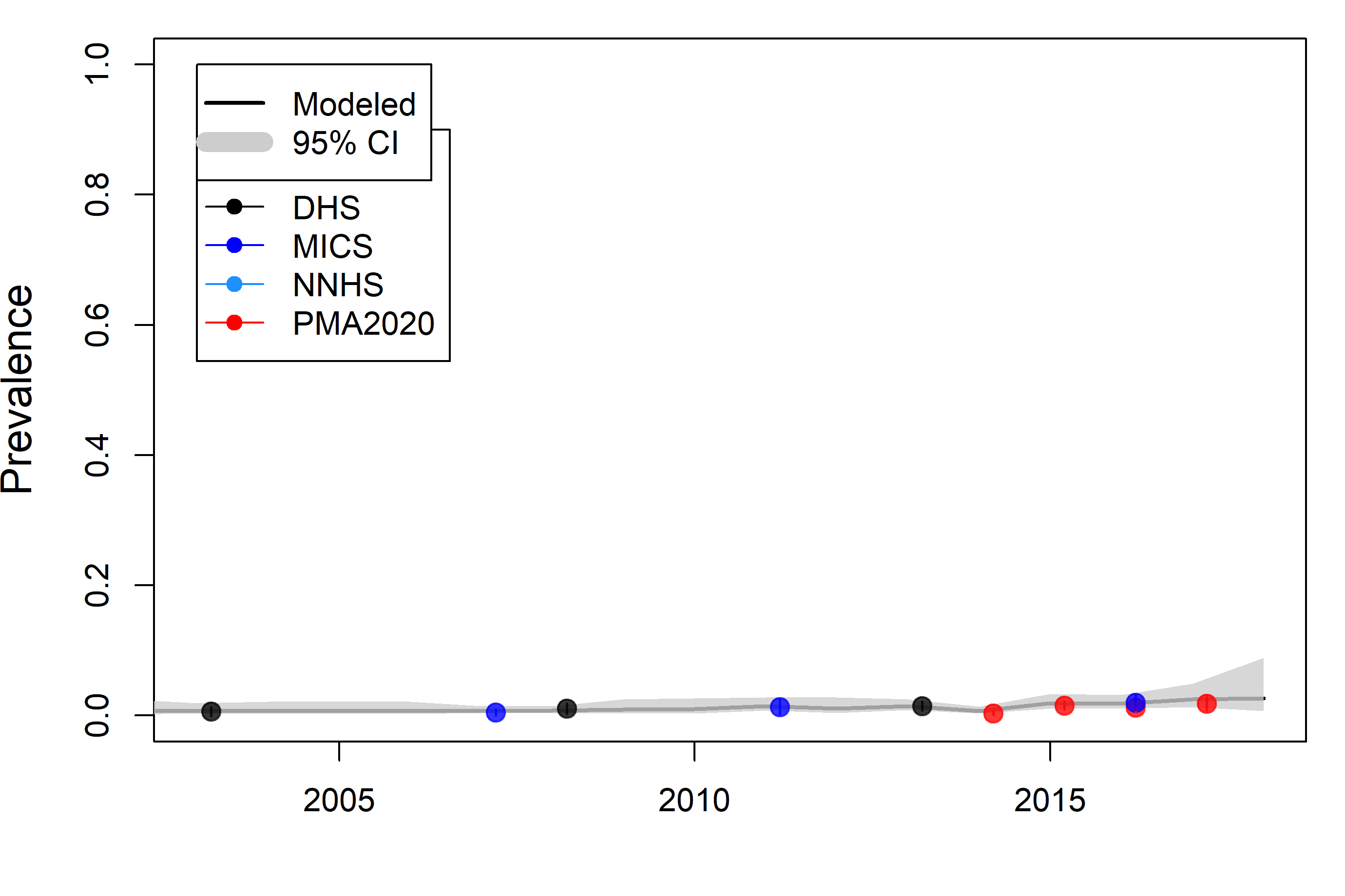}} 
   & \subfloat[Demand Satisfied]{\includegraphics[width=9cm]{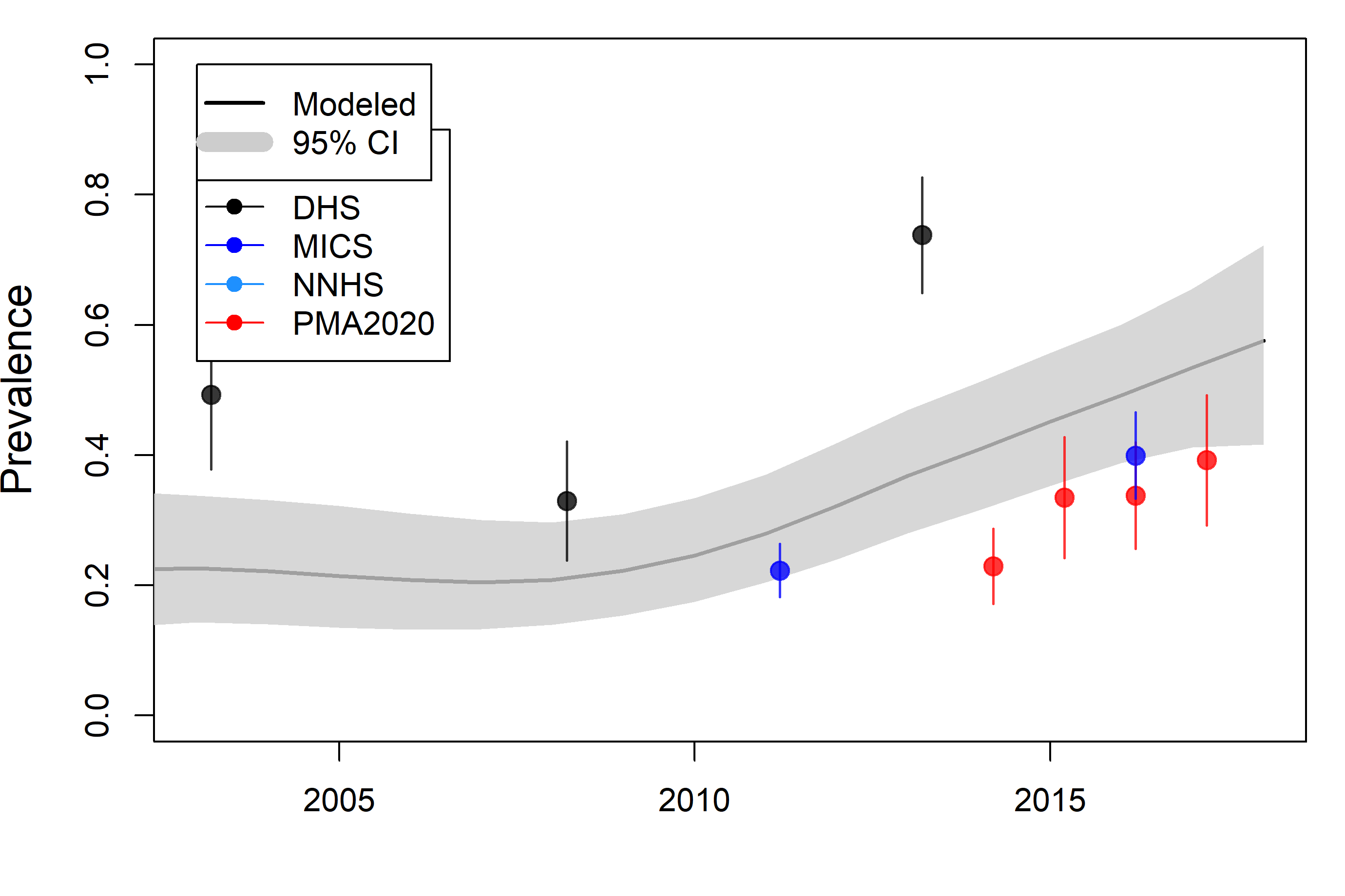}}\\
\end{tabular}

\caption{Data and smoothed estimates for family planning indicators in Kaduna state.}\label{kaduna}
\end{figure*}

\begin{figure*}[t]
	\centering
\def\tabularxcolumn#1{m{#1}}

\begin{tabular}{cc}
\subfloat[mCPR]{\includegraphics[width=9cm]{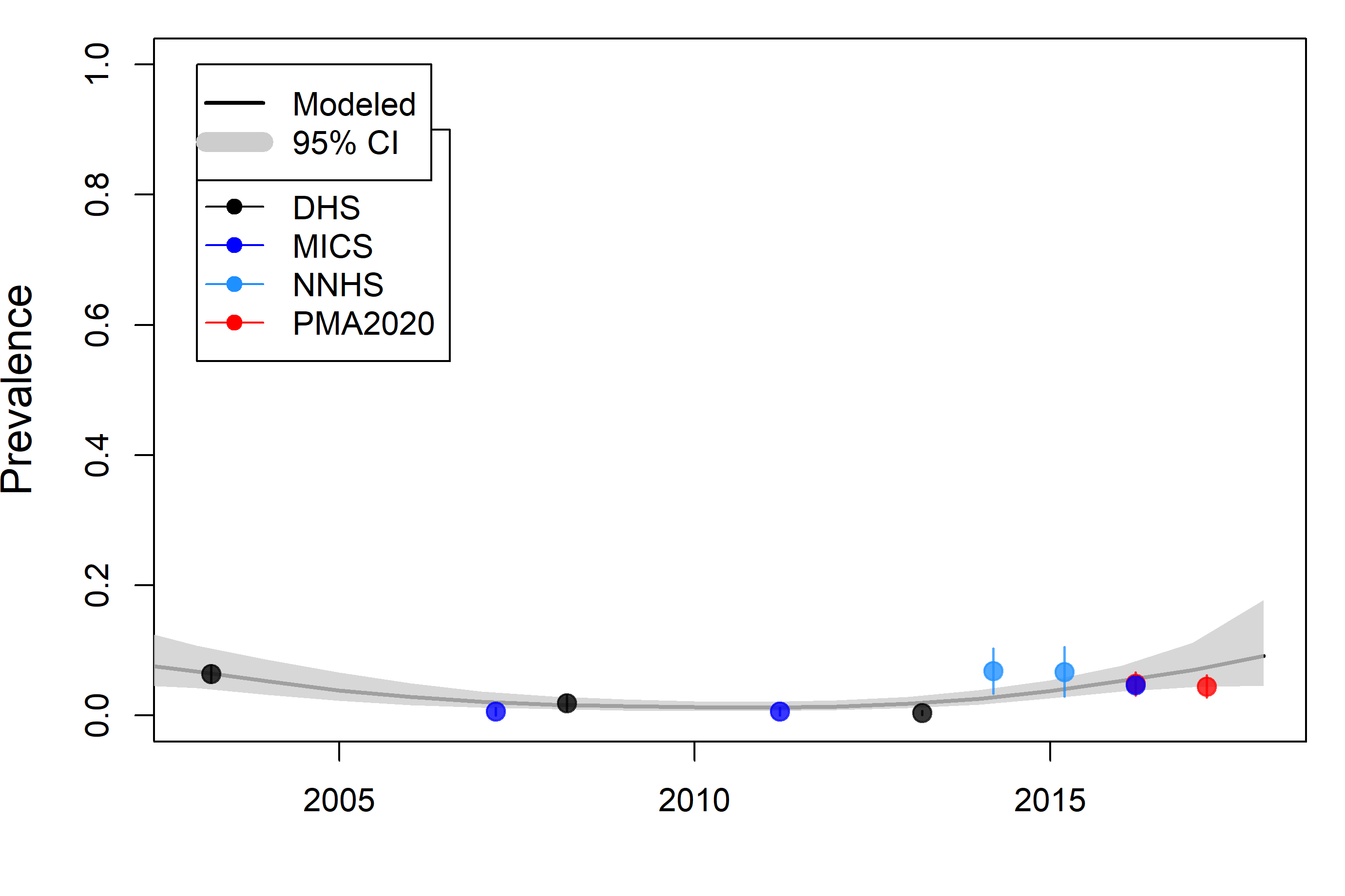}} 
   & \subfloat[Unmet Need]{\includegraphics[width=9cm]{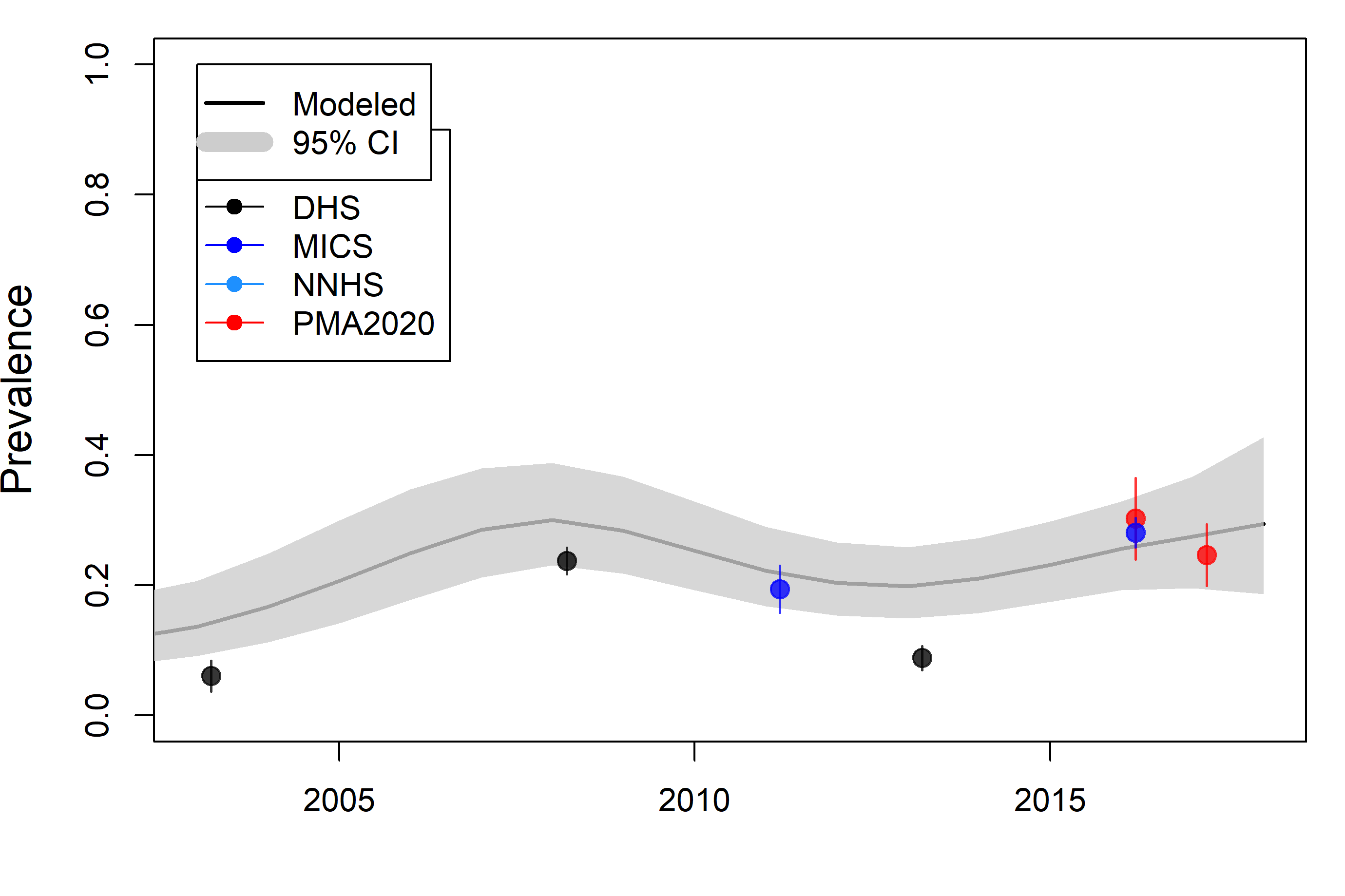}}\\
\subfloat[Traditional CPR]{\includegraphics[width=9cm]{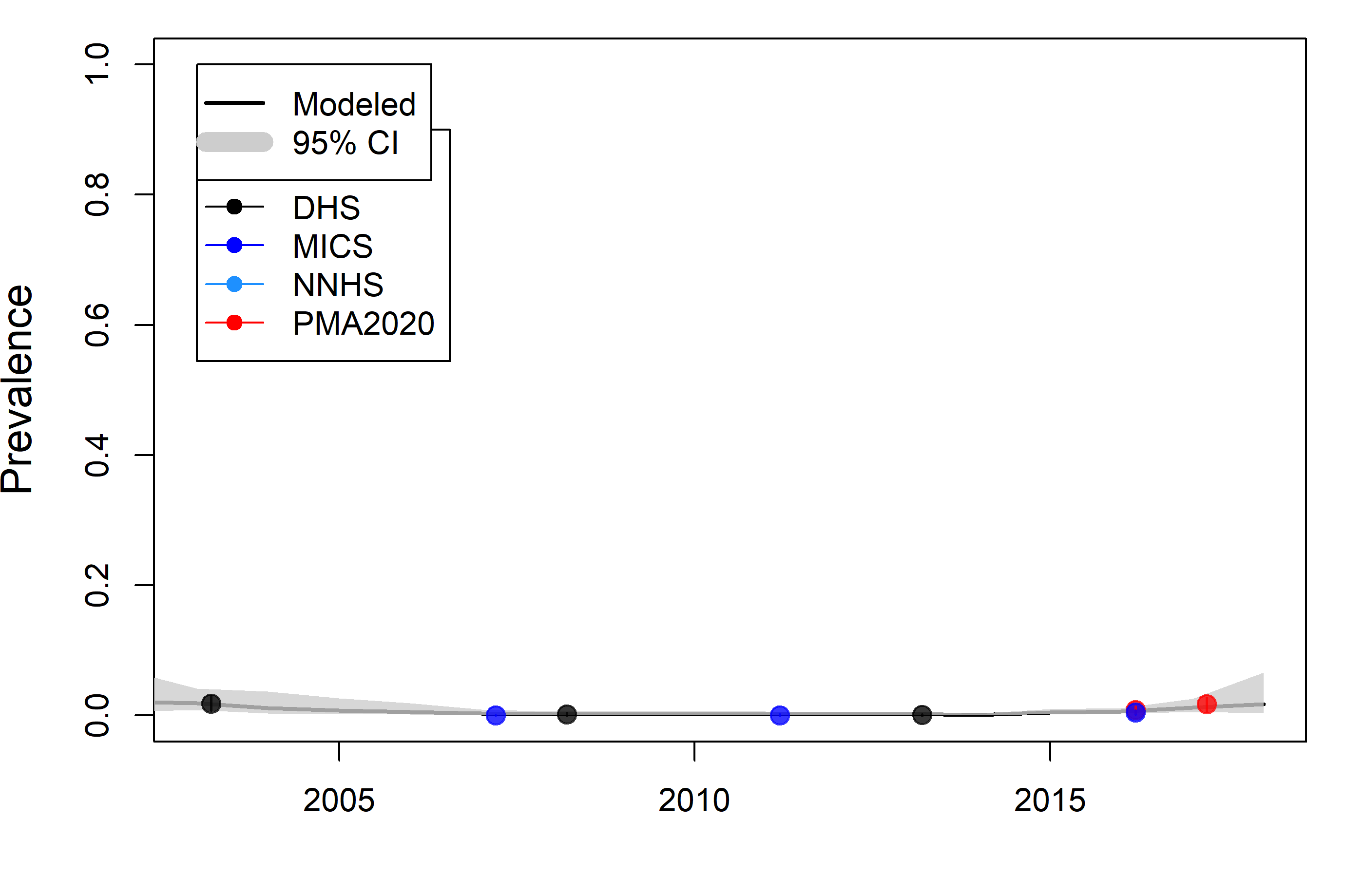}} 
   & \subfloat[Demand Satisfied]{\includegraphics[width=9cm]{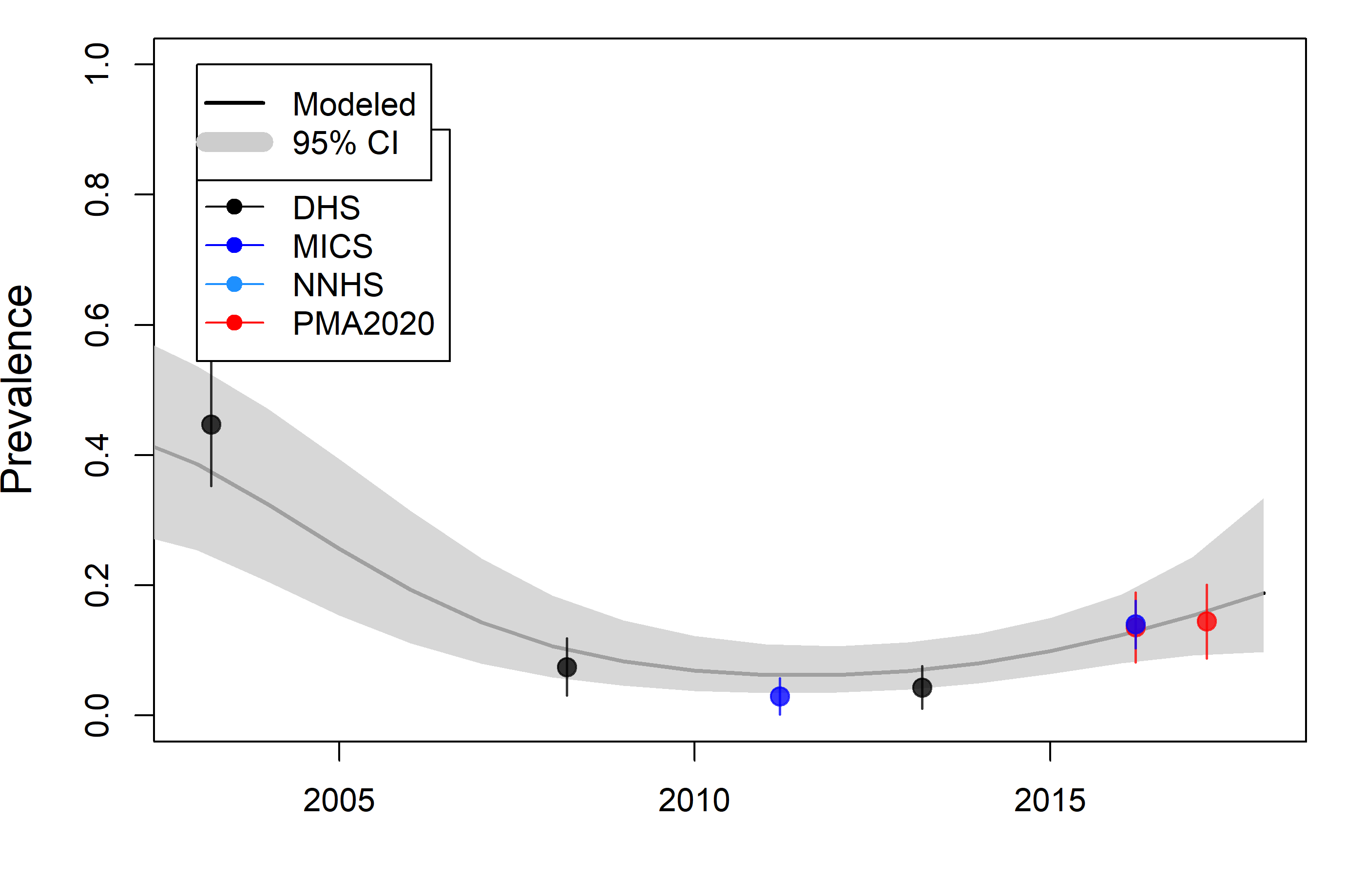}}\\
\end{tabular}

\caption{Data and smoothed estimates for family planning indicators in Kano state.}\label{kano}
\end{figure*}

\begin{figure*}[t]
	\centering
\def\tabularxcolumn#1{m{#1}}

\begin{tabular}{cc}
\subfloat[mCPR]{\includegraphics[width=9cm]{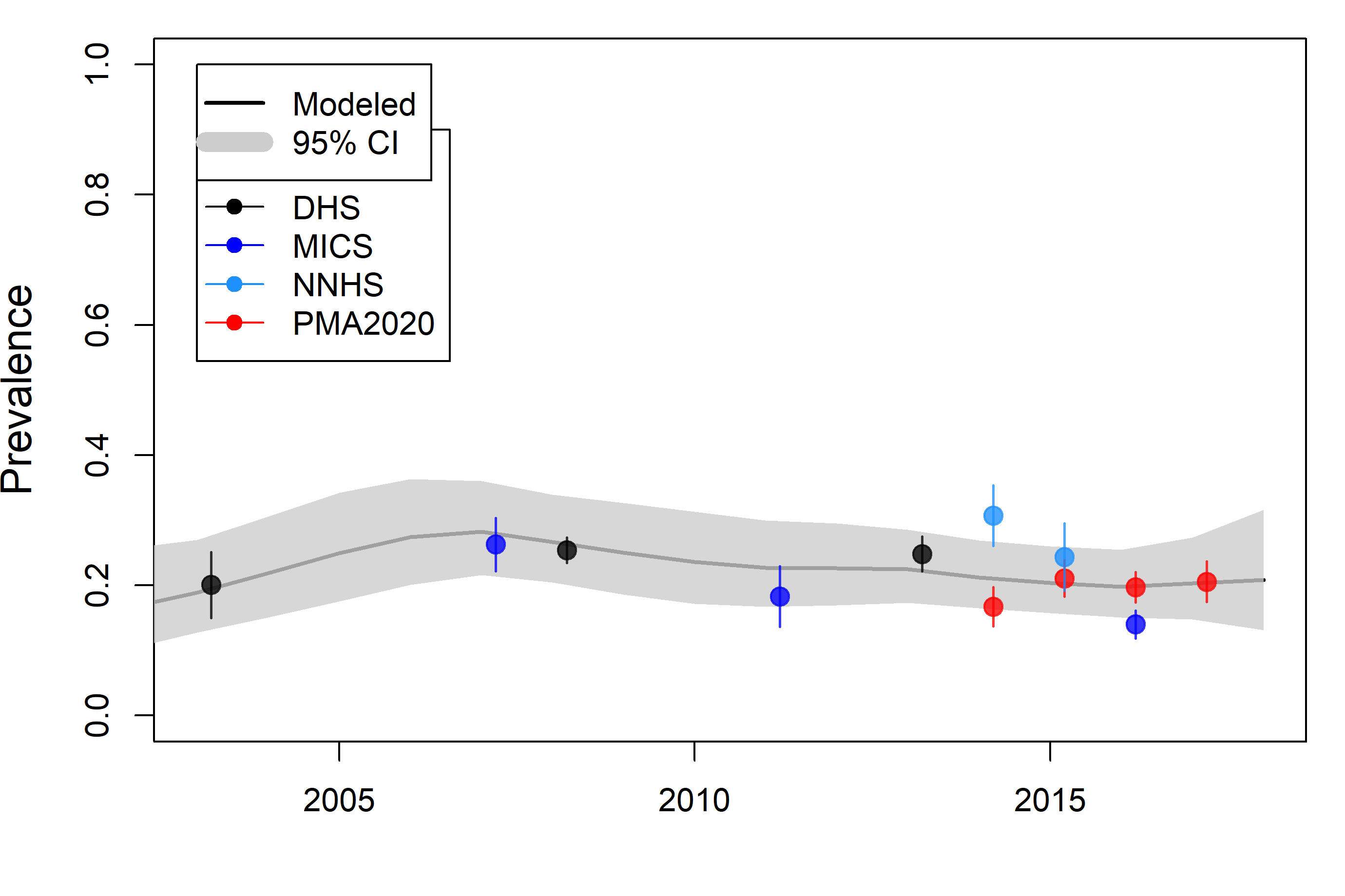}} 
   & \subfloat[Unmet Need]{\includegraphics[width=9cm]{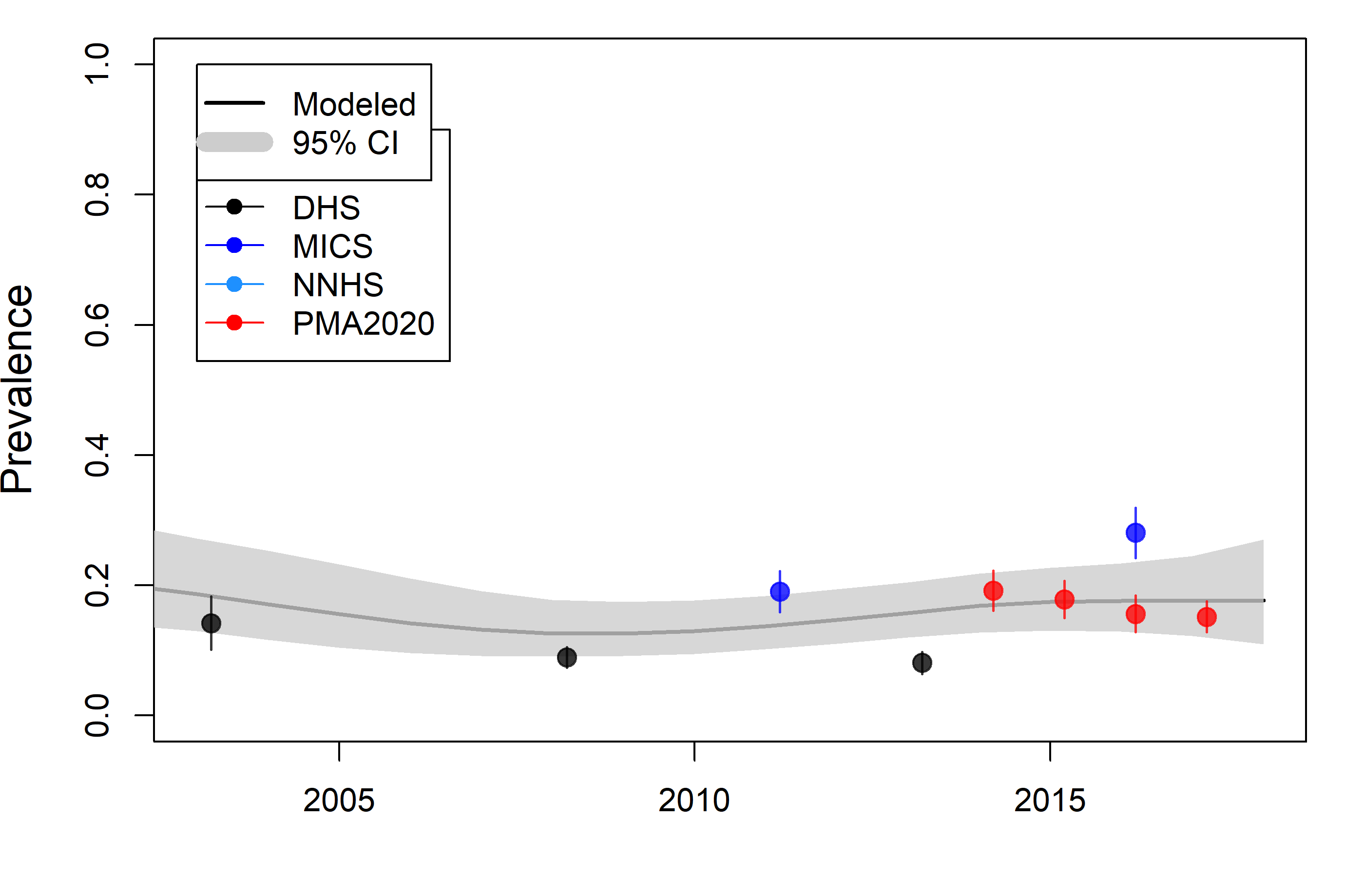}}\\
\subfloat[Traditional CPR]{\includegraphics[width=9cm]{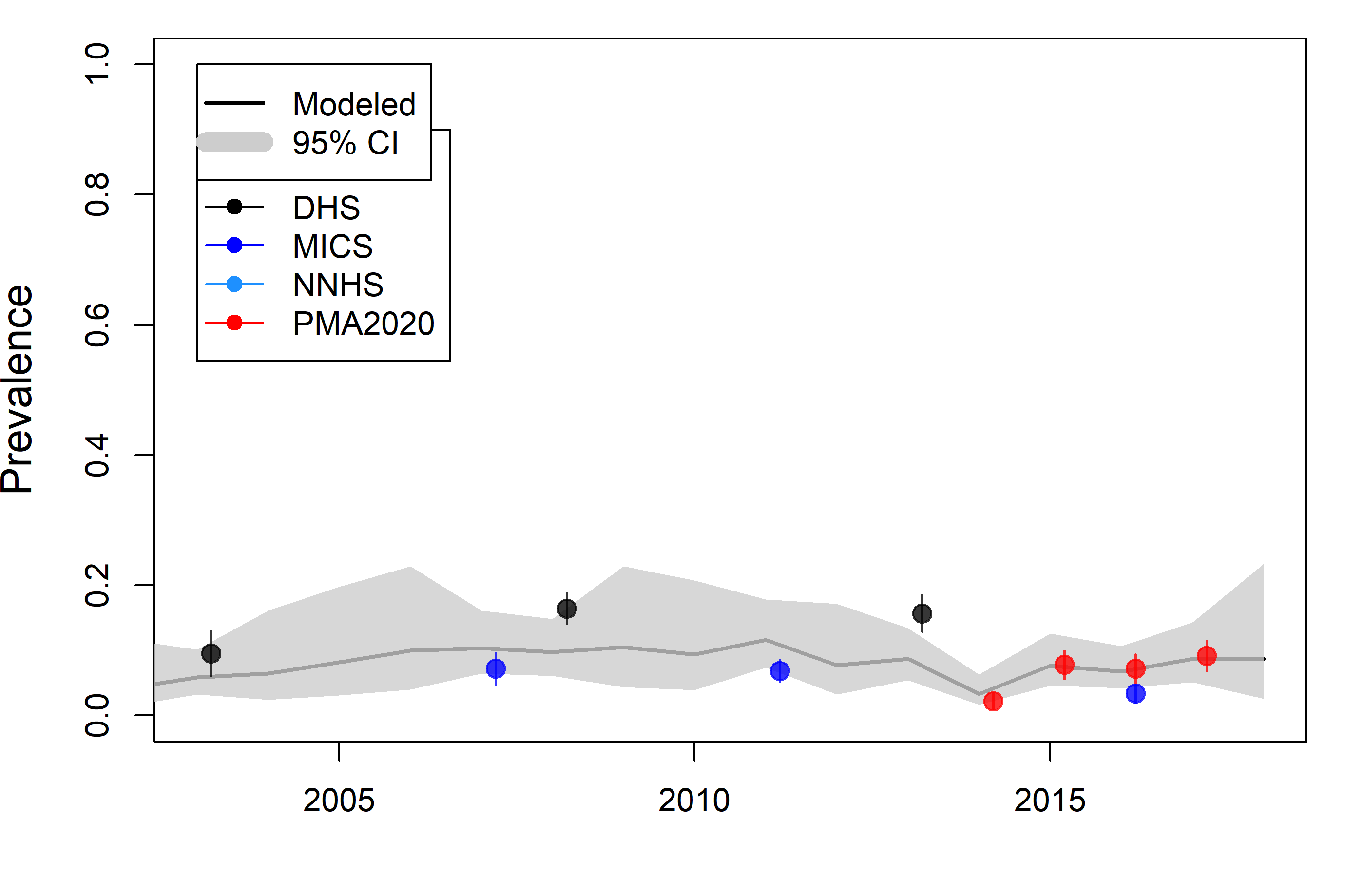}} 
   & \subfloat[Demand Satisfied]{\includegraphics[width=9cm]{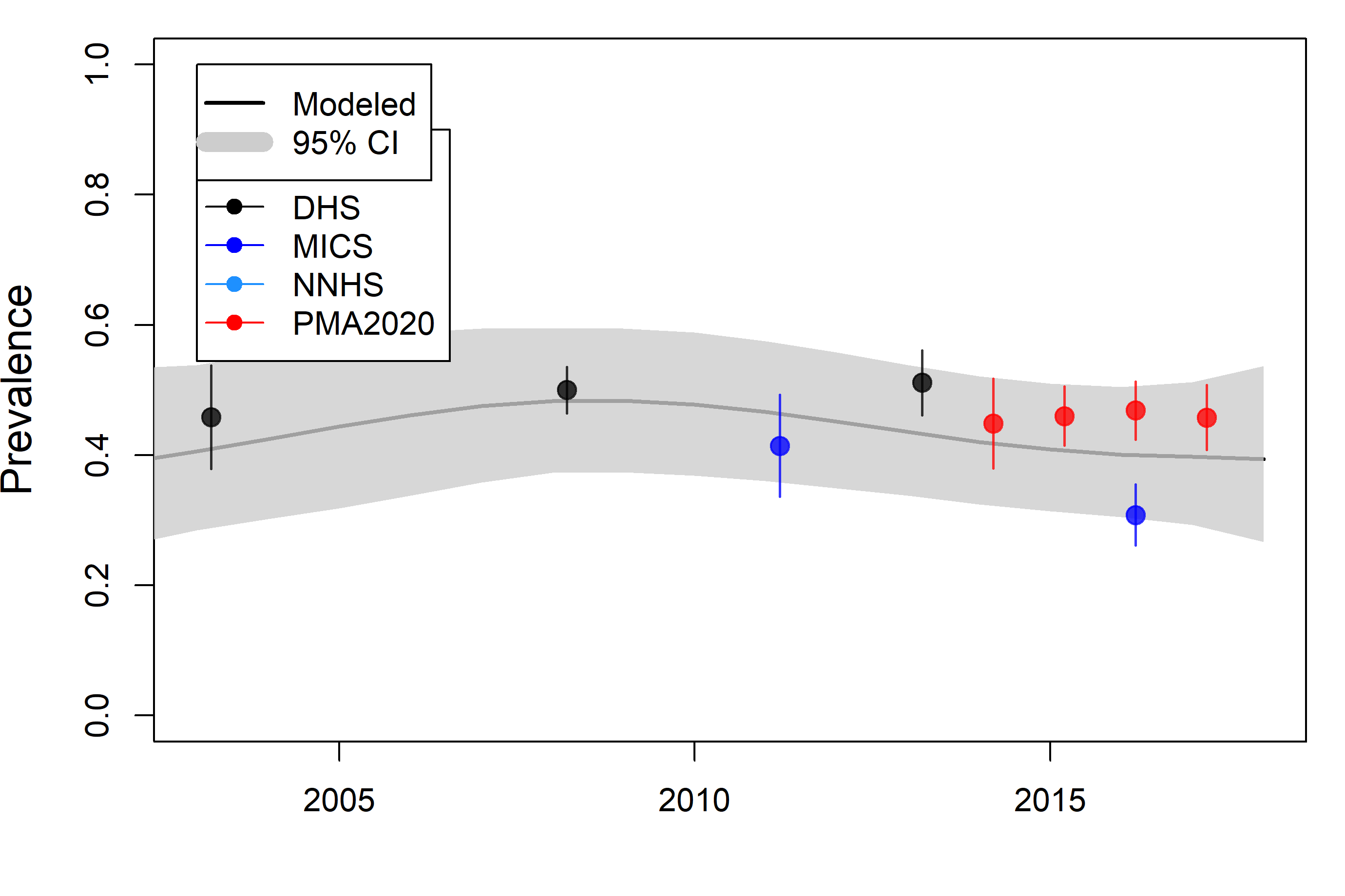}}\\
\end{tabular}

\caption{Data and smoothed estimates for family planning indicators in Lagos state.}\label{lagos}
\end{figure*}

\begin{figure*}[t]
	\centering
\def\tabularxcolumn#1{m{#1}}

\begin{tabular}{cc}
\subfloat[mCPR]{\includegraphics[width=9cm]{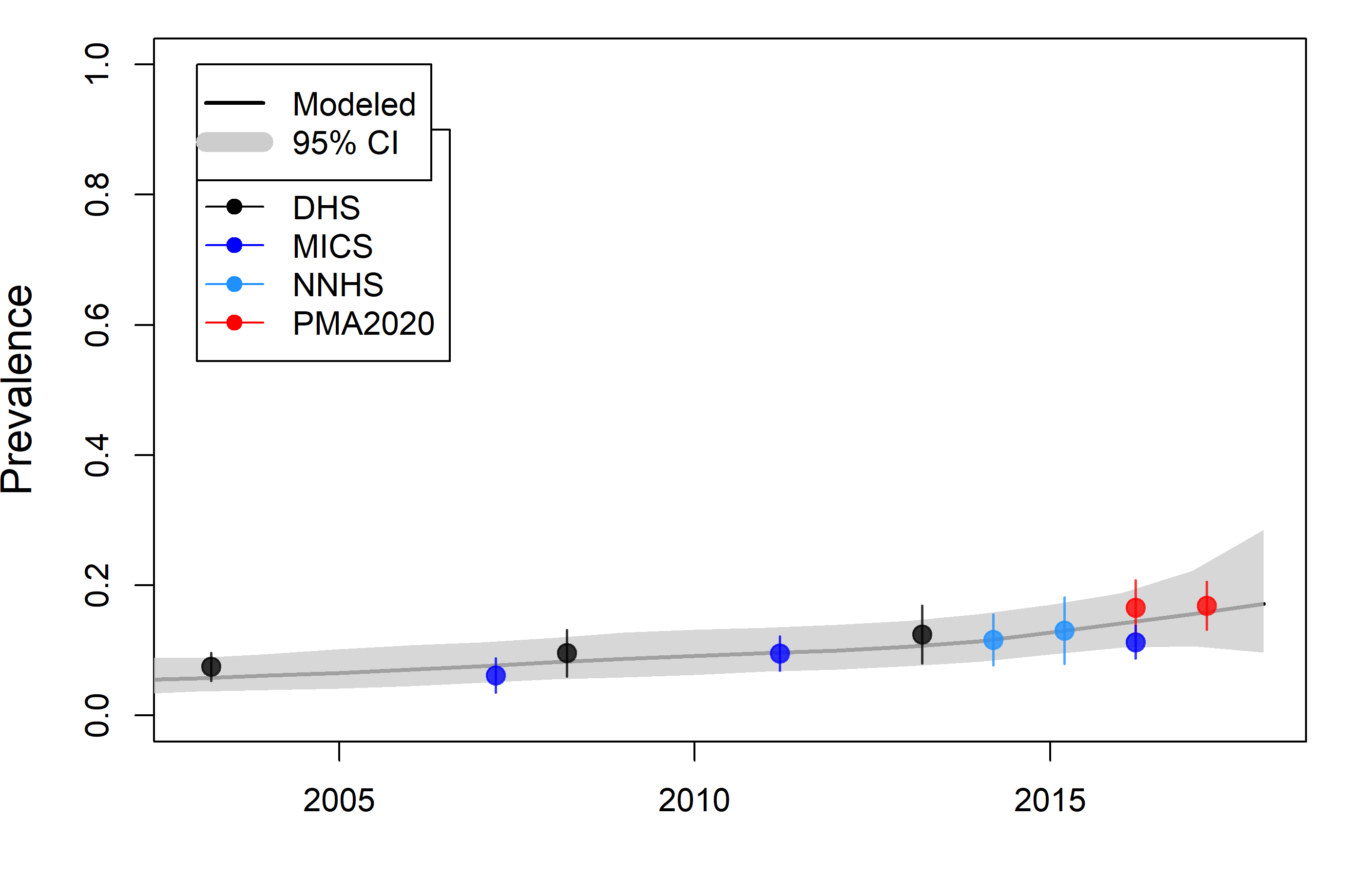}} 
   & \subfloat[Unmet Need]{\includegraphics[width=9cm]{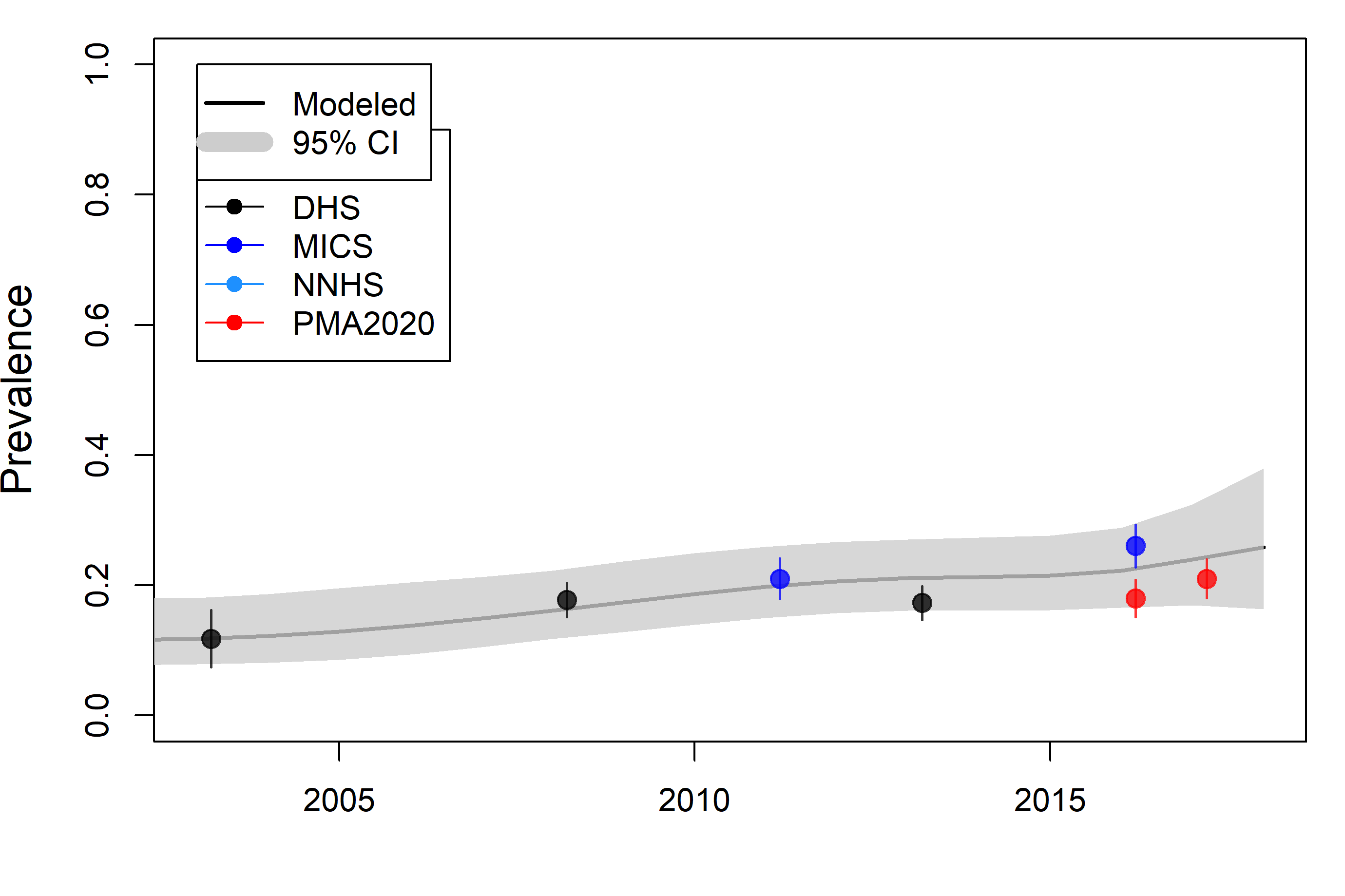}}\\
\subfloat[Traditional CPR]{\includegraphics[width=9cm]{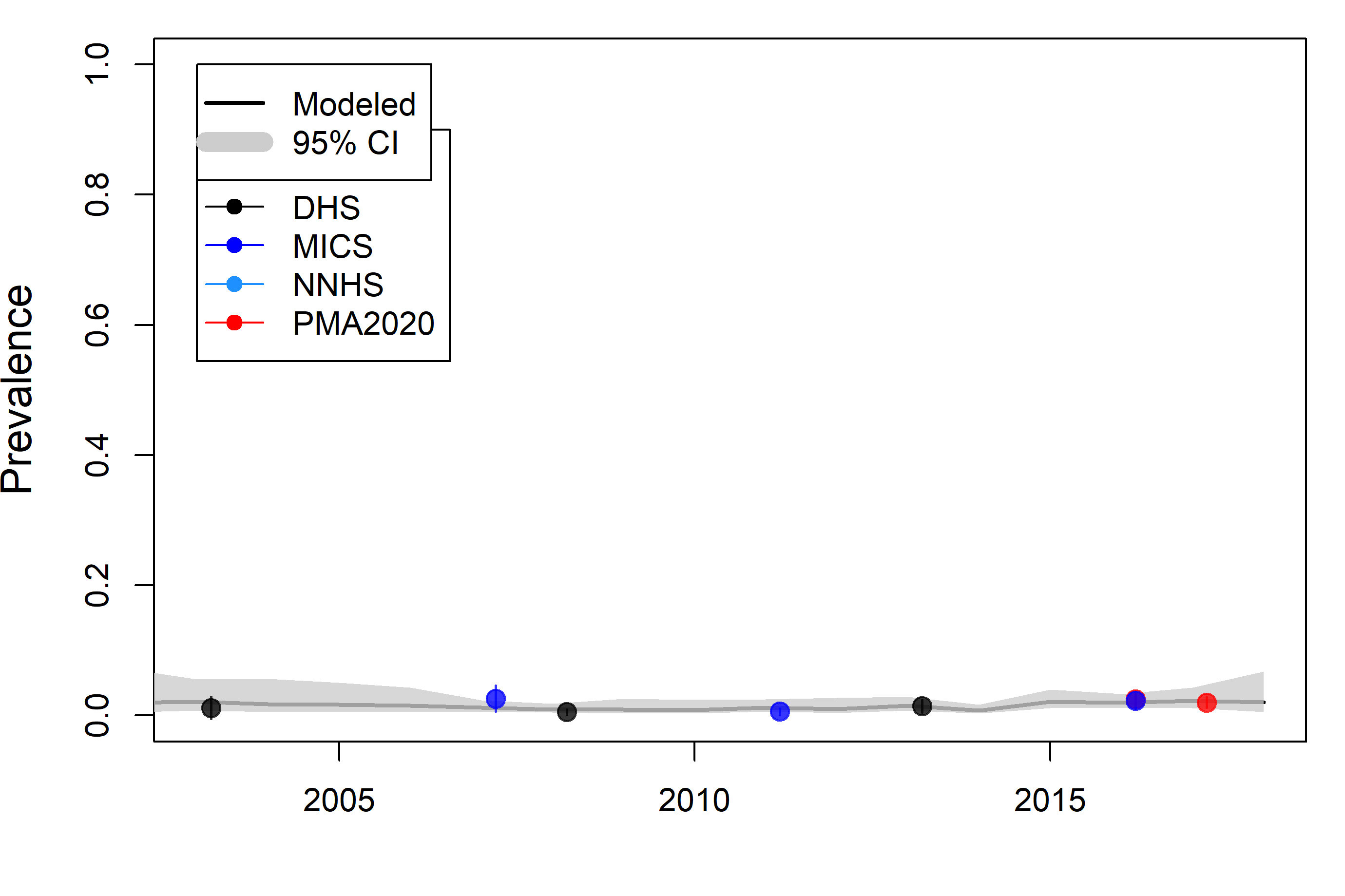}} 
   & \subfloat[Demand Satisfied]{\includegraphics[width=9cm]{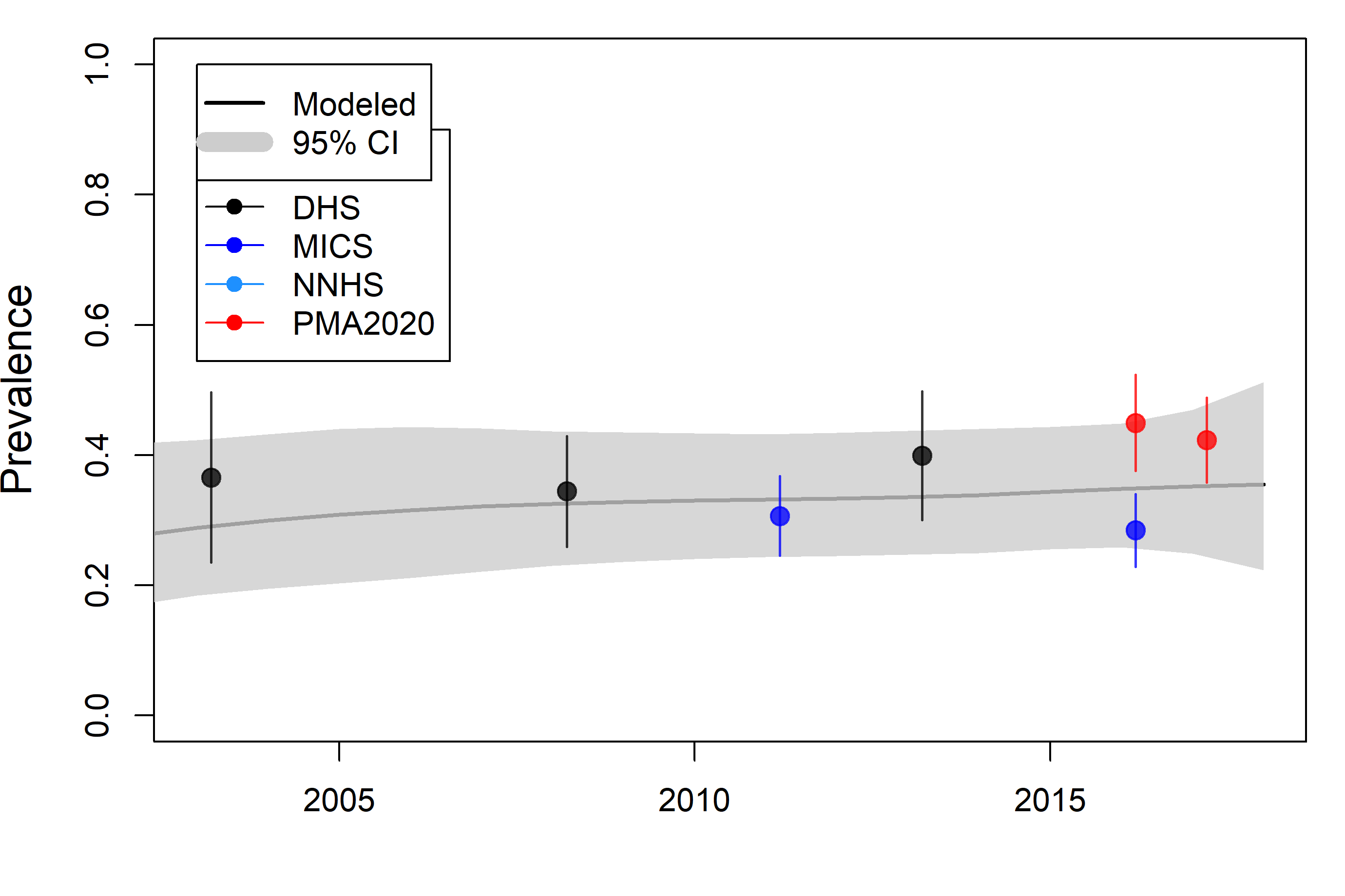}}\\
\end{tabular}

\caption{Data and smoothed estimates for family planning indicators in Nasarawa state.}\label{nasarawa}
\end{figure*}

\begin{figure*}[t]
	\centering
\def\tabularxcolumn#1{m{#1}}

\begin{tabular}{cc}
\subfloat[mCPR]{\includegraphics[width=9cm]{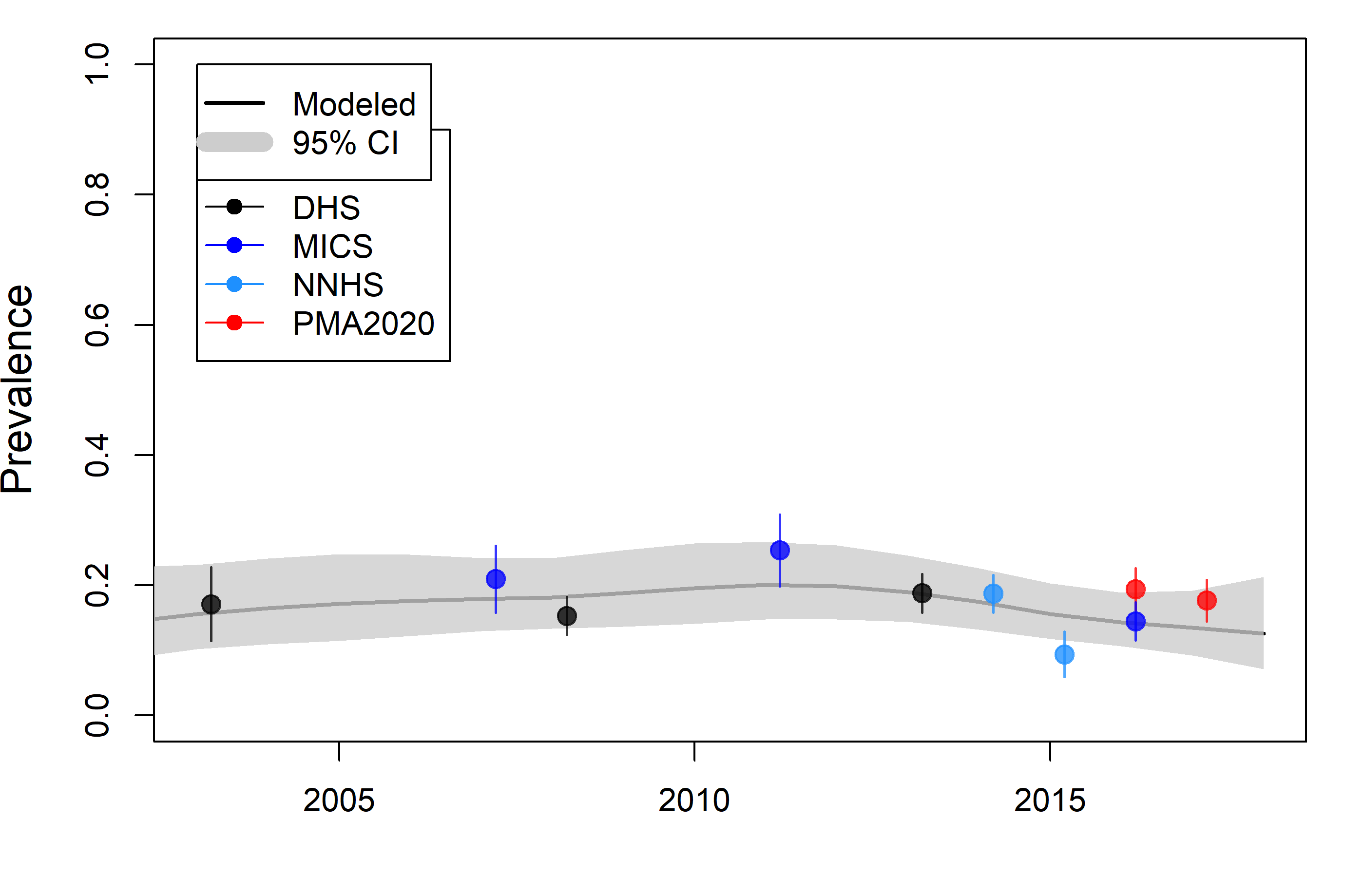}} 
   & \subfloat[Unmet Need]{\includegraphics[width=9cm]{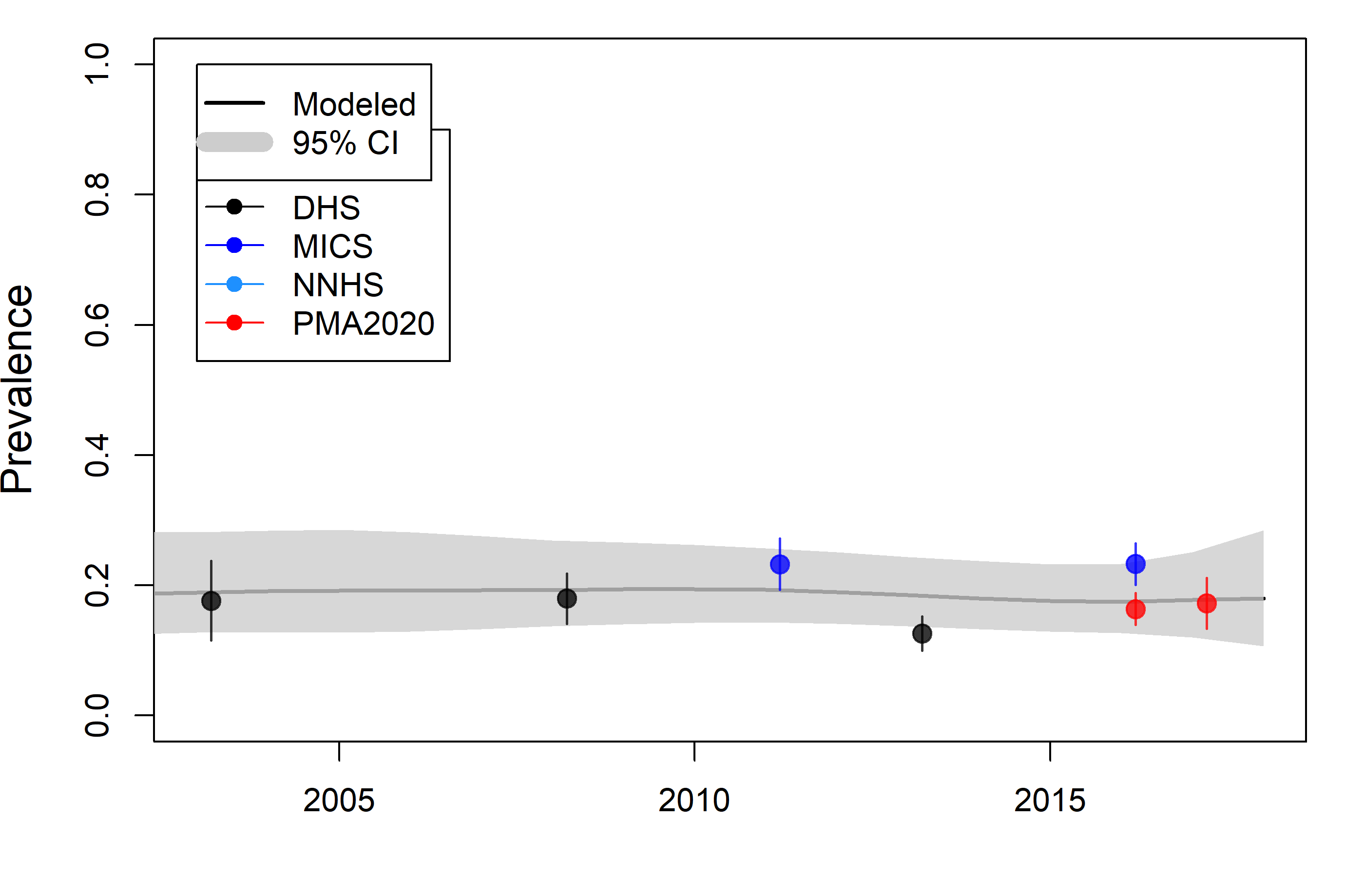}}\\
\subfloat[Traditional CPR]{\includegraphics[width=9cm]{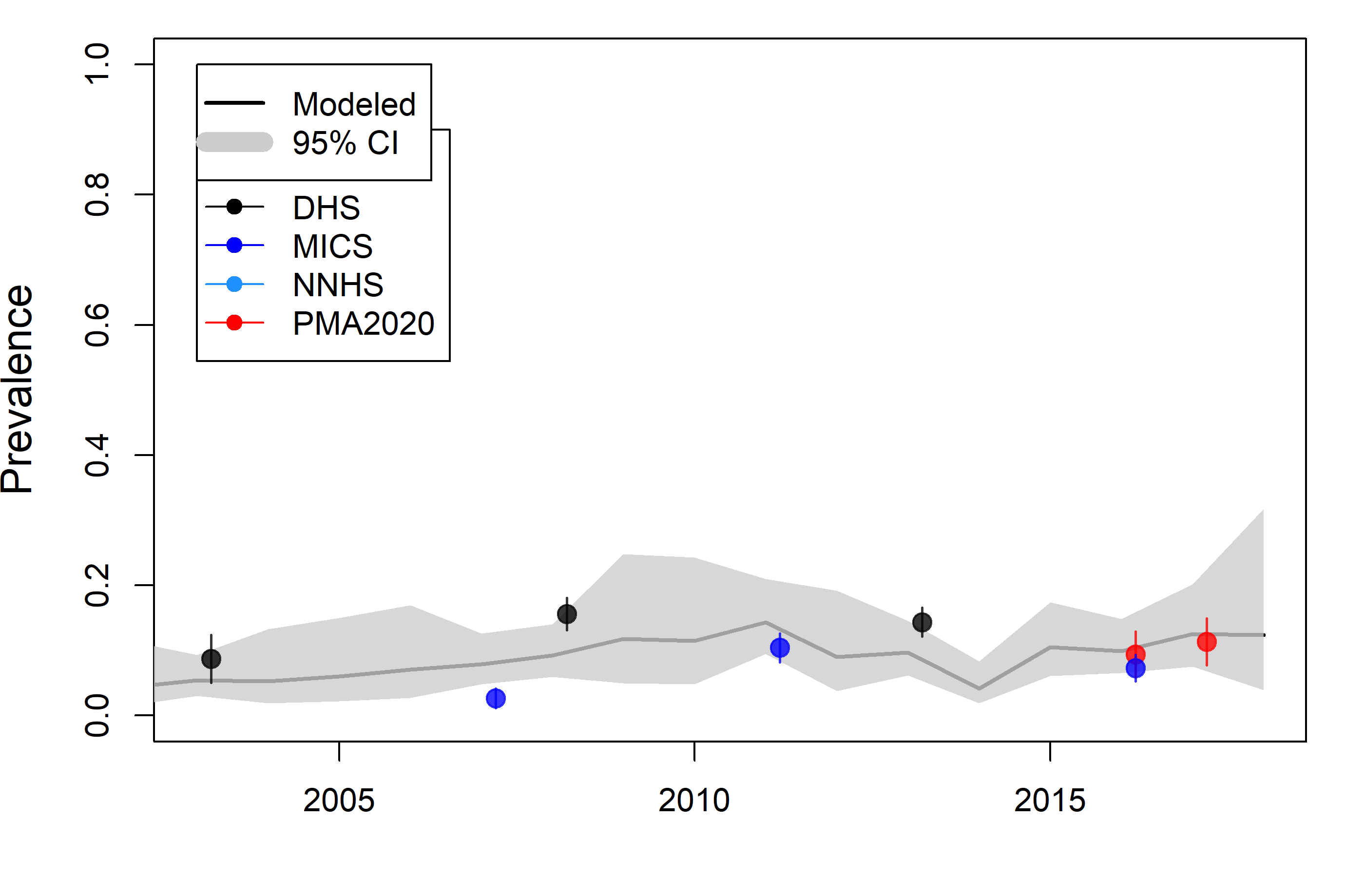}} 
   & \subfloat[Demand Satisfied]{\includegraphics[width=9cm]{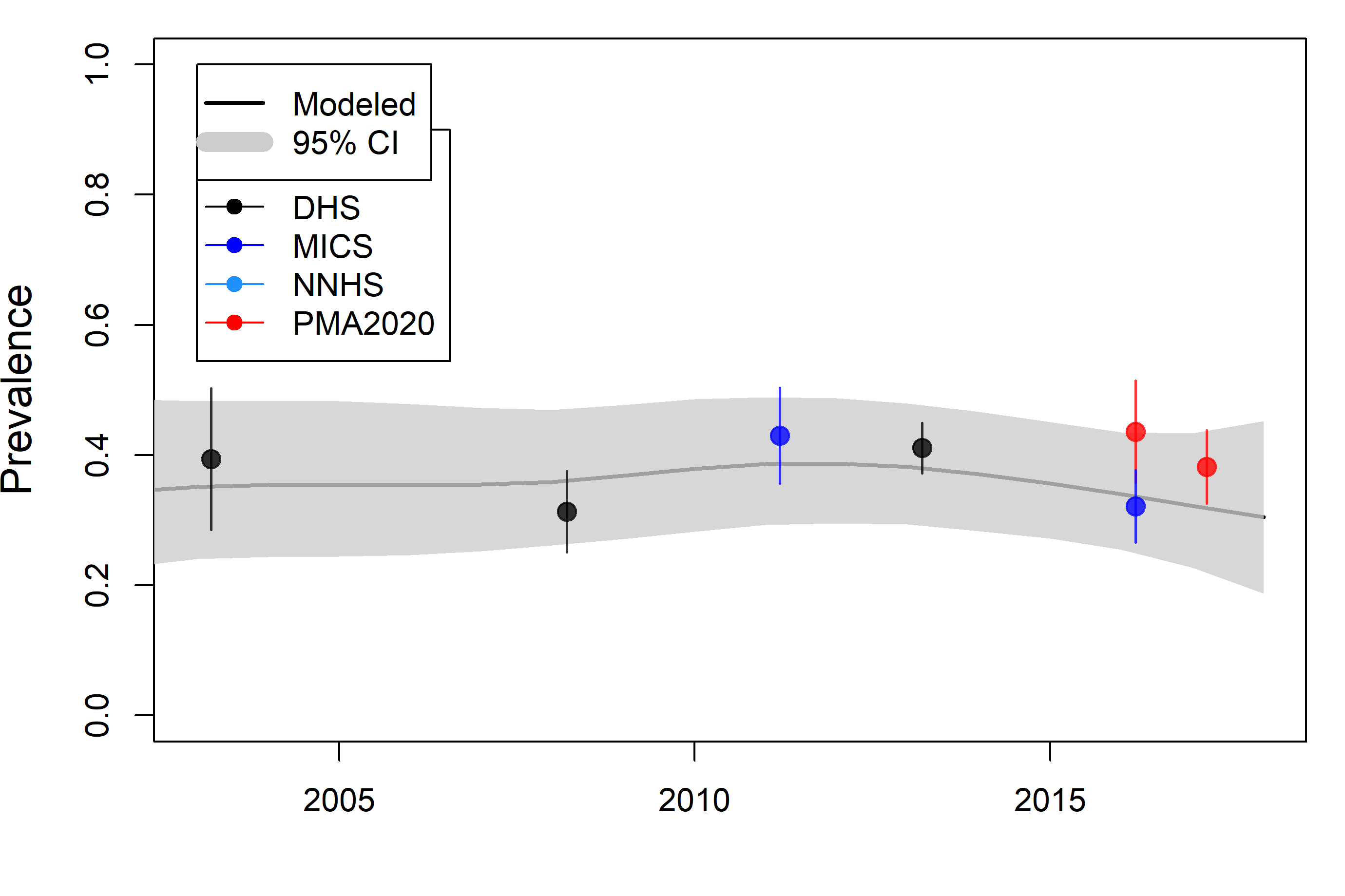}}\\
\end{tabular}

\caption{Data and smoothed estimates for family planning indicators in Rivers state.}\label{rivers}
\end{figure*}

\begin{figure*}[t]
	\centering
\def\tabularxcolumn#1{m{#1}}

\begin{tabular}{cc}
\subfloat[mCPR]{\includegraphics[width=9cm]{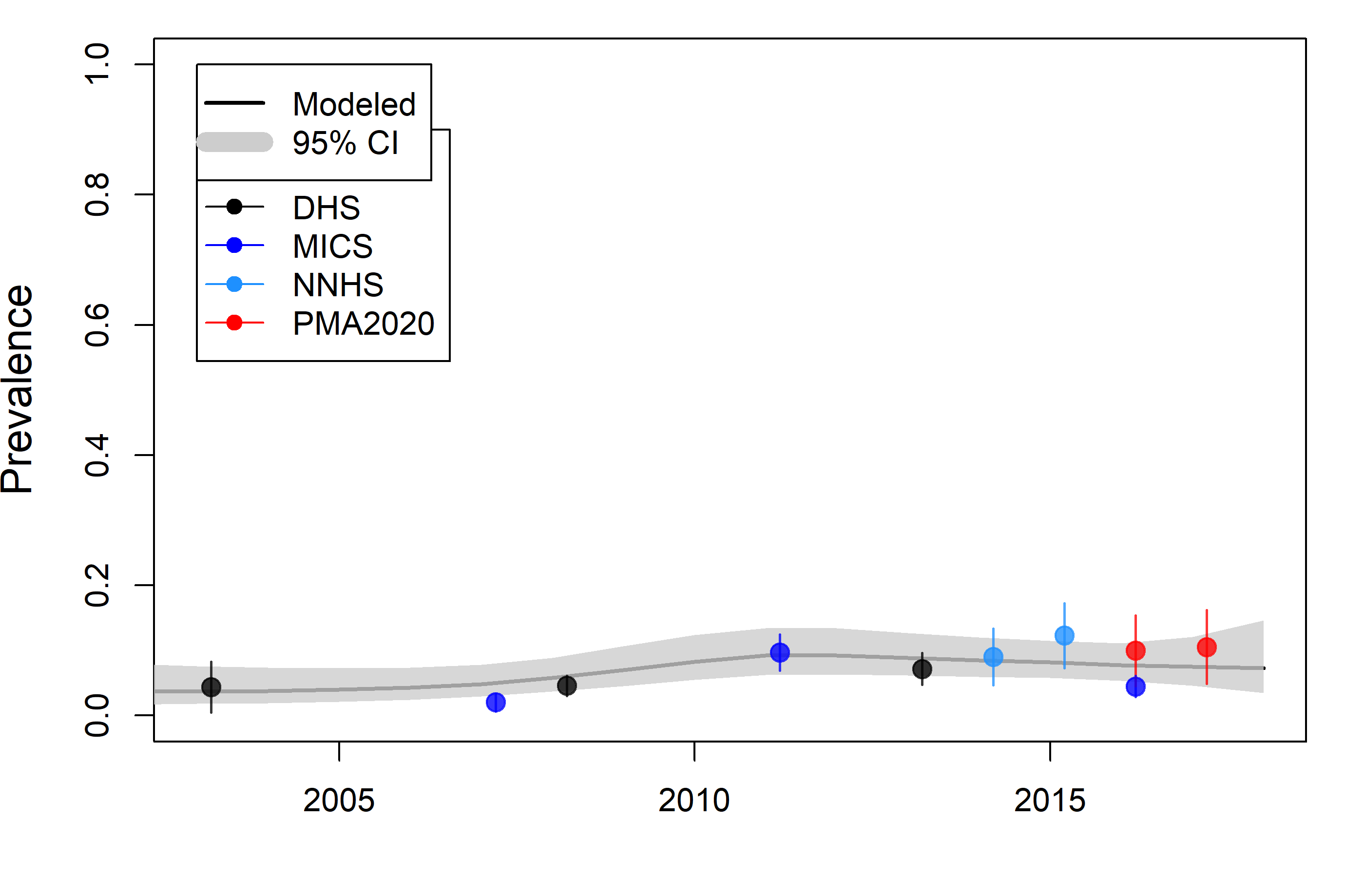}} 
   & \subfloat[Unmet Need]{\includegraphics[width=9cm]{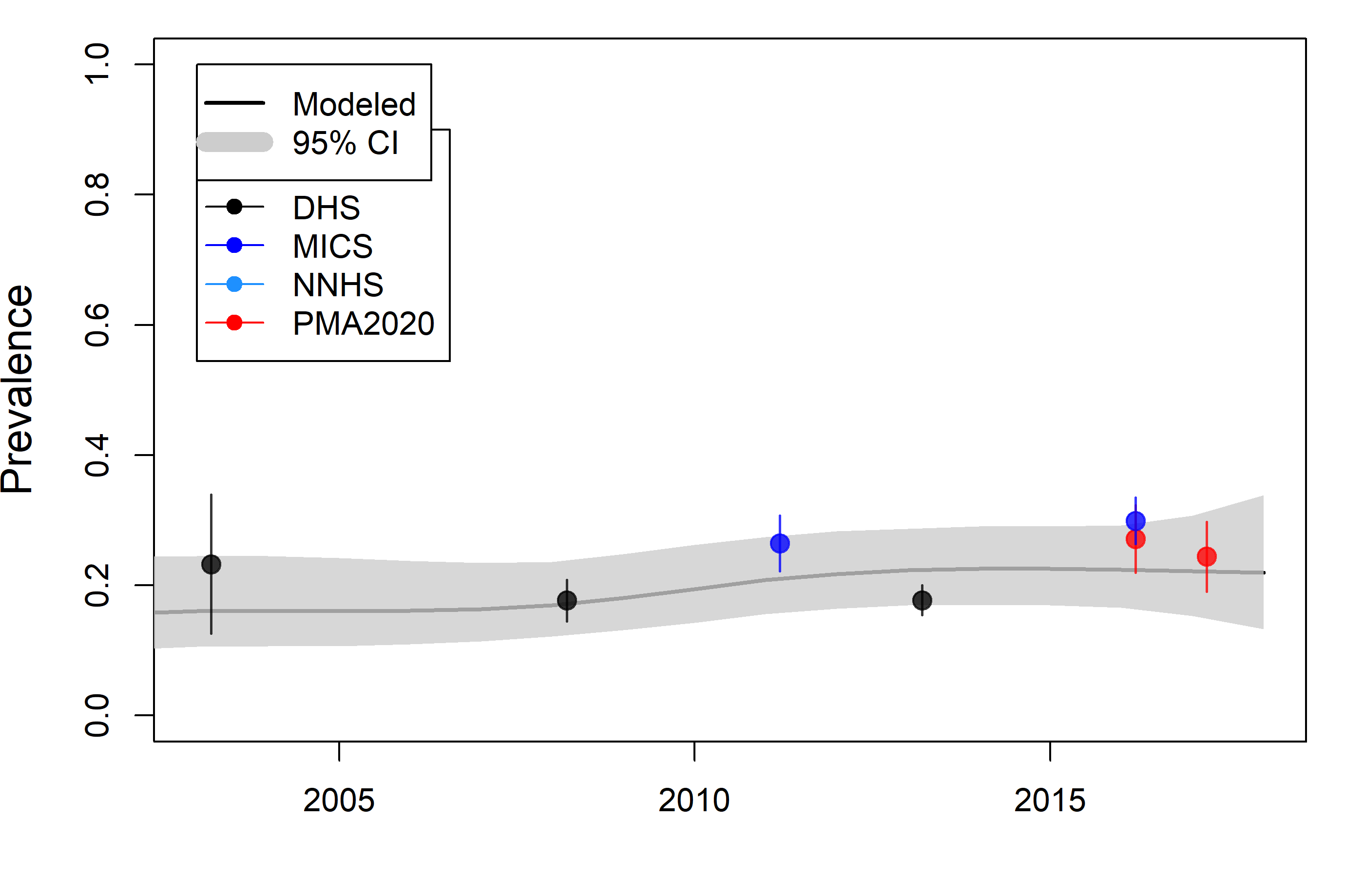}}\\
\subfloat[Traditional CPR]{\includegraphics[width=9cm]{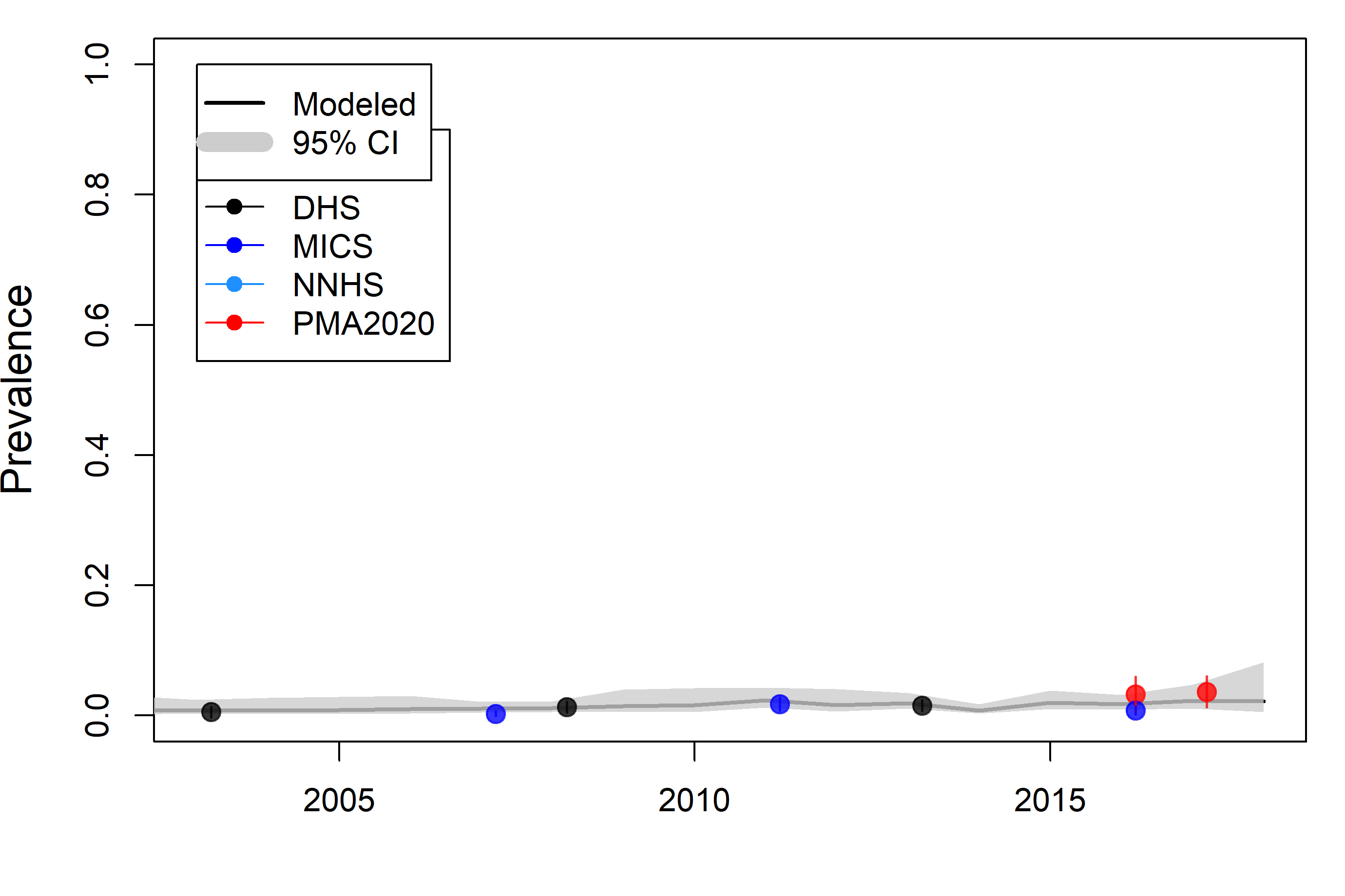}} 
   & \subfloat[Demand Satisfied]{\includegraphics[width=9cm]{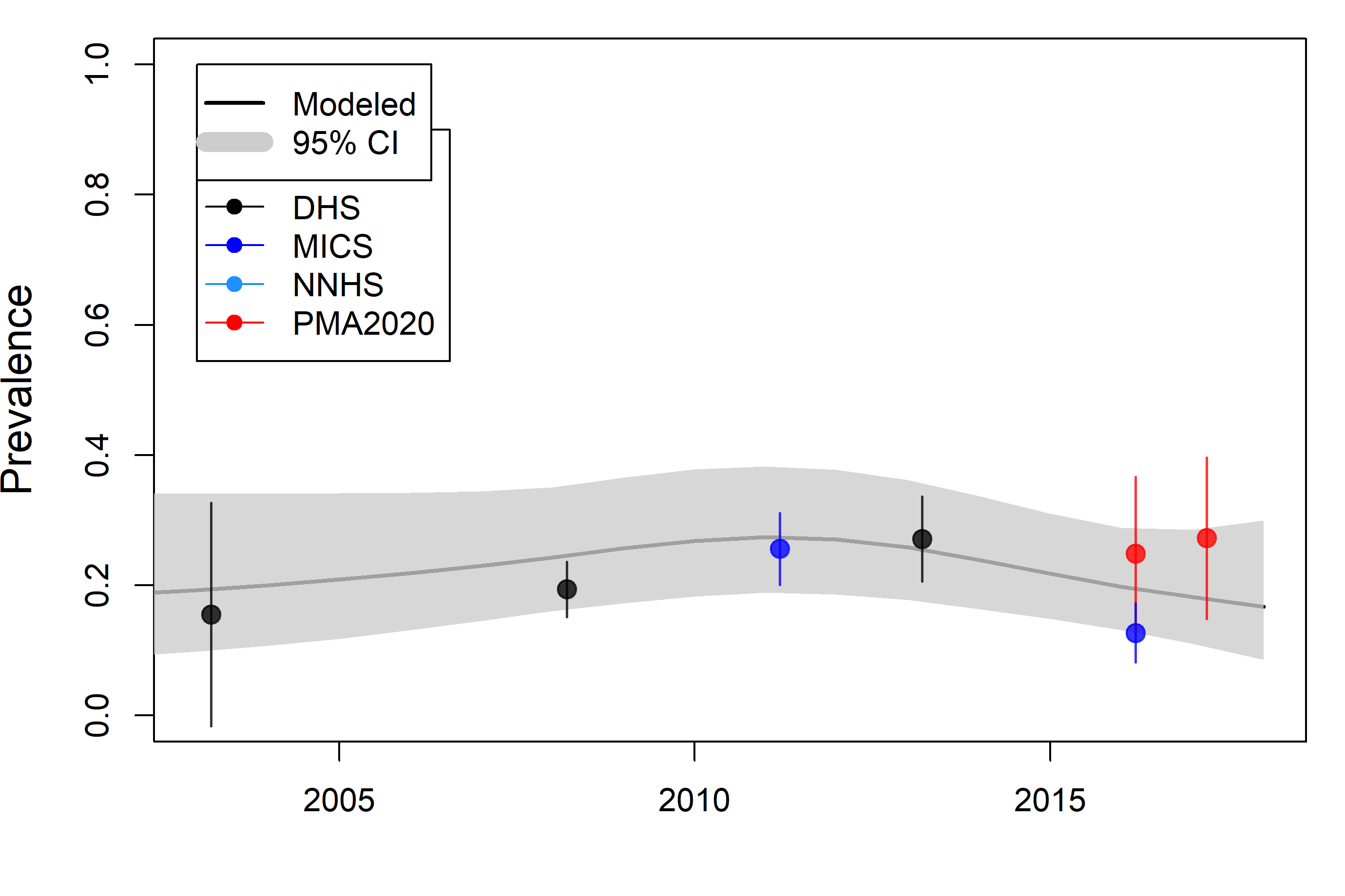}}\\
\end{tabular}

\caption{Data and smoothed estimates for family planning indicators in Taraba state.}\label{taraba}
\end{figure*}

The observed data and smoothed estimates for mCPR, traditional contraceptive prevalence, unment need, and demand satistified in the seven states (Anambra, Kaduna, Kano, Lagos, Nasarawa, Rivers, and Taraba) are shown in Figures \ref{anambra} through \ref{taraba}.


\section{Maps of family planning indicators for all age-parity subgroups}
\label{s:map2017}

\begin{figure*}[t]
	\centering
\def\tabularxcolumn#1{m{#1}}

\begin{tabular}{cc}
\subfloat[Age 15-24 years and parity 0.]{\includegraphics[width=9cm]{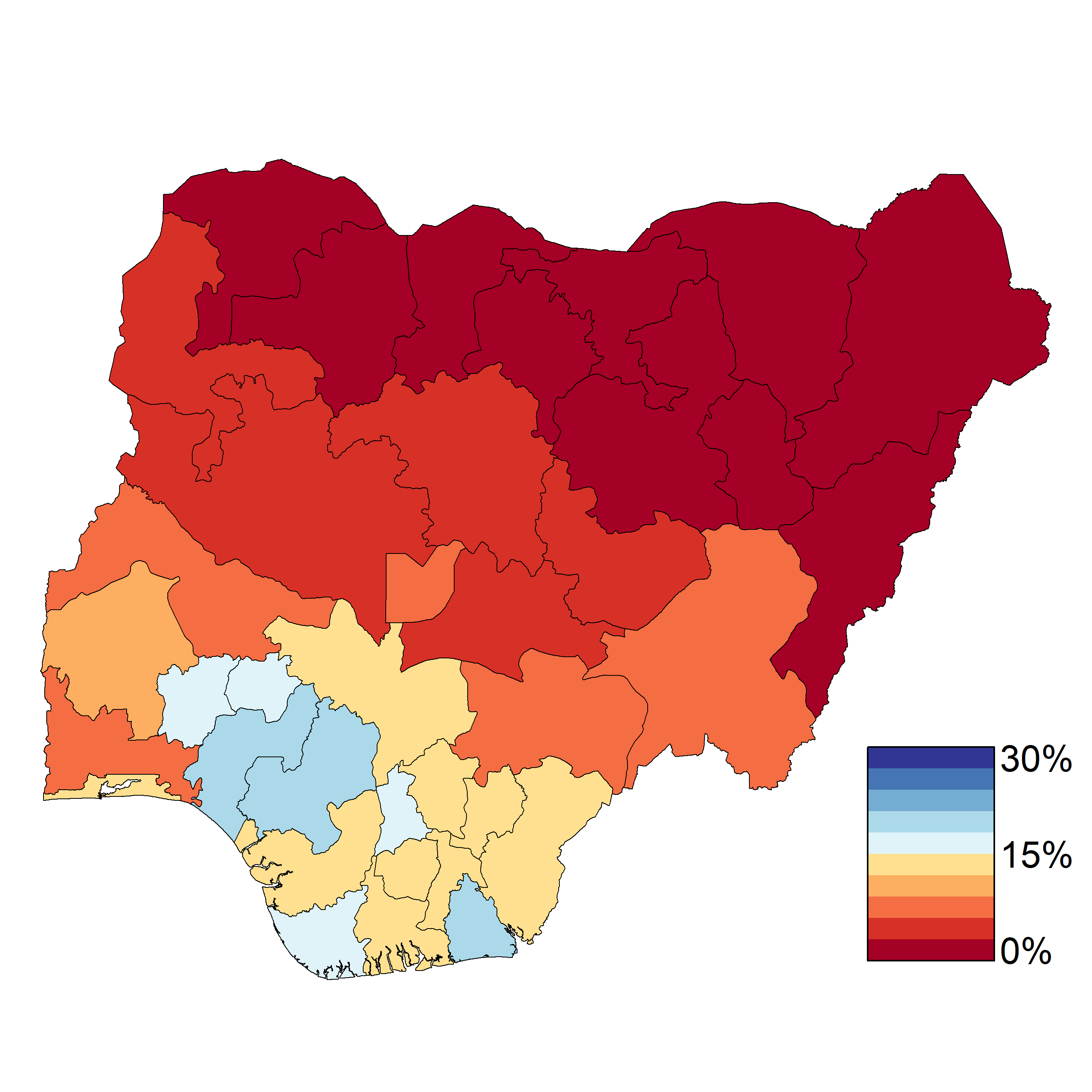}} 
   & \subfloat[Age 15-24 years and parity 1+.]{\includegraphics[width=9cm]{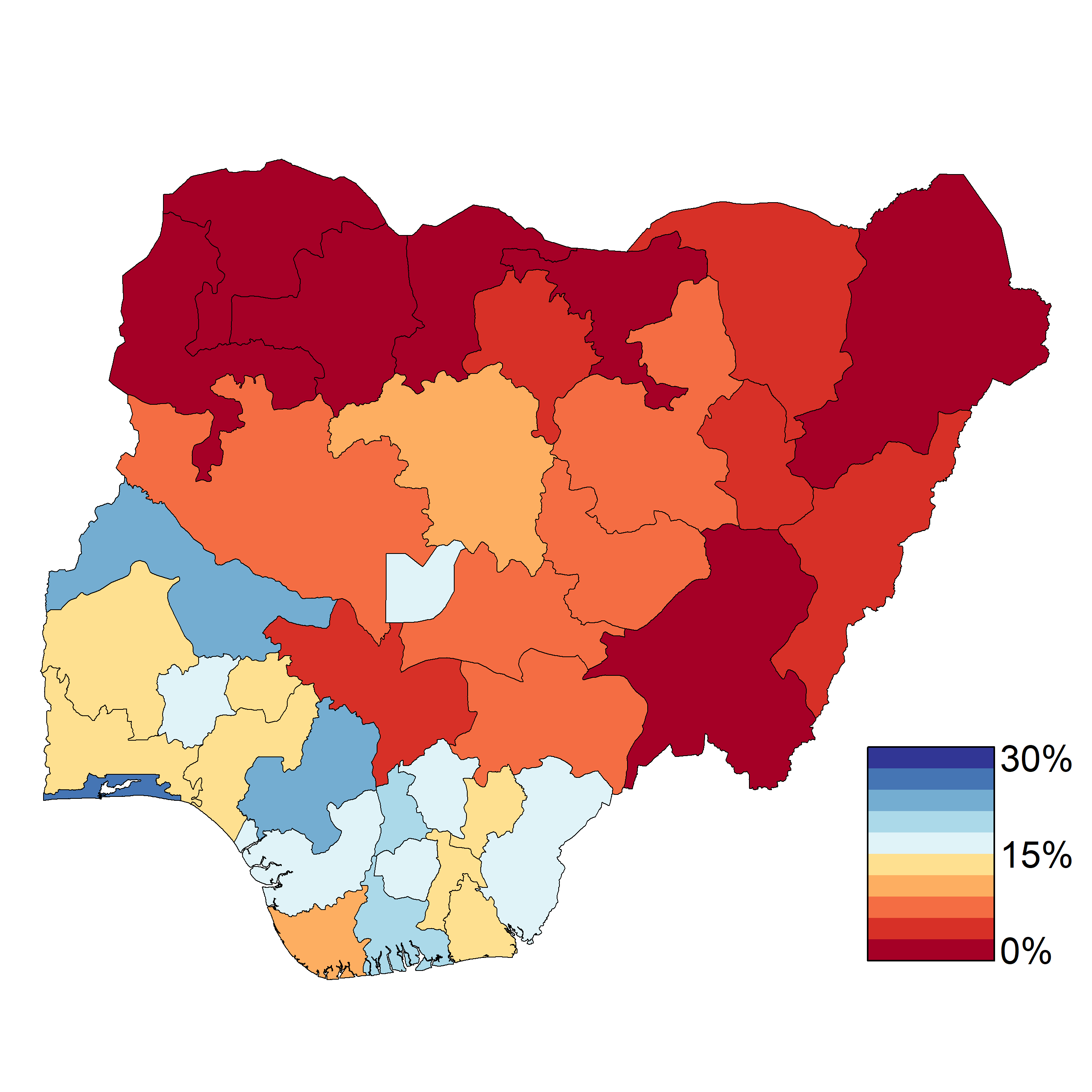}}\\
\subfloat[Age 25-49 years and parity 0.]{\includegraphics[width=9cm]{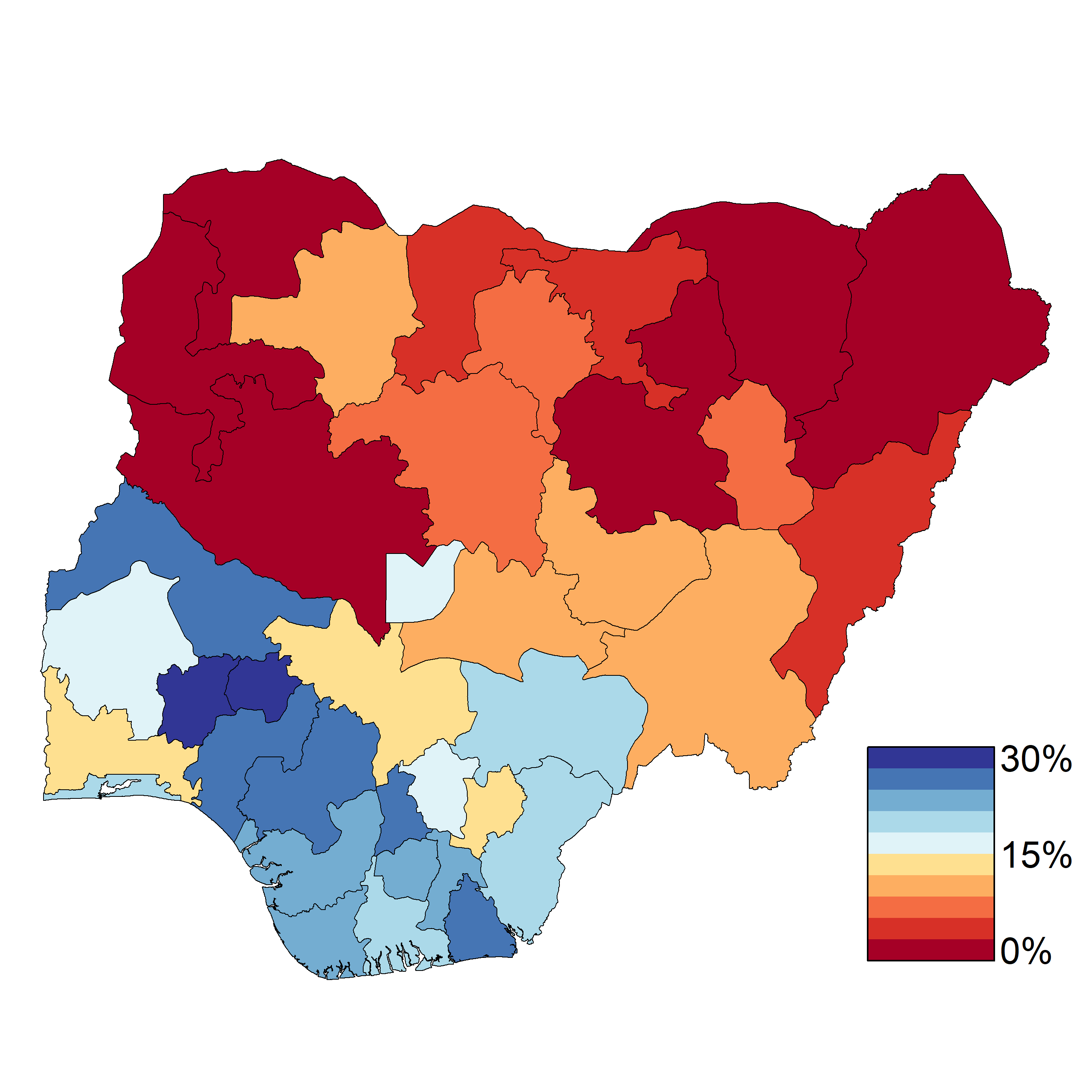}} 
   & \subfloat[Age 25-49 years and parity 1+.]{\includegraphics[width=9cm]{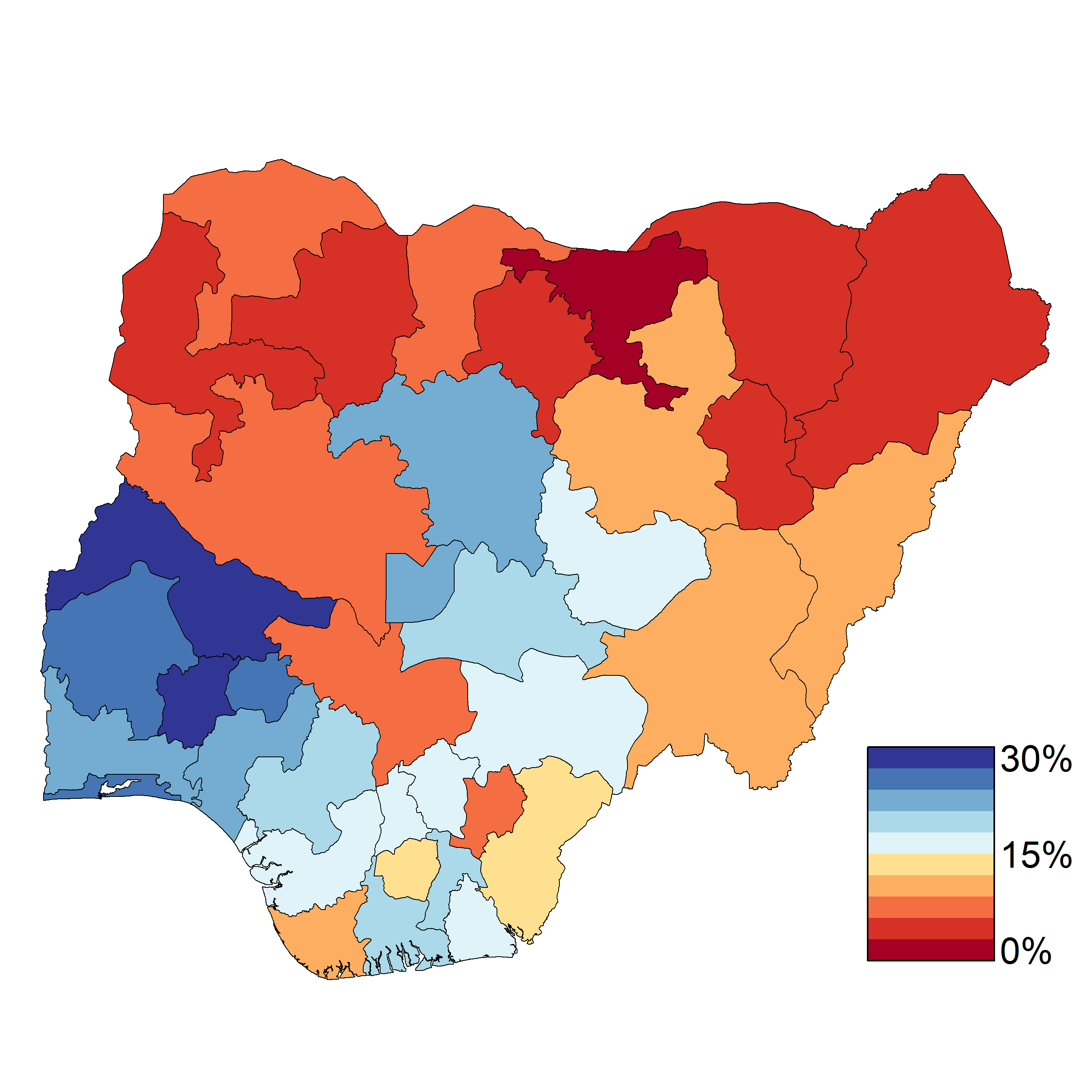}}\\
\end{tabular}

\caption{Smoothed estimates of mCPR by age and parity group for 2017.}\label{fig:mcpr_map}
\end{figure*}

\begin{figure*}[t]
	\centering
\def\tabularxcolumn#1{m{#1}}

\begin{tabular}{cc}
\subfloat[Age 15-24 years and parity 0.]{\includegraphics[width=9cm]{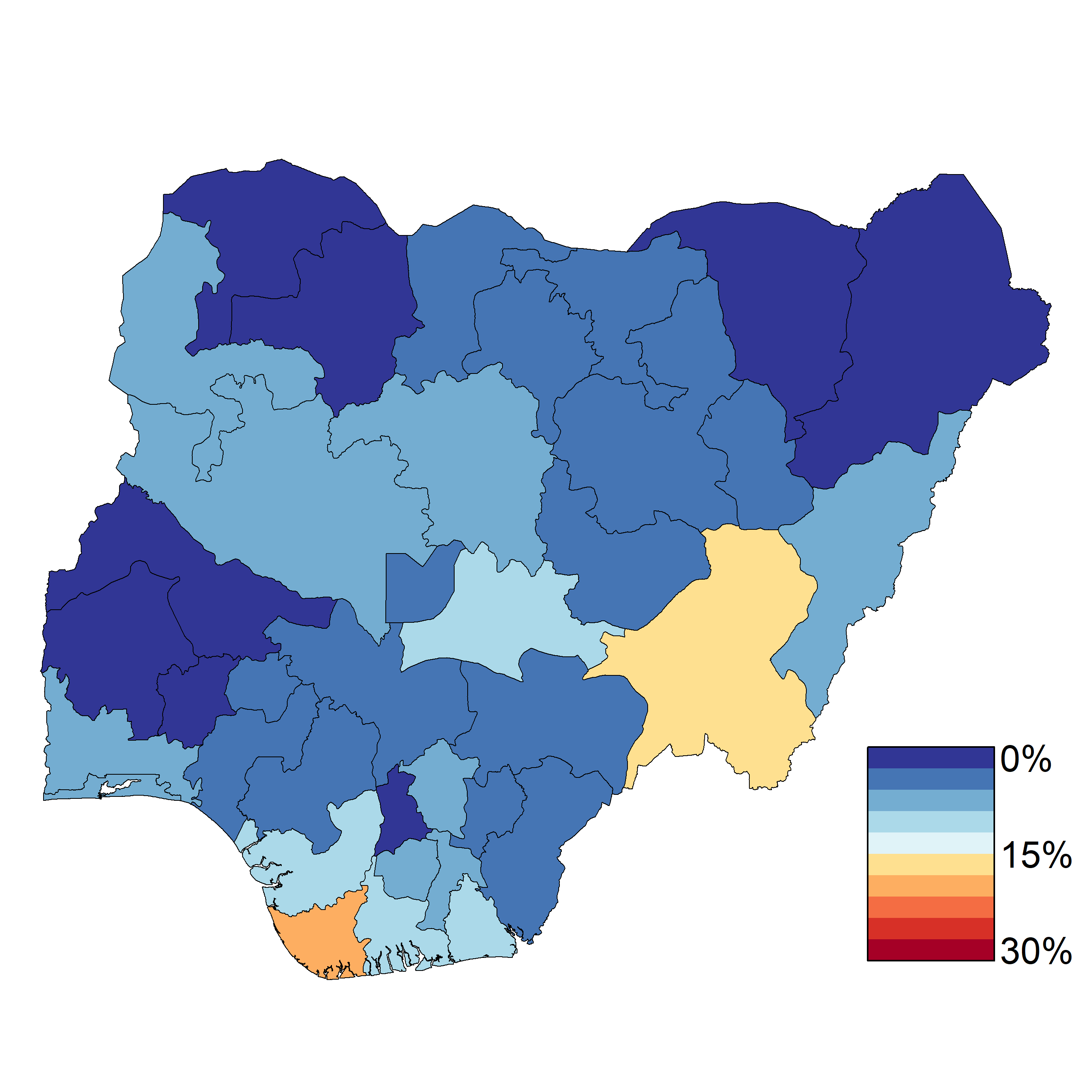}} 
   & \subfloat[Age 15-24 years and parity 1+.]{\includegraphics[width=9cm]{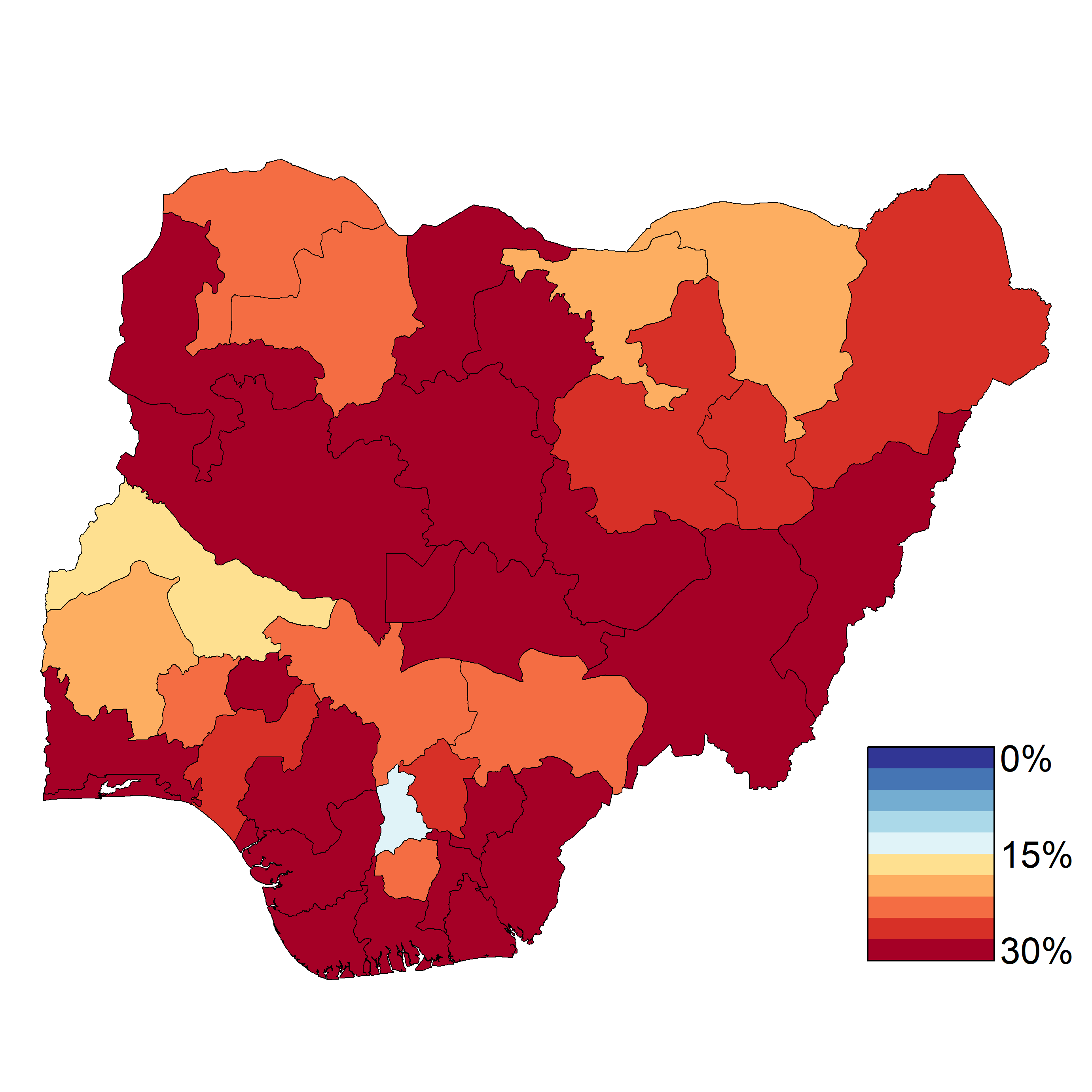}}\\
\subfloat[Age 25-49 years and parity 0.]{\includegraphics[width=9cm]{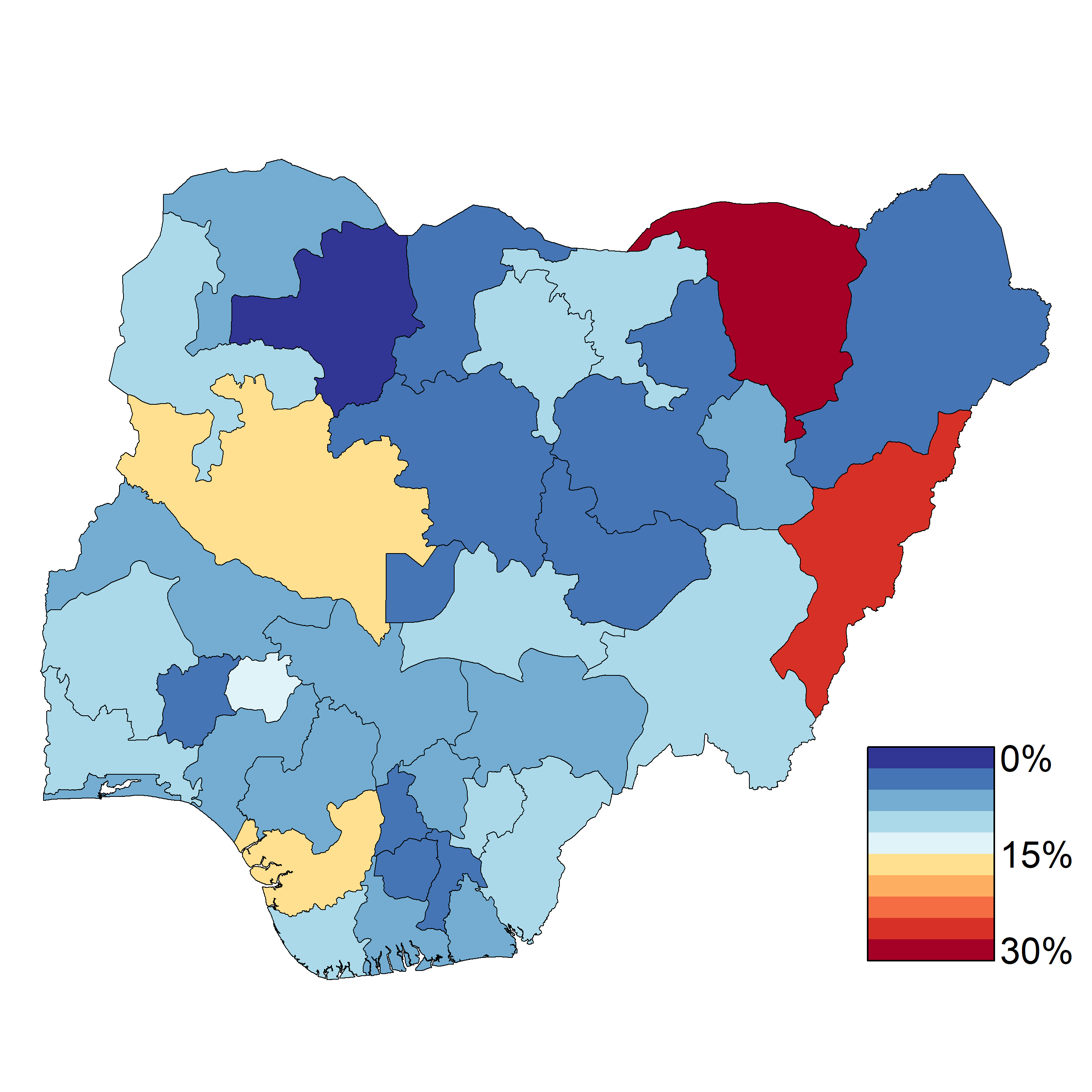}} 
   & \subfloat[Age 25-49 years and parity 1+.]{\includegraphics[width=9cm]{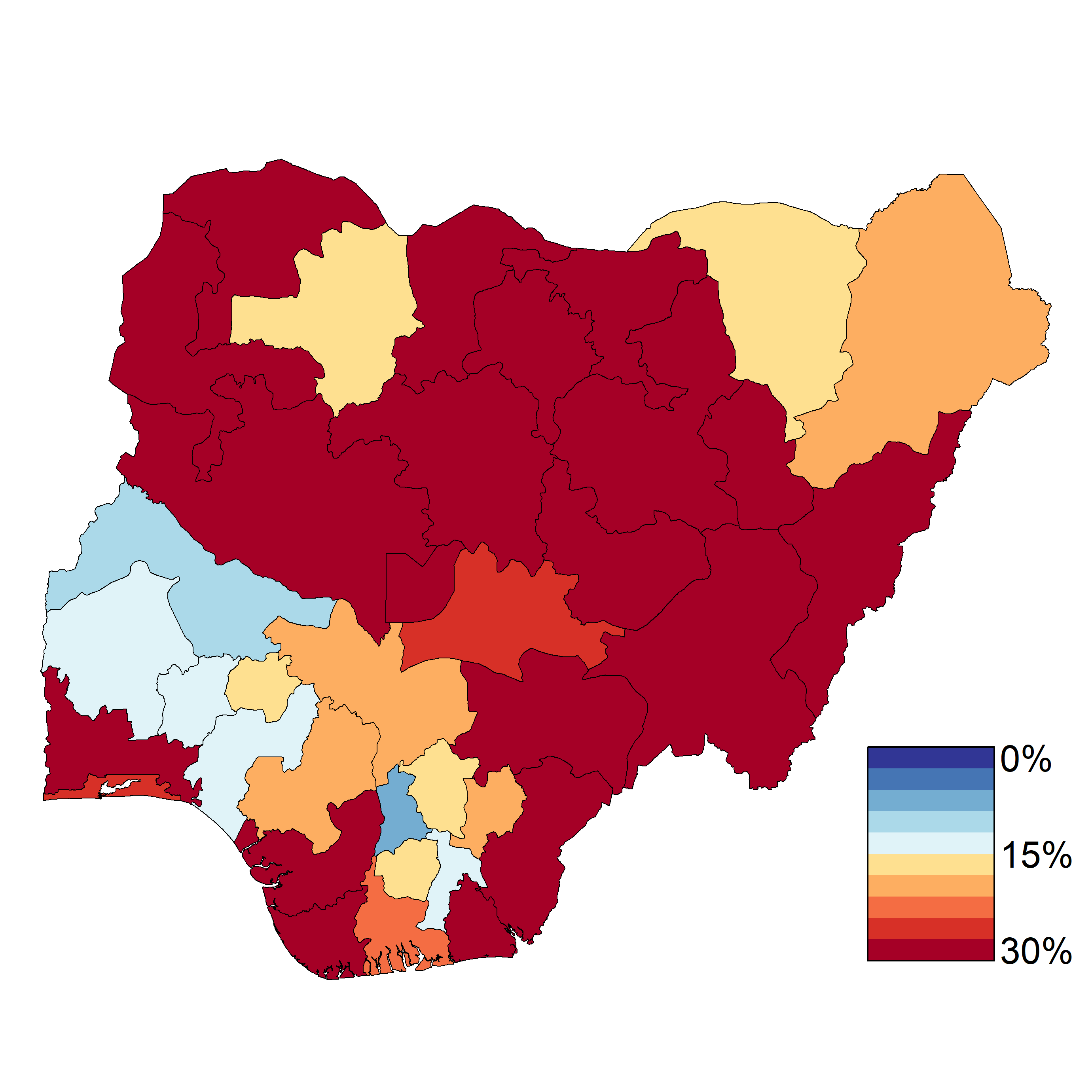}}\\
\end{tabular}

\caption{Smoothed estimates of unmet need by age and parity group for 2017.}\label{fig:unmet_map}
\end{figure*}

\begin{figure*}[t]
	\centering
\def\tabularxcolumn#1{m{#1}}

\begin{tabular}{cc}
\subfloat[Age 15-24 years and parity 0.]{\includegraphics[width=9cm]{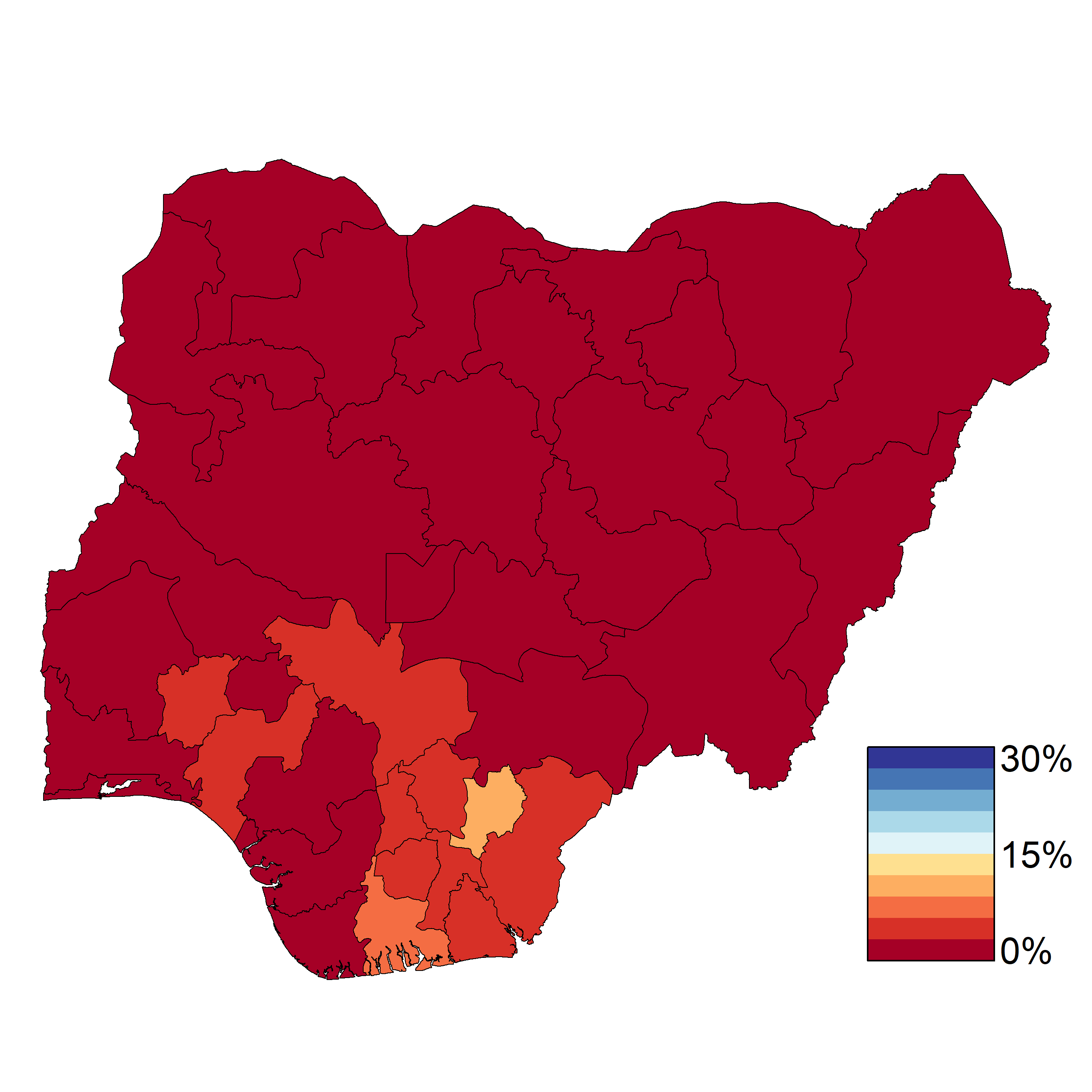}} 
   & \subfloat[Age 15-24 years and parity 1+.]{\includegraphics[width=9cm]{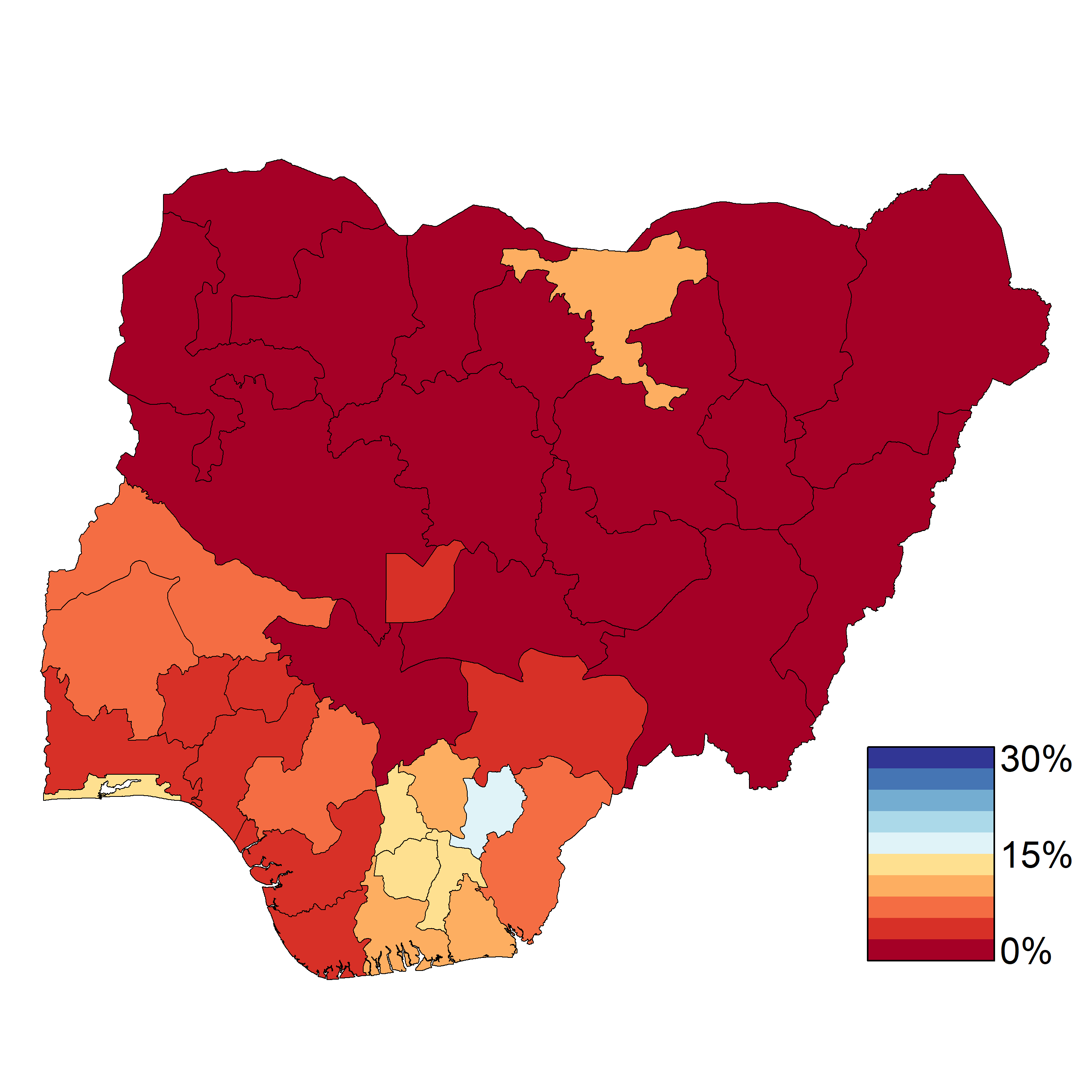}}\\
\subfloat[Age 25-49 years and parity 0.]{\includegraphics[width=9cm]{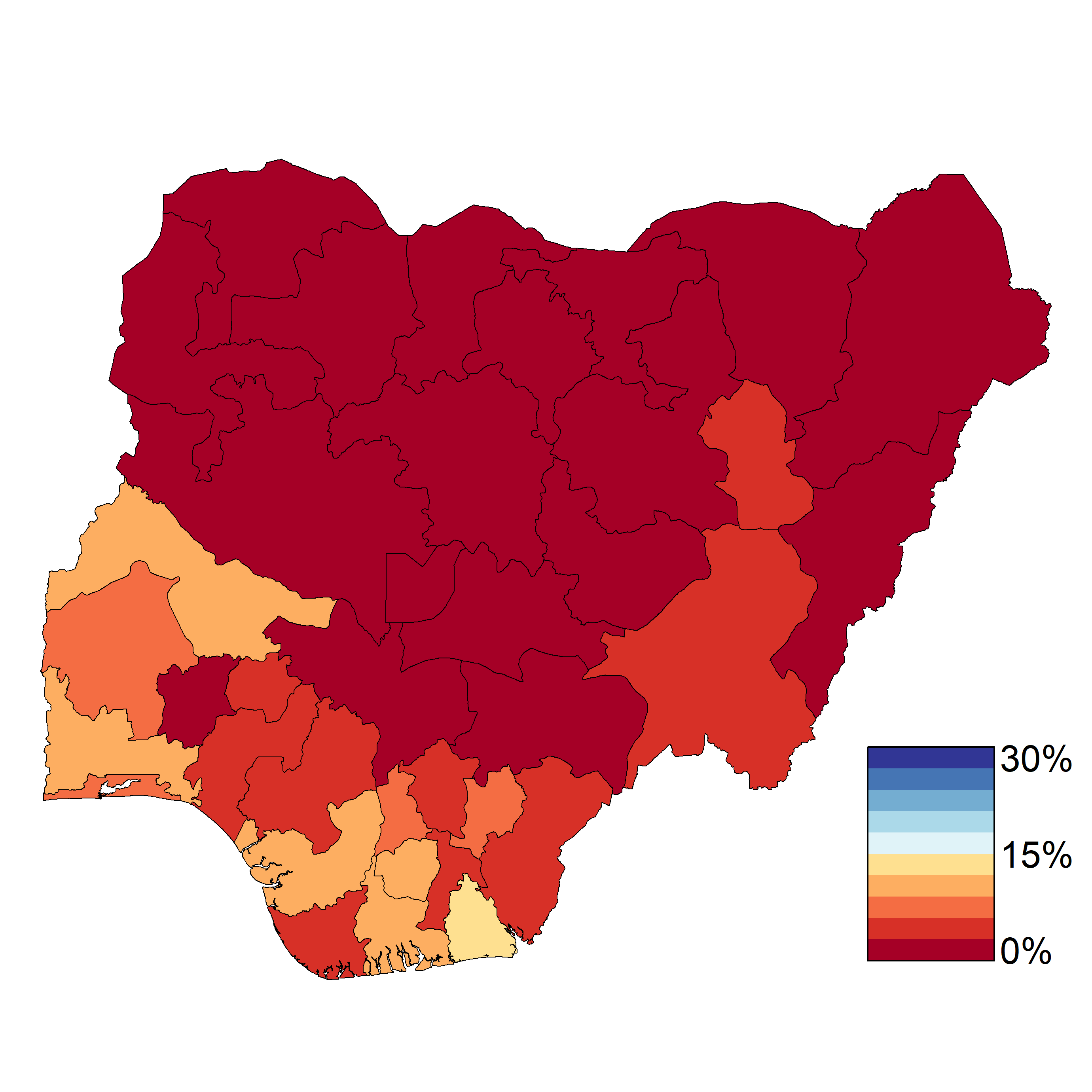}} 
   & \subfloat[Age 25-49 years and parity 1+.]{\includegraphics[width=9cm]{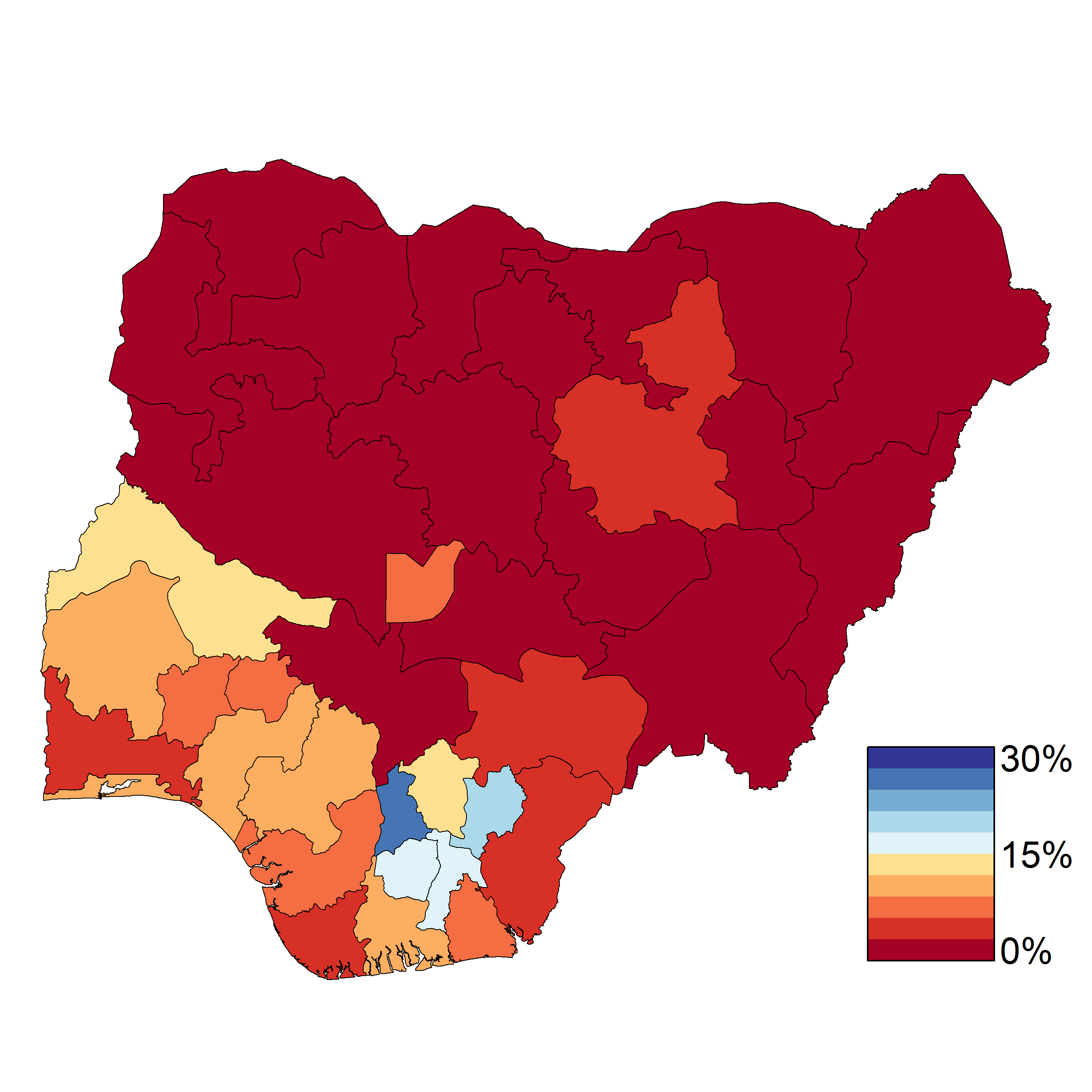}}\\
\end{tabular}

\caption{Smoothed estimates of traditional CPR by age and parity group for 2017.}\label{fig:trad_map}
\end{figure*}

\begin{figure*}[t]
	\centering
\def\tabularxcolumn#1{m{#1}}

\begin{tabular}{cc}
\subfloat[Age 15-24 years and parity 0.]{\includegraphics[width=9cm]{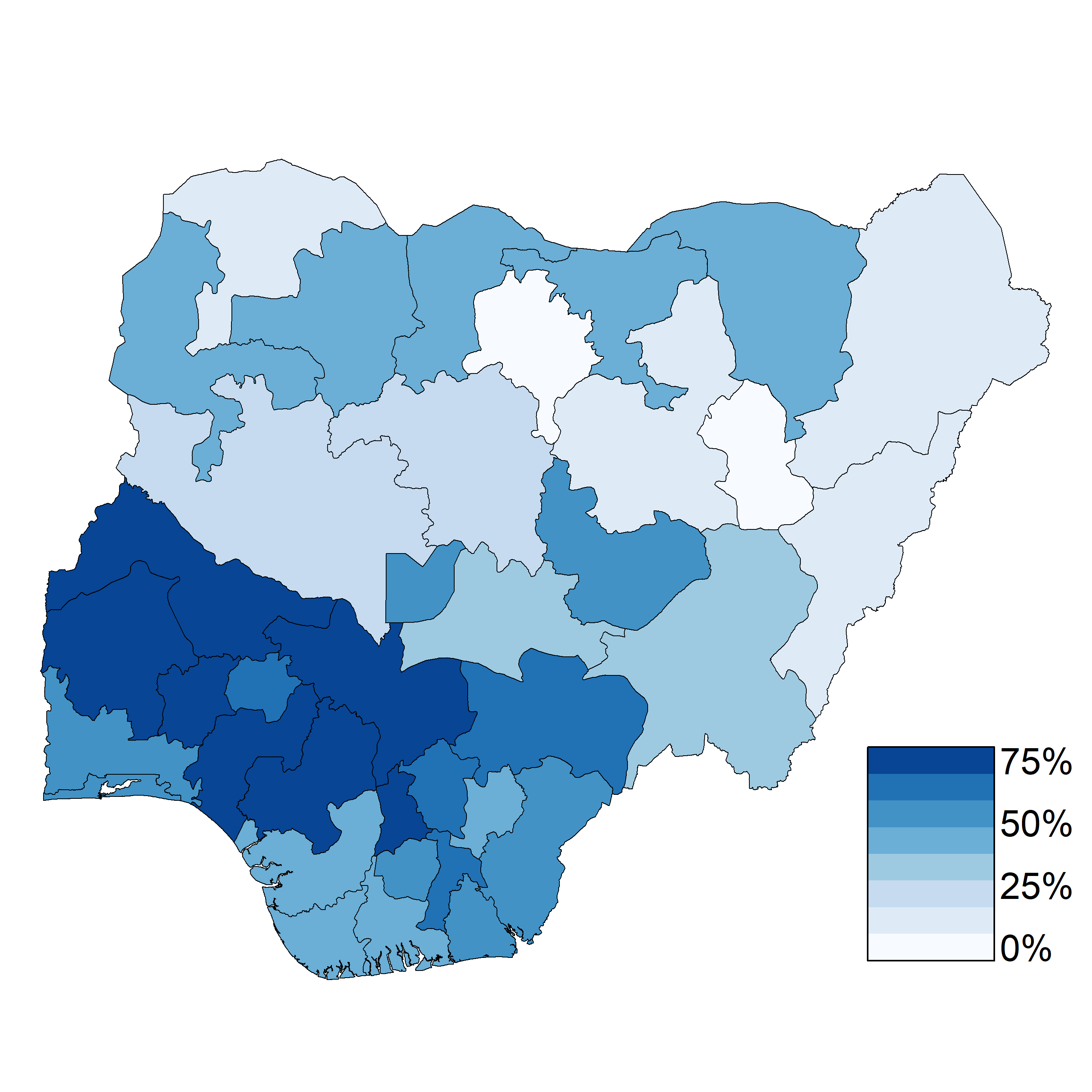}} 
   & \subfloat[Age 15-24 years and parity 1+.]{\includegraphics[width=9cm]{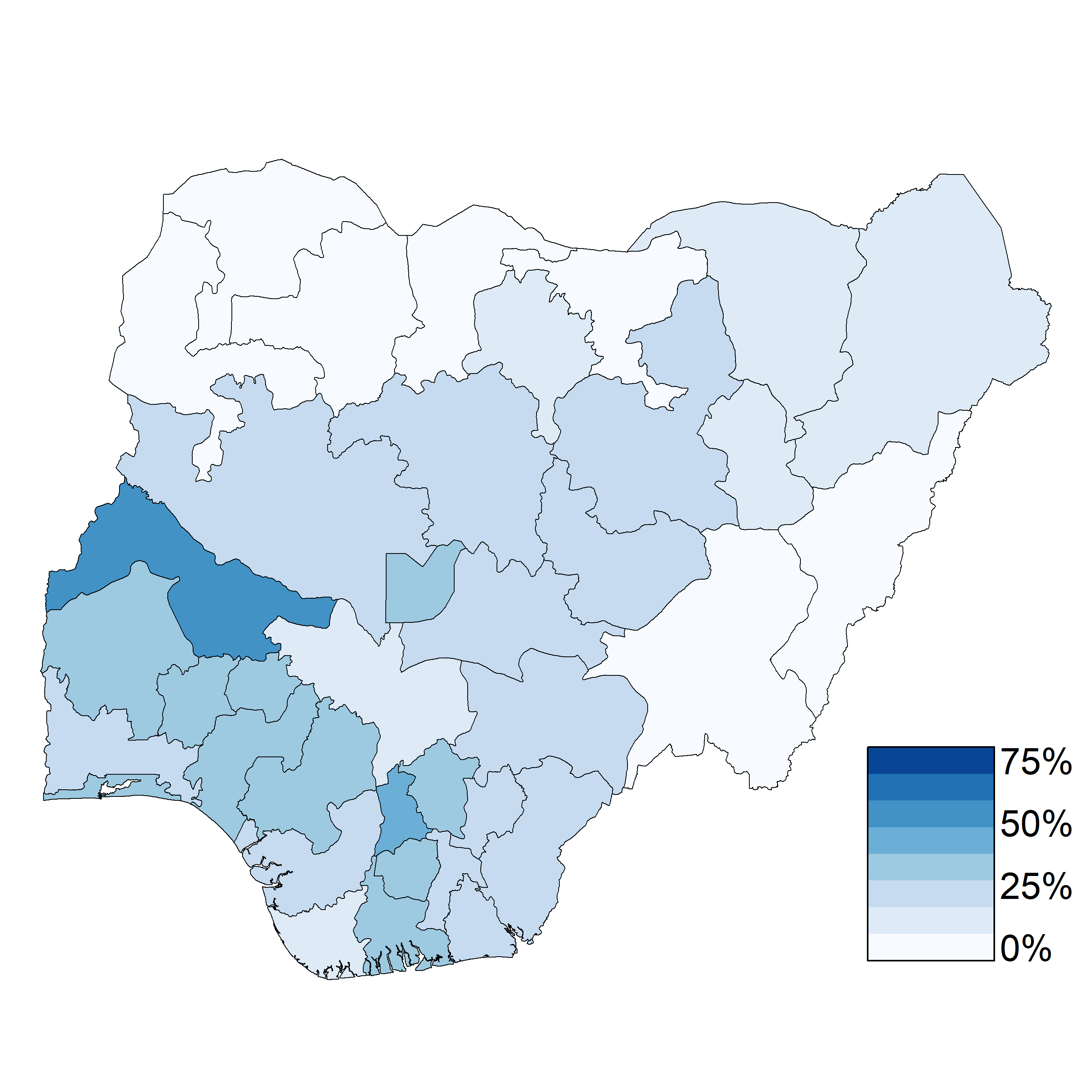}}\\
\subfloat[Age 25-49 years and parity 0.]{\includegraphics[width=9cm]{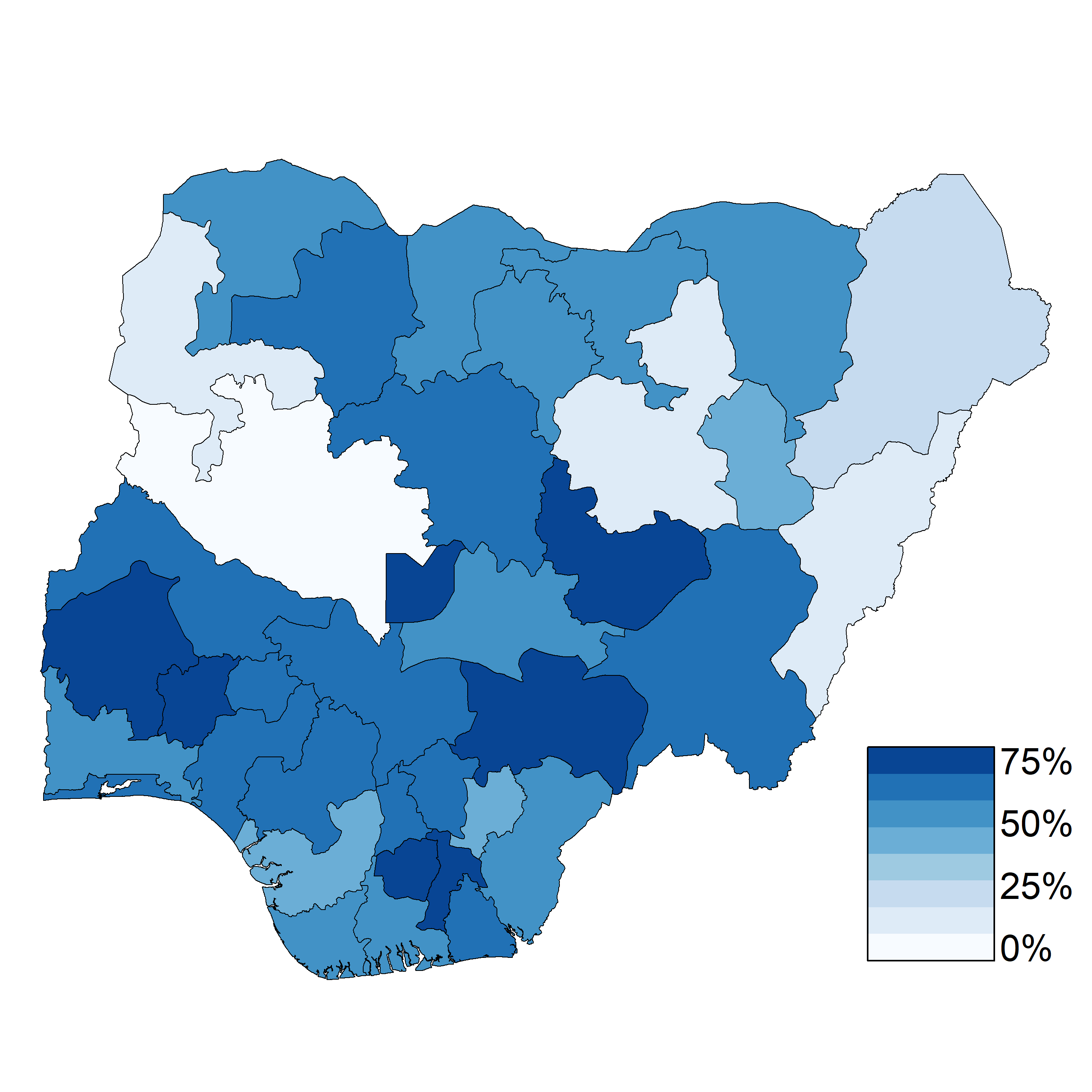}} 
   & \subfloat[Age 25-49 years and parity 1+.]{\includegraphics[width=9cm]{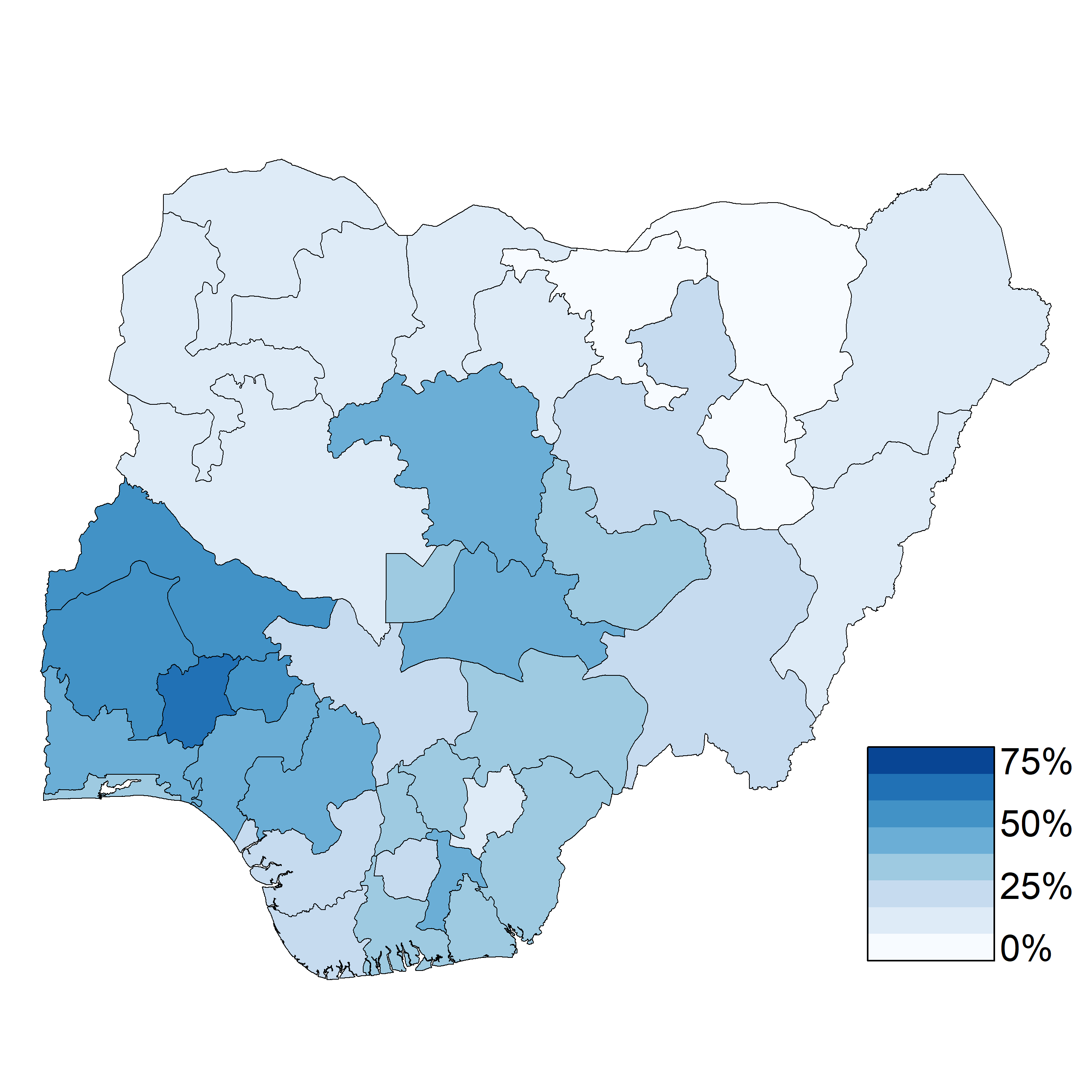}}\\
\end{tabular}

\caption{Smoothed estimates of demand satisfied by age and parity group for 2017.}\label{fig:ds_map}
\end{figure*}

Figures \ref{fig:mcpr_map}, \ref{fig:unmet_map}, \ref{fig:trad_map}, and \ref{fig:ds_map} show the 2017 state-level estimates  or the four age-parity subgroups for mCPR, unmet need, traditional contraceptive prevalence, and demand satisfied, respectively.

\section{Small area estimation at the second administrative level}


The health policies are often enacted at the second adminstrative level, called local government areas (LGAs) in Nigeria, and ideally we would provide estimates at the LGA-level.  Unfortunately, MICS, DHS, PMA2020, and NNHS do not inlcude the LGA locations for the sampled clusters.  However, the DHS have provided jittered~\cite{burgert2011incorporating} GPS locations which can be assigned LGA membership.  In this section we investigate and compare the precision attained at the state- and LGA-level.  We apply the methods described in Section \ref{s:model}, removing the survey-related random effects, to the DHS data from 1990. 2003, 2008, and 2013.

\begin{figure*}[t]
	\centering
	\includegraphics[width=1.0\textwidth]{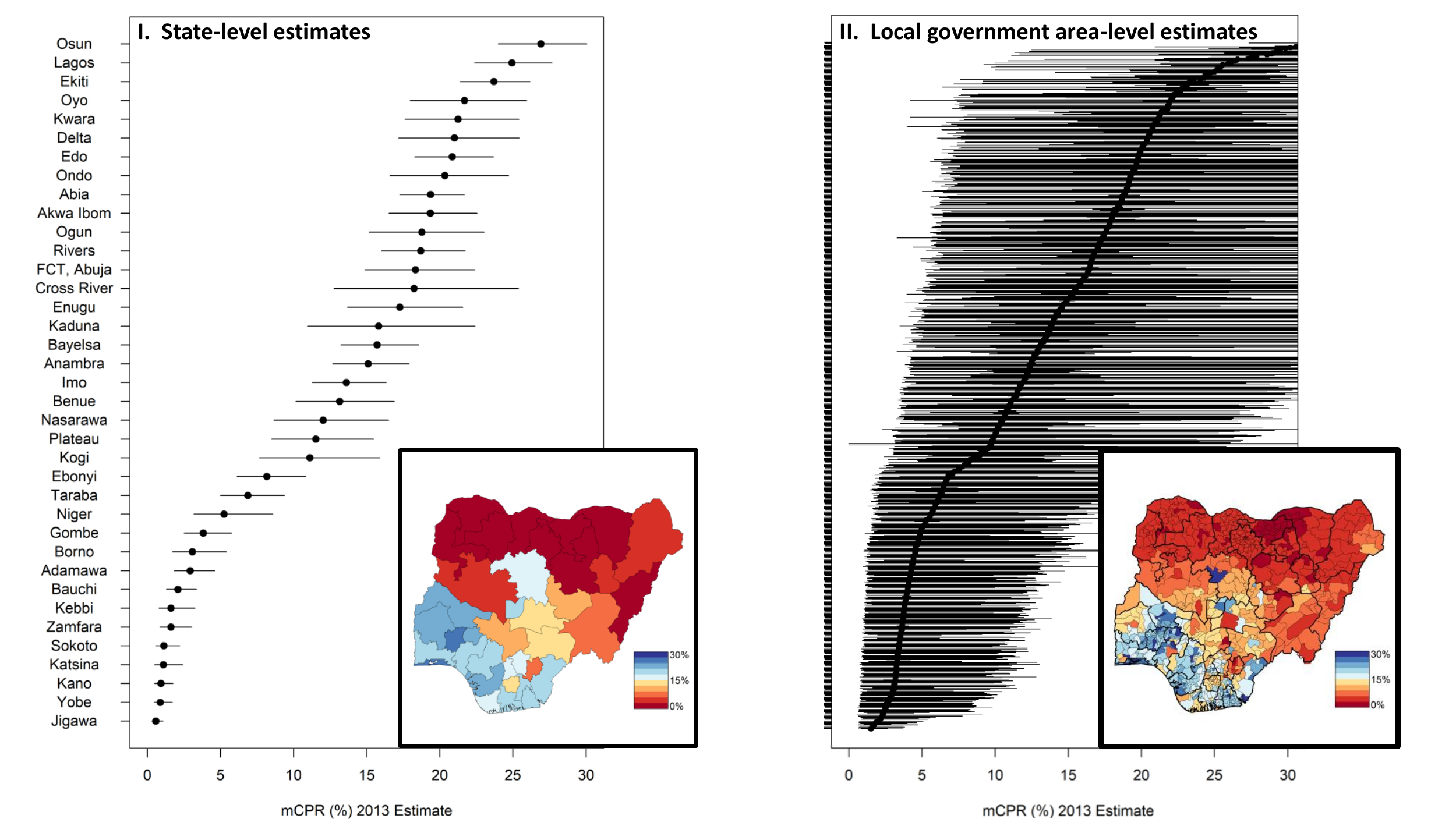}
	\vspace*{-.3in}
	\caption{Maps and ranked posterior medians and 95\% credible intervals for 2013 at the I. State- and II. Local goverment area-level.}
	\label{fig:lgaEstimates}
\end{figure*}

The left panel of Figure~\ref{fig:lgaEstimates} display the state-level estimates and 95\% credible intervals for mCPR and the right panel displays the same, but at the LGA-level. 
From the figure it is clear that in many of the LGAs the uncertainty is intolerably large. 
%
This is not a surprising result, as many LGAs have few or no sampled clusters.
By contrast differences between states and within states over time can be reliably estimated; see Figure~\ref{fig:lgaEstimates} in the Appendix and Figure $4$ in the main article, respectively. 
With currently available data it is not possible to generate sub-state level estimates for FP indicators with reasonable precision.
FP estimates at the LGA rely exclusively on DHS data and are thus not timely nor precise. 



\end{document}